\documentclass{aa}  
\usepackage{graphicx}
\usepackage{txfonts}
\usepackage{natbib,twoopt}
\usepackage{xcolor}
\usepackage{threeparttable}  
\usepackage{booktabs}

\bibpunct{(}{)}{;}{a}{}{,}           
\definecolor{xlinkcolor}{cmyk}{1,1,0,0}
\usepackage[bookmarks=true,
    pdfnewwindow=true,
   colorlinks=true,
  linkcolor=xlinkcolor,
 citecolor=xlinkcolor,
filecolor=xlinkcolor,
urlcolor=xlinkcolor,
final=true]{hyperref}

\def\Fig{\mbox{Figure~}}

\def\Tab{\mbox{Table~}}

\def\Sec{\mbox{Section~}}
\def\Secs{\mbox{Sections~}}
\def\Eq{\mbox{Equation~}}

\titlerunning{Dark progenitors and massive descendants}
\authorrunning{F. Gentile et al.}

\begin{document}

   \title{Dark progenitors and massive descendants:\\ A first ALMA perspective on Radio-Selected NIRdark galaxies in the COSMOS field}

   \author{Fabrizio Gentile
          \inst{1,2}
          \and
          Margherita Talia\inst{1,2}
          \and
          Emanuele Daddi\inst{3}
          \and 
          Marika Giulietti\inst{4,2}
          \and
          Andrea Lapi\inst{4,5,6}
          \and
          Marcella Massardi\inst{7,8,4}
          \and
          Francesca Pozzi\inst{1,2}
          \and
          Giovanni Zamorani\inst{2}
          \and
          Meriem Behiri\inst{4,2}
          \and
          Andrea Enia\inst{1,2}
          \and
          Matthieu Bethermin\inst{9,10}
          \and
          Daniele Dallacasa\inst{1,7}
          \and
          Ivan Delvecchio\inst{11}
          \and
          Andreas L. Faisst\inst{12}
          \and
          Carlotta Gruppioni\inst{2}
          \and
          Federica Loiacono\inst{1,2}
          \and
          Alberto Traina\inst{1,2}
          \and
          Mattia Vaccari\inst{13,14,7}
          \and
          Livia Vallini\inst{2}
          \and
          Cristian Vignali\inst{1,2}
          \and 
          Vernesa Smol{\v{c}}i{\'c}\inst{15}
          \and
          Andrea Cimatti\inst{1,2}
          }

   \institute{University of Bologna, Department of Physics and Astronomy (DIFA), Via Gobetti 93/2, I-40129, Bologna, Italy\\
              \email{fabrizio.gentile3@unibo.it}
         \and
             INAF -- Osservatorio di Astrofisica e Scienza dello Spazio, via Gobetti 93/3 - 40129, Bologna - Italy
        \and
            CEA, IRFU, DAp, AIM, Université Paris-Saclay, Université Paris Cité, Sorbonne Paris Cité, CNRS, 91191, Gif-sur-Yvette, France
        \and
            SISSA, Via Bonomea 265, I-34136 Trieste, Italy
        \and 
            IFPU - Institute for fundamental physics of the Universe, Via Beirut 2, 34014 Trieste, Italy
        \and
            INFN-Sezione di Trieste, via Valerio 2, 34127 Trieste,  Italy
        \and
            INAF/IRA, Istituto di Radioastronomia, Via Piero Gobetti 101, 40129 Bologna, Italy
        \and 
            INAF, Istituto di Radioastronomia - Italian ARC, Via Piero Gobetti 101, I-40129 Bologna, Italy
        \and
            Universit\'e de Strasbourg, CNRS, Observatoire astronomique de Strasbourg, UMR 7550, 67000 Strasbourg, France
        \and 
            Aix Marseille Univ, CNRS, CNES, LAM, Marseille, France
        \and
            INAF - Osservatorio Astronomico di Brera, via Brera 28, I-20121, Milano, Italy \& via Bianchi 46, I-23807, Merate, Italy
        \and 
            Caltech/IPAC, MS314-6, 1200 E. California Blvd. Pasadena, CA, 91125, USA
        \and
            Inter-University Institute for Data Intensive Astronomy, Department of Astronomy, University of Cape Town, 7701 Rondebosch, Cape Town, South Africa
        \and 
            Inter-University Institute for Data Intensive Astronomy, Department of Physics and Astronomy, University of the Western Cape, 7535 Bellville, Cape Town, South Africa
        \and
            Department of Physics, University of Zagreb, Bijenicka cesta 32, 10002 Zagreb, Croatia}

   \date{Received ?? ; accepted ??}
 
  \abstract
  {We present the first spectroscopic ALMA follow-up for a pilot sample of nine Radio-Selected NIRdark galaxies in the COSMOS field. These sources were initially selected as radio-detected sources ($S_{\rm 3GHz}>12.65\mu$Jy) lacking an optical/NIR counterpart in the COSMOS2015 catalog ($Ks\gtrsim24.7$ mag), {with just three of them subsequently detected in the deeper COSMOS2020}. Several studies highlighted how this selection could provide a population of highly dust-obscured, massive and star-bursting galaxies. With these new ALMA observations, we assess the spectroscopic redshifts of this pilot sample of sources and improve the quality of the physical properties estimated through SED-fitting. Moreover, we measure the quantity of molecular gas present inside these galaxies and forecast their potential evolutionary path, finding that the RS-NIRdark galaxies could represent a likely population of high-\textit{z} progenitors of the massive and passive galaxies discovered at $z\sim3$. Finally, we present some initial constraints on the kinematics of the ISM within the analysed galaxies, reporting a high fraction ($\sim55\%$) of double-peaked lines that can be interpreted as the signature of a rotating structure in our targets or with the presence of major mergers in our sample. Our results presented in this paper showcase the scientific potential of (sub)mm observations for this elusive population of galaxies and highlight the potential contribution of these sources in the evolution of the massive and passive galaxies at high-\textit{z}.}


   \maketitle

\section{Introduction}
\label{sec:intro}
One of the biggest puzzles of modern astronomy concerns the recent discovery of a significant population of massive ($M_\star\sim10^{11}M_\odot$) and passive (sSFR$<10^{-11}$ yr$^{-1}$) galaxies already in place at $z\sim3$ (see some examples in \citealt{Straatman_14,Schreiber_18,Valentino_20}). 

This result challenges our galaxy evolution models for two main reasons. On the one hand, these galaxies assembled most of their stellar mass at $z>3$, in a period of the cosmic time when – according to studies based at optical/NIR wavelengths (see \citealt{Madau_14} and references therein) – the cosmic Star Formation Rate Density (SFRD; i.e. the average amount of stellar mass created in the Universe per each year and each comoving Mpc$^3$) was at least one order of magnitude lower than at the so-called \textit{cosmic noon} ($z\sim3$). On the other hand, looking at higher redshifts, we should detect a significant population of massive and star-forming galaxies on the way to quench their star-formation (the “\textit{progenitors}” of these massive and passive galaxies; see e.g. \citealt{Valentino_20}). However, when looking at the galaxies selected at these redshifts in the optical/NIR regimes (mostly Lyman Break Galaxies; LBGs, see \citealt{Giavalisco_02} and references therein), we see a population of galaxies with stellar masses and star formation rates too low to be likely progenitors of these sources. In addition, the number density of LBGs at $z>3$ is generally found to be one or two orders of magnitude lower than the massive and passive galaxies at $z=3$ \citep[e.g.][]{Stark_09,Toft_14,Valentino_20}.

Before invoking a dramatic change in our galaxy evolution models, we must exclude the presence of any bias in our samples of high-\textit{z} galaxies. A possible issue -- for instance -- could reside in the wavelength in which we select these sources. In the last decades, several studies \citep[e.g.][]{Smail_99,Smail_02,Frayer_04, Simpson_14,Franco_18,Wang_19,Gruppioni_20,Smail_21,Talia_21,Manning_22,Enia_22,Behiri_23} unveiled the existence of a significant population of "dark" galaxies constantly missed by optical/NIR surveys: the so-called Dusty Star Forming Galaxies (DSFGs; see e.g. the review by \citealt{casey_14}). These sources are characterized by significant amounts of dust in their interior, making them extremely faint (or even undetected) at short wavelengths.

Several studies selecting these sources at longer wavelengths (i.e. where the effect of dust is negligible or where we can take advantage of its bright thermal emission, mainly in the FIR/(sub)mm), assessed how they could represent a population of massive and star-forming galaxies, with estimated number densities comparable with those reported for the massive and passive galaxies at $z\sim3$ \citep[e.g.][]{Toft_14,Valentino_20,Talia_21,Behiri_23}. Moreover, the inclusion of these sources in the cosmic census of high-\textit{z} galaxies could be significant enough to change the behavior of the cosmic SFRD at $z>3$ \citep[see e.g.][]{RowanRobinson_16,Gruppioni_20,Talia_21,Behiri_23,Traina_23}.

The main drawback of these studies, when performed at FIR/sub(mm) wavelengths, resides in the size of the analysed samples. When employing old-generation instruments such as the SCUBA camera equipped on the \textit{James Clerk Maxwell Telescope} \citep[e.g.][]{Smail_97,Hughes_98,Dunlop_04} or the PACS and SPIRE cameras on the \textit{Herschel Space Observatory} \citep[e.g.][]{Gruppioni_13,Burgarella_13}, the low sensitivity of these instruments biases these samples towards the most extreme objects. In addition, their coarse resolution makes it difficult to associate the right multi-wavelength counterpart to the FIR emission without a high-resolution follow-up \citep[see e.g.][]{Simpson_19,Simpson_20,Stach_19,Dudzevicute_21}. 

All these issues could be solved, in principle, with the employment of state-of-the-art facilities observing at these wavelengths such as the \textit{Atacama Large (sub)Millimeter Array} (ALMA) or the \textit{Northern Extended Millimeter Array} (NOEMA), with higher sensitivity and better spatial resolution. Nevertheless, these instruments are not designed to perform wide blind surveys: their small field of view makes incredibly time-consuming observing statistically significant volumes of the Universe \citep[see some noteworthy examples in][]{Dunlop_17,Franco_18,Casey_21}.

A possible solution to all these problems could reside in a radio selection. Since radio photons can be generated by free-free emission in HII regions and synchrotron emission from relativistic electrons accelerated in supernovae remnants, they represent a dust unbiased tracer of star formation (see e.g. \citealt{Kennicutt_12} and references therein). {As shown for the first time by \citet{Chapman_01}, the selection of faint radio sources lacking an optical/NIR counterpart can provide a sample of likely DSFGs \citep{Chapman_02,Chapman_04}, but with a selection taking advantage of the large FOVs and optimal spatial resolution of modern radio interferometers. {Moreover, the high sensitivity reached by deep radio surveys can unveil samples of galaxies with less-extreme properties than what is commonly obtained through FIR/(sub)mm selections (see e.g. \citealt{Chapman_02,Chapman_04,Talia_21,Behiri_23,Gentile_24}}). 

Nevertheless, the drawback of this radio selection is represented by the possible contribution by AGN, since nuclear activity can also produce radio emission. This issue can be partly solved with a multi-wavelength approach, by focusing on extra-galactic fields where a broad photometric coverage is available and -- therefore -- where one can look for the characteristic signatures of AGN at other frequencies than radio (see e.g. the discussions in \citealt{Enia_22} and \citealt{Gentile_24}; see also the review by \citealt{Hickox_18} and references therein).

Focusing on deep radio surveys and requiring the lack of counterparts in deeper NIR surveys (than what initially employed by \citealt{Chapman_01}; see e.g. \citealt{Talia_21}, \citealt{Enia_22}, \citealt{Behiri_23}, \citealt{VanDerVlugt_23}, and \citealt{Gentile_24}), we can collect the so-called Radio-Selected NIRdark galaxies\footnote{\url{https://sites.google.com/inaf.it/rsnirdark/}} (RS-NIRdark galaxies hereafter).} The first studies analyzing these sources (\citealt{Talia_21,Enia_22,Behiri_23,Gentile_24}) reported a series of interesting results:
\begin{itemize}
    \item The RS-NIRdark galaxies represent a population of highly dust-obscured ($A_{\rm v}\sim 4$ mag), massive ($M_\star\sim10^{11}M_\odot$) and star-forming (SFR$\sim 500$ M$_\odot$ yr$^{-1}$) galaxies. The bulk of the population is located at $z\sim3$ and there is a significant tail of high-\textit{z} sources at $z>4.5$.
    \item When compared with other galaxies in the same redshift ranges, the RS-NIRdark galaxies always lie above the main sequence of star-forming galaxies, with nearly half of them in the starburst regime.
    \item Their number density at $z>3.5$ (not corrected for incompleteness or for the expected duty cycle) is higher than n=($3.3\pm0.9)\times10^{-6}$ Mpc$^{-3}$, just in moderate tension with that reported by \citet{Straatman_14}, \citet{Schreiber_18}, and \citet{Valentino_20} for the massive and passive galaxies at $z\sim3$.
    \item Their contribution to the cosmic SFRD at $z>4.5$ could be as high as 20-40\% of that obtained only considering optically/NIR selected galaxies.
\end{itemize}

If, on the one side, these results increase the scientific potential of the RS-NIRdark galaxies, they make new questions to arise. Firstly, all these results are based on photometric redshifts and SED-fitting. Since the photometry of these sources is mostly constrained by upper limits in the optical/NIR regimes, a spectroscopic confirmation of the redshift is necessary to decrease the degeneracies affecting the physical properties estimated through this procedure. Secondly, if the number density of the RS-NIRdark galaxies is compatible with the passive galaxies at $z\sim3$, we need to constrain their evolutionary path to establish a possible relation between the progenitors and the descendants. Finally, we must explain their location in the SFR-stellar mass plane and unveil the physical processes taking place in their ISM responsible for such an intense star formation.

Clearly, most of these questions cannot be addressed just relying on the existing photometry already analysed in the previous studies on these sources: we need to collect more data. Given the elusive nature of the RS-NIRdark galaxies, our choice is limited to the new facilities observing at longer wavelengths: ALMA, NOEMA, and the \textit{James Webb Space Telescope} (JWST). {As noted by several previous studies, (sub)mm observations can be incredibly useful in assessing the true nature of "dark" galaxies selected at other wavelengths, allowing us to constrain their dust and gas content \citep[e.g.][]{Chapman_01,Chapman_02,Chapman_04} as well as prove their obscured star formation \citep[e.g.][]{Wang_19}.}

This paper is focused on the first ALMA follow-up of a pilot sample of nine RS-NIRdark galaxies selected in the COSMOS field by \citet{Talia_21} and analysed by \citet{Behiri_23} and \citet{Gentile_24}. The first results involving an accepted NOEMA follow-up and the first JWST data obtained thanks to the COSMOS-Web survey \citep{Casey_22} will be described in following papers (Gentile et al., in prep.).

This study follows this structure. In \Sec\ref{sec:data} we introduce our targets, the ancillary photometry already available for them, and the new ALMA observations. In \Sec\ref{sec:methods}, we describe the analysis of the ALMA cubes, the identification of any bright emission line in our targets, and our modeling of the spectroscopic redshifts. Moreover, we present some initial insights on the ISM kinematics and derive the physical properties of our pilot sample of galaxies through SED-fitting. Then, in \Sec\ref{sec:results}, we discuss our results, estimate the gas mass within our sources, and forecast a possible evolutionary path for them. Finally, we draw our conclusions in \Sec\ref{sec:conclusions}. Throughout this paper, we assume a \citet{Chabrier_03} Initial Mass Function (IMF) and a flat $\Lambda$CDM cosmology with $[\Omega_m,\Omega_\Lambda,h]=[0.3,0.7,0.7]$.

\section{Data}
\label{sec:data}

\begin{table*}
\centering
\begin{threeparttable}
\renewcommand{\arraystretch}{1.1}
\centering
\caption{Main observational properties of the nine targets analysed in this work. For each galaxy, we report the ID employed in \citet{Gentile_24} and throughout this paper. For completeness, we also report the original ID employed by \citet{Talia_21} that can be used to retrieve the observations from the ALMA science archive. Finally, we report the coordinates {(of the radio counterpart, i.e. that with the higher spatial resolution)} and a flag signaling if each source  has a counterpart or not in the COSMOS2020 catalog \citep{Weaver_22}. }
\label{tab:coords}
\begin{tabular}{ccccccc}
\toprule
ID & ID (Talia+21) & RA & Dec & C20 & Ks & IRAC Ch 2\\
 &  & [\textit{hh:mm:ss}] & [\textit{dd:mm:ss}] & & [mag] & [mag] \\
\midrule
RSN-41 & COSMOSVLA3-49 & 09:58:17.869 & +02:30:38.305 & - & $>25.7$ & $22.47\pm0.04$\\
RSN-84 & COSMOSVLA3-106 & 09:58:43.440 & +02:45:18.135 & - & $>25.8$ & $23.21\pm0.07$ \\
RSN-121 & COSMOSVLA3-152 & 09:59:14.234 & +02:35:26.432 & \checkmark & $25.0\pm0.3$ & $22.54\pm0.02$ \\
RSN-182 & COSMOSVLA3-225 & 09:59:46.699 & +02:48:41.215 & - & $>25.0$ & $23.29\pm0.05$ \\
RSN-235 & COSMOSVLA3-291 & 10:00:09.550 & +01:42:50.834 & - & $>26.2$ & $23.40\pm0.04$ \\
RSN-247 & COSMOSVLA3-308 & 10:00:23.785 & +01:41:59.273 & \checkmark & $24.9\pm0.3$ & $23.02\pm0.03$ \\
RSN-298 & COSMOSVLA3-370 & 10:00:57.970 & +01:38:26.078 & - & $>25.7$ & $23.43\pm0.04$ \\
RSN-361 & COSMOSVLA3-442 &10:01:28.390 & +02:21:27.857 & \checkmark & $25.0\pm0.3$ & $23.84\pm0.07$ \\
RSN-456 & COSMOSVLA3-576 &10:02:48.219 & +02:24:30.629 & - & $>25.7$ & $22.97\pm0.06$ \\
\bottomrule
\end{tabular}
\end{threeparttable}
\end{table*}

\subsection{ALMA observations and data reduction}
\label{sec:alma}
The main focus of this study is on the observations carried out by ALMA during its cycle 8 as a part of the observing program 2021.1.01467.S (PI: M. Talia). The required observations consisted of a spectroscopic follow-up at mm wavelengths for a pilot sample of 9 RS-NIRdark galaxies (\Tab\ref{tab:coords}). These sources were initially selected in the COSMOS field by \citet{Talia_21} among those located in the high-\textit{z} tail of the redshift distribution (photo-$z>4.5$), with the best-sampled SEDs (i.e., with at least one significant detection at $S/N>3$ at FIR/(sub)mm wavelengths), and with a reliable SED-fitting. Following the selection by \citet{Talia_21}, these objects were part of a sample of 476 galaxies robustly detected ($S_{\rm 3GHz}>12.65 \mu$Jy; $S/N>5.5$) in the catalog of the VLA-COSMOS 3GHz Large Project \citep{Smolcic_17} and lacking an optical/NIR counterpart in the COSMOS2015 catalog \citep{Laigle_16}, i.e. the most recent NIR-selected catalog of the COSMOS field at the time of that study. As highlighted by \citet{Gentile_24}, the public release of the deeper COSMOS2020 catalog ($\sim 1$ mag deeper in the $Ks$ band {and with the detection operated in a more complete detection image; see} \citealt{Weaver_22}), allowed us to associate an optical/NIR counterpart for $\sim150$ sources analysed in \citet{Talia_21}. The targets with a counterpart in the COSMOS2020 catalog are highlighted with an appropriate flag in \Tab\ref{tab:coords}.\footnote{Since these sources were detected in the COSMOS2020 catalog, they are not part of the sample analysed in \citet{Gentile_24}. For these sources, we performed the same analysis (photometry extraction and SED-fitting) as described in that study.}

\begin{figure}
    \centering
 \includegraphics[width=\columnwidth]{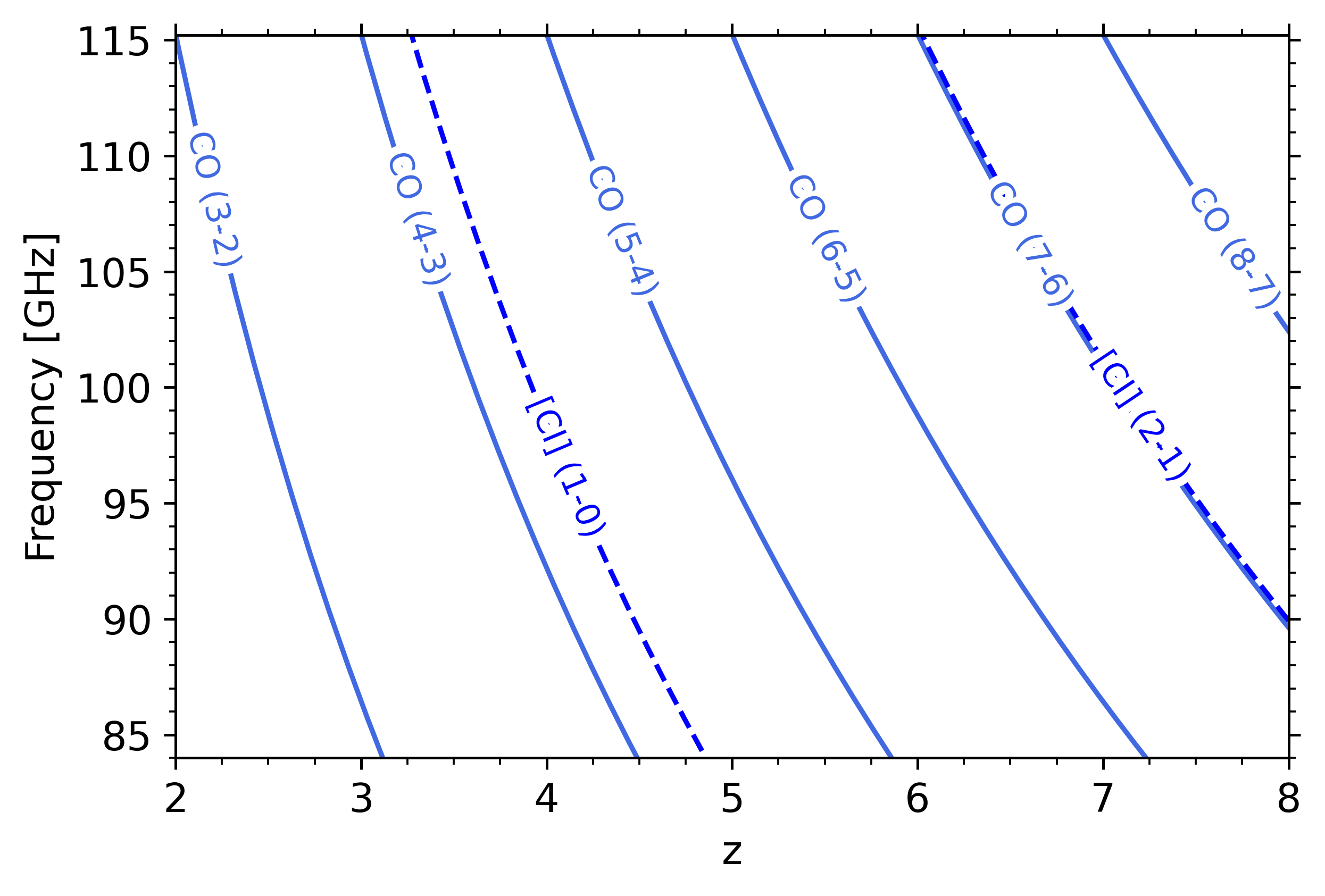} 
    \caption{Spectral setup adopted for the ALMA observations described in this study. This configuration allows us to observe at least one line among CO and [CI] transition for all the redshifts below 8, with most of the redshifts higher than 3 having two lines detectable at the observed frequencies.} 
    \label{fig:setup}
\end{figure}

The main scientific goal of the observing program was to assess the spectroscopic redshifts of the nine targets. Therefore, we requested a spectral setup covering the whole band 3 of ALMA (i.e., all the frequencies between $\sim$84 and $\sim$115 GHz). This setup, analogous to those employed in similar studies in the current literature (see e.g. \citealt{Walter_16,Jin_19,Jin_22,Cox_23}), ensures that at least one line among the CO and [CI] transitions should be detected for almost all the redshifts in the range $0<z<8$. Moreover, this spectral scan provides the possible detection of two lines for most of the redshifts higher than 3, allowing an unambiguous determination of the spec-\textit{z} (see \Fig\ref{fig:setup}). To cover the whole band 3 with ALMA, five settings were required. By estimating the integrated fluxes of the expected CO/[CI] lines observable in our setup, we requested a sensitivity of 0.32 mJy/beam per stacked channel for a total of 27h of ALMA observing time. The observations were performed in service mode between March and September 2022, when the interferometer was in its C-4/C-3 configurations (i.e. with baselines between 15 and 500/784 m, an expected beam size of 1.4''/0.92'' and a maximum recoverable scale of 16.2''/11.2''). The calibration was performed by the Alma Regional Center through the ALMA standard pipeline. Once obtained the calibrated Measurement Sets, we merged the multiple observations through the \textit{Common Astronomy Software Applications} package (\textsc{CASA} v6.1; \citealt{CASA_22}). Finally, to achieve the sensitivity originally requested in the proposal, we re-sampled the native spectral resolution to obtain $\sim0.02$ GHz channels ($\sim$ 50 km/s at the reference frequency of 100 GHz). After a first cleaning (performed with the "\textsc{tclean}" task and employing natural weighting to maximize the sensitivity; see \citealt{Hogbom_74}), we verified that the median rms across the stacked channels was 0.2 mJy/beam (i.e. slightly better than what requested in the original proposal), increasing towards higher frequencies due to the decreasing transmissivity of the ALMA band 3 up to 0.4 mJy/beam. The various settings have some overlap in frequency, therefore those (overlapping) ranges have a better rms. The beam shape is quite uniform across the channels: it can be modeled as an ellipse with a Half Power Beam Width equal to 1.45” x 1.31” and a Position Angle of $\sim -70^\circ$.

\subsection{Ancillary data}
\label{sec:ancillary}

Since the nine galaxies analyzed in this work are located in the Cosmic Evolution Survey field (COSMOS; one of the most famous extra-galactic fields in modern astronomy, observed by most of the main telescopes in the last decades, see, e.g. \citealt{Scoville_07,Koekemoer_07,Laigle_16,Civano_16,Weaver_22,Moneti_22,Casey_22}), they have an almost complete multi-wavelength coverage from the radio to the X-rays. Most of the photometry for these sources is described in \citet{Gentile_24} and briefly summarized here. The fluxes in the optical/NIR/MIR regimes are extracted from the scientific maps employed by \citet{Weaver_22} to build the COSMOS2020 catalog. These maps were produced by the HSC@Subaru, VIRCAM@VISTA, and IRAC@Spitzer instruments and telescope. \citet{Gentile_24} analysed these maps through the Photometry Extractor for Blended Objects (\textsc{PhoEBO}). This pipeline implements a modified version of the algorithm introduced by \citet{Labbe_06} and already employed in several studies (see e.g. \citealt{Endsley_21,Whitler_22}), but optimized for the deblending of RS-NIRdark galaxies. It mainly relies on a double prior coming from the radio and NIR maps, employing a PSF-matching between the high-resolution maps (radio and NIR) and the low-resolution ones (mainly those produced by IRAC) to deblend the different sources and extract forced photometry.
Further details on the algorithm and its validation are presented in \citet{Gentile_24}, while the algorithm is freely available in a GitHub repository\footnote{\url{https://github.com/fab-gentile/PhoEBO}}. Additional photometry at longer wavelengths is retrieved through cross-matching with pre-existing catalogs. More in detail, the photometry at FIR wavelengths is obtained from the SuperDeblended catalog (v20201010; \citealt{Jin_18}), containing deblended photometry from MIPS@Spitzer, PACS/SPIRE@Herschel, and {SCUBA-2@JCMT} instruments and telescopes. The fluxes at (sub)mm wavelengths are retrieved through cross-matching with the catalog from the \textit{Automated Mining of the ALMA Archive in the COSMOS Field} (A3COSMOS) survey (v.20200310; \citealt{Liu_19}). Finally, radio fluxes at 1.28, 1.4, and 3 GHz are obtained from the catalogs of the COSMOS-VLA large program \citep{Schinnerer_10,Smolcic_17} and of the MIGHTEE Early Science Data Release \citep{Jarvis_16,Heywood_22}. A shallow X-ray coverage is also available for the COSMOS field thanks to the public catalogs by \citet{Elvis_09} and \citet{Civano_16}. This latter information was employed in \citet{Gentile_24} to ensure that none of the sources analyzed in this work are hosting a powerful and un-obscured AGN ($L_{\rm x}>10^{42}$ erg s$^{-1}$).

\section{Analysis of the datacubes}
\label{sec:methods}

\subsection{Continuum images}
\label{sec:continuum}
The first analysis performed on the calibrated MSs consists in the production of a continuum image, useful to study the properties of dust in our targets. This procedure is performed through the \textsc{CASA} task "\textsc{tclean}" in "multi-frequency synthesis" (mfs) mode, after masking any bright line that could contaminate the continuum emission. To maximize the sensitivity of the cleaned image, we employ a natural weighting throughout this procedure. To estimate the continuum fluxes, we perform an aperture photometry with \textsc{CARTA} \citep{CARTA_21}, by employing an aperture corresponding to the $2\sigma$ contour of the continuum image. {We verify that this estimate is compatible -- within the estimated uncertainties -- with the flux estimated through a 2D profile-fitting performed with the \textsc{CASA} task "\textsc{imfit}"}. The results of this procedure are reported in \Tab\ref{tab:cont}. We obtain that 6 out of 9 targets are robustly detected (S/N$>$3) in the continuum images.

\begin{table}
\centering
\begin{threeparttable}
\renewcommand{\arraystretch}{1.1}
\caption{Continuum fluxes for the RS-NIRdark galaxies analysed in this work. The values at $\lambda=3$ mm are obtained through aperture photometry on the continuum images obtained in \Sec\ref{sec:continuum}. We consider a source as robustly detected with a S/N$>$3, therefore we report a $3\sigma$ upper limit for the undetected targets. {The value at 850 $\mu$m are retrieved from the SuperDeblended catalog \citep{Jin_18} for the sources at S/N>3 (reported in bold), while for the other galaxies, we report the best-fitting flux at 850 $\mu$m computed with \textsc{Cigale}.} For two sources, we also report an additional (sub)mm flux measured with ALMA through cross-matching with the A3COSMOS catalog \citep{Liu_19}.}
\label{tab:cont}
\begin{tabular}{cccc}
\toprule
ID & $S_{3\rm mm}$ & $S_{850\rm \mu m}$ & Other flux \\
&  [mJy] & [mJy] & [mJy]\\
\midrule
41 & $(0.04\pm0.01)$& $(2.9\pm0.3)$ & - \\
84 & $(0.11\pm0.03)$ & ${\bf(7\pm2)}$ & 0.873 mm: $(4.6\pm0.8)$ \\
121 & $(0.09\pm0.03)$& $(3.9\pm0.4)$ & -\\
182 & $(0.14\pm0.02)$& ${\bf (5\pm1)}$& - \\
235 & $(0.18\pm0.04)$& ${\bf (5\pm1)}$ & -\\
247 & $(0.09\pm0.01)$& ${\bf (5\pm1)}$ & - \\
298  & $< 0.036$ & $(3.3\pm0.3)$& - \\
361 & $< 0.036$ & $(3.3\pm0.4)$ &  1.249 mm: $(2.0\pm0.4)$\\
456 & $< 0.036$ & $(1.9\pm0.2)$ & -\\
\bottomrule
\end{tabular}
\end{threeparttable}
\end{table}

\begin{figure*}
    \centering
    \includegraphics[width=0.16\textwidth]{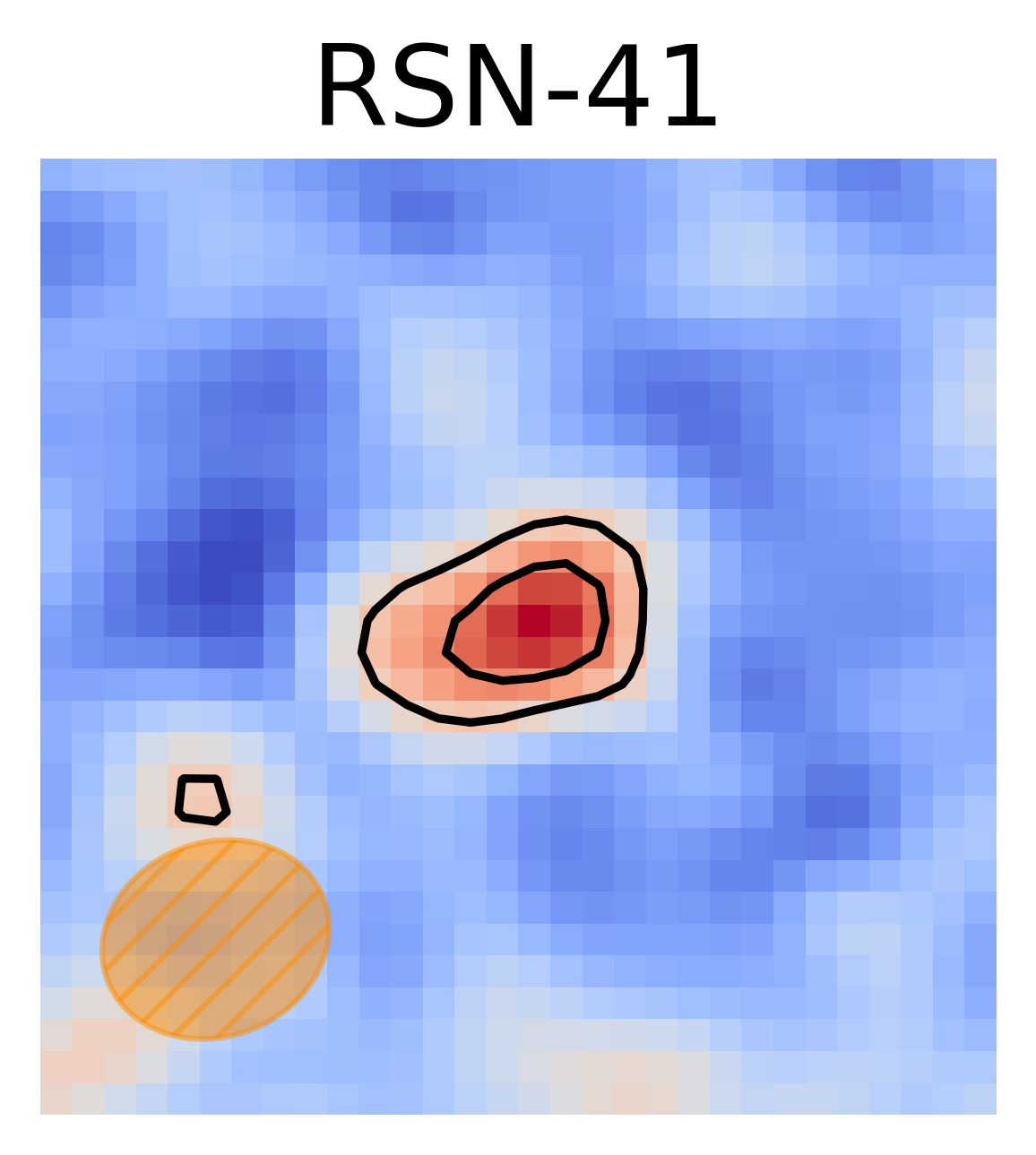} \includegraphics[width=0.16\textwidth]{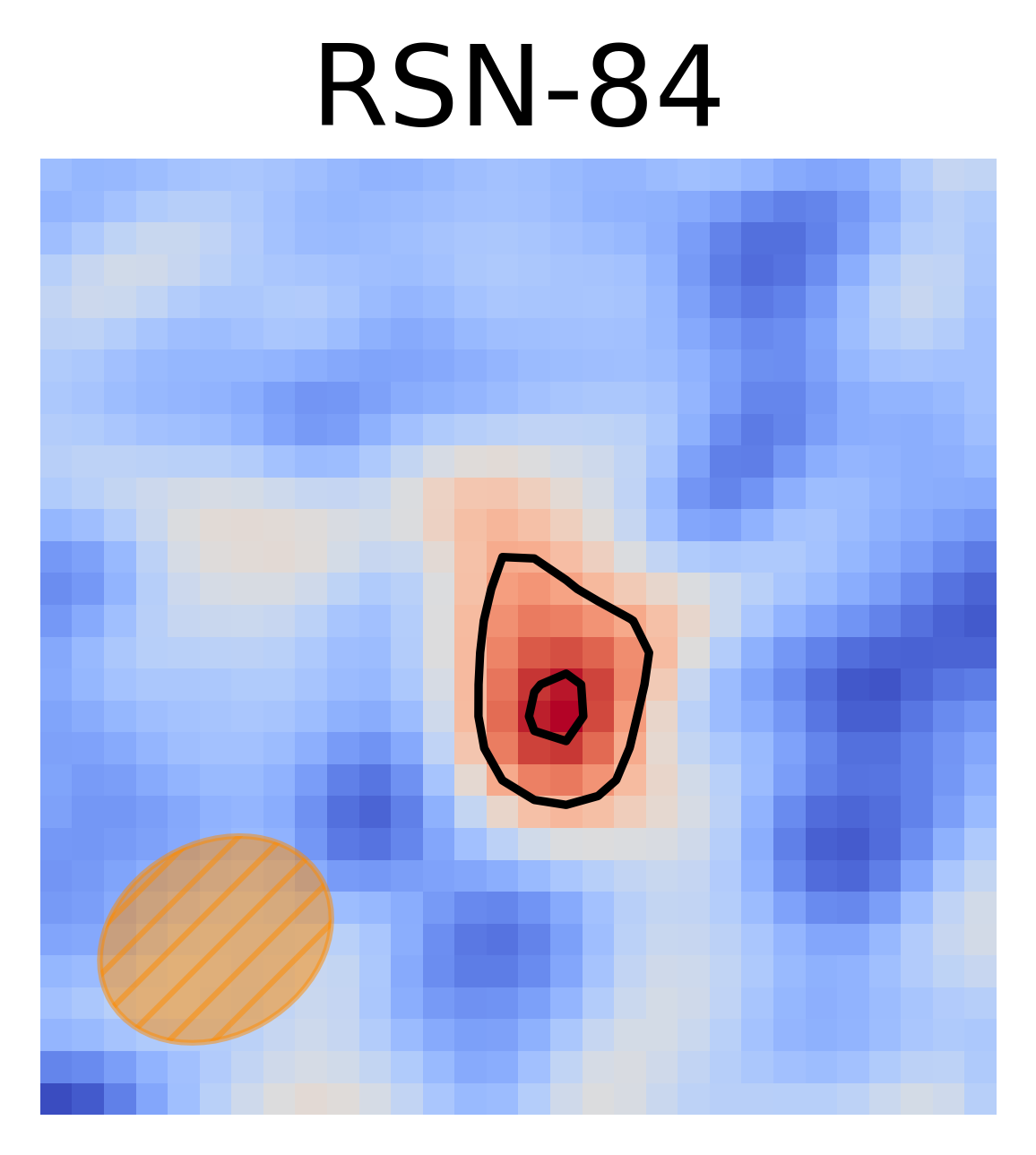} \includegraphics[width=0.16\textwidth]{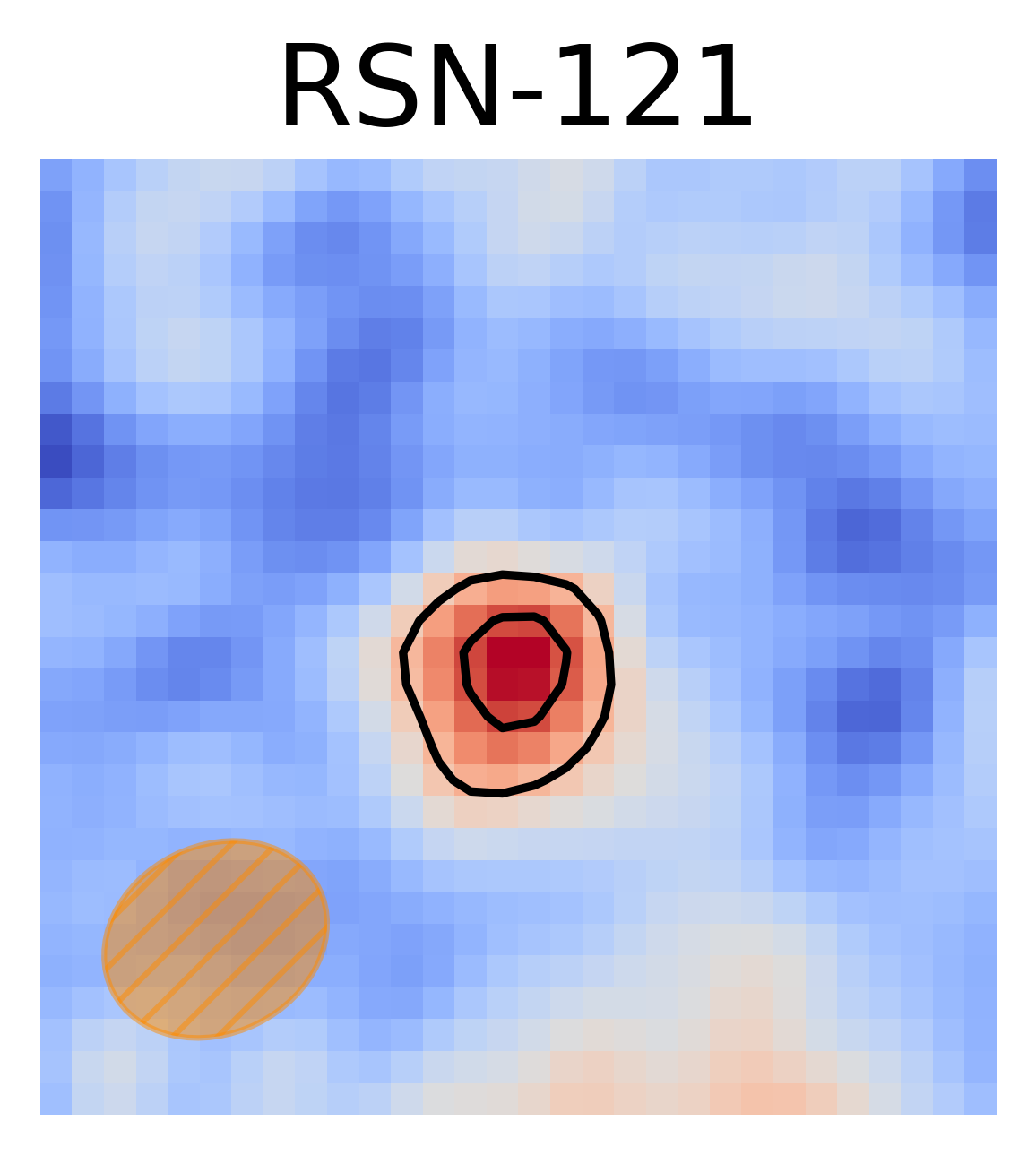} \includegraphics[width=0.16\textwidth]{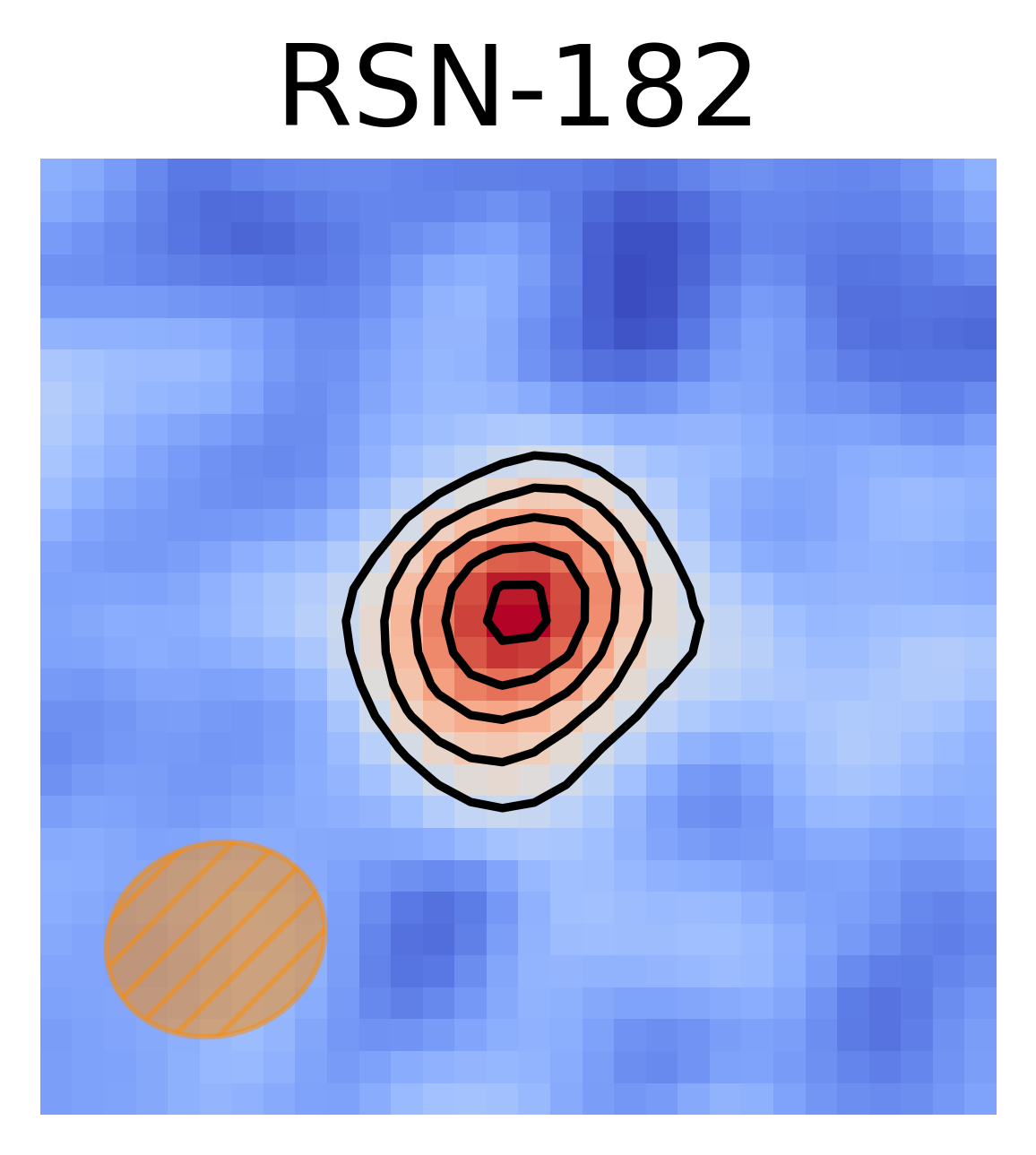} \includegraphics[width=0.16\textwidth]{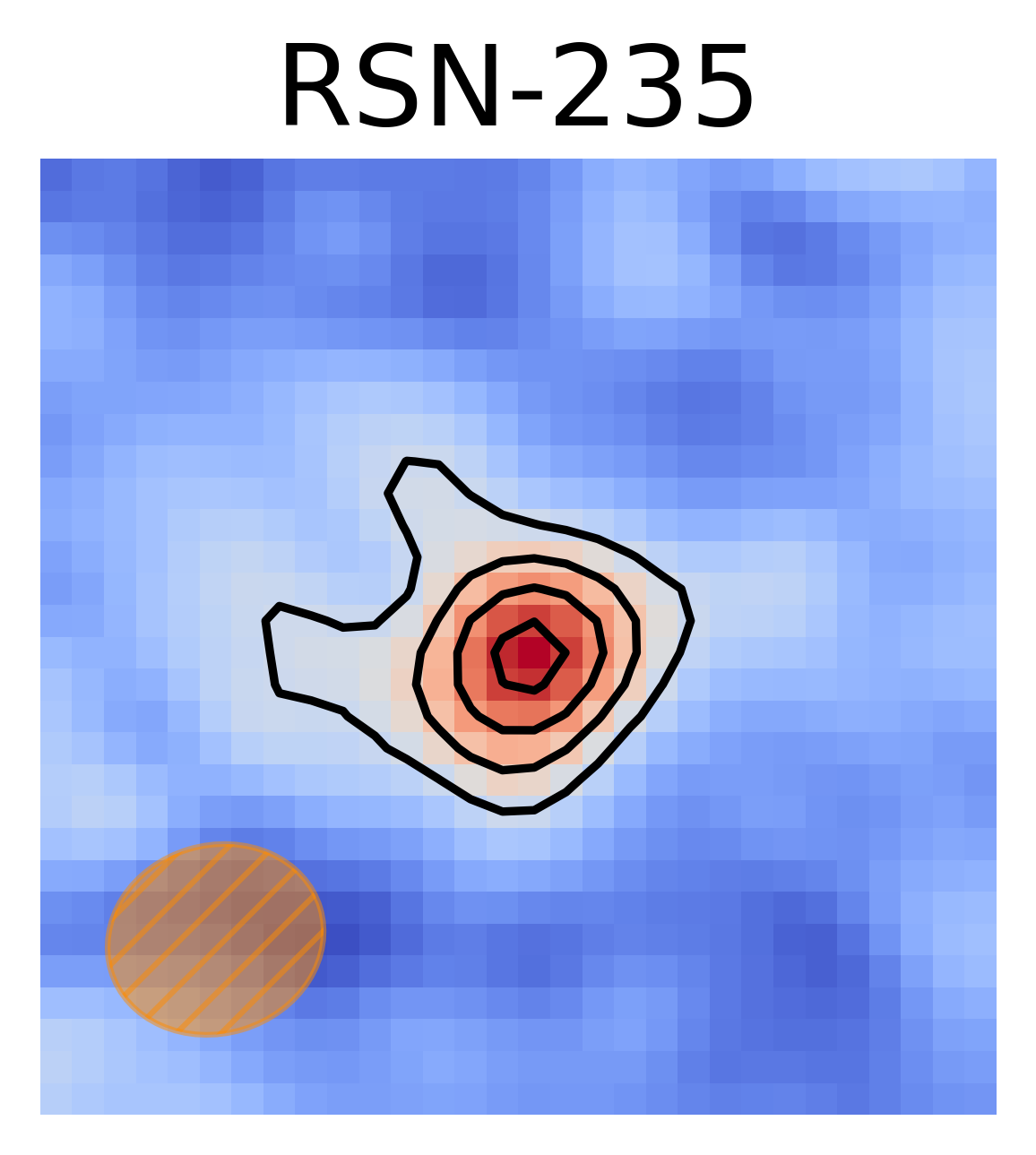} \\
    \includegraphics[width=0.16\textwidth]{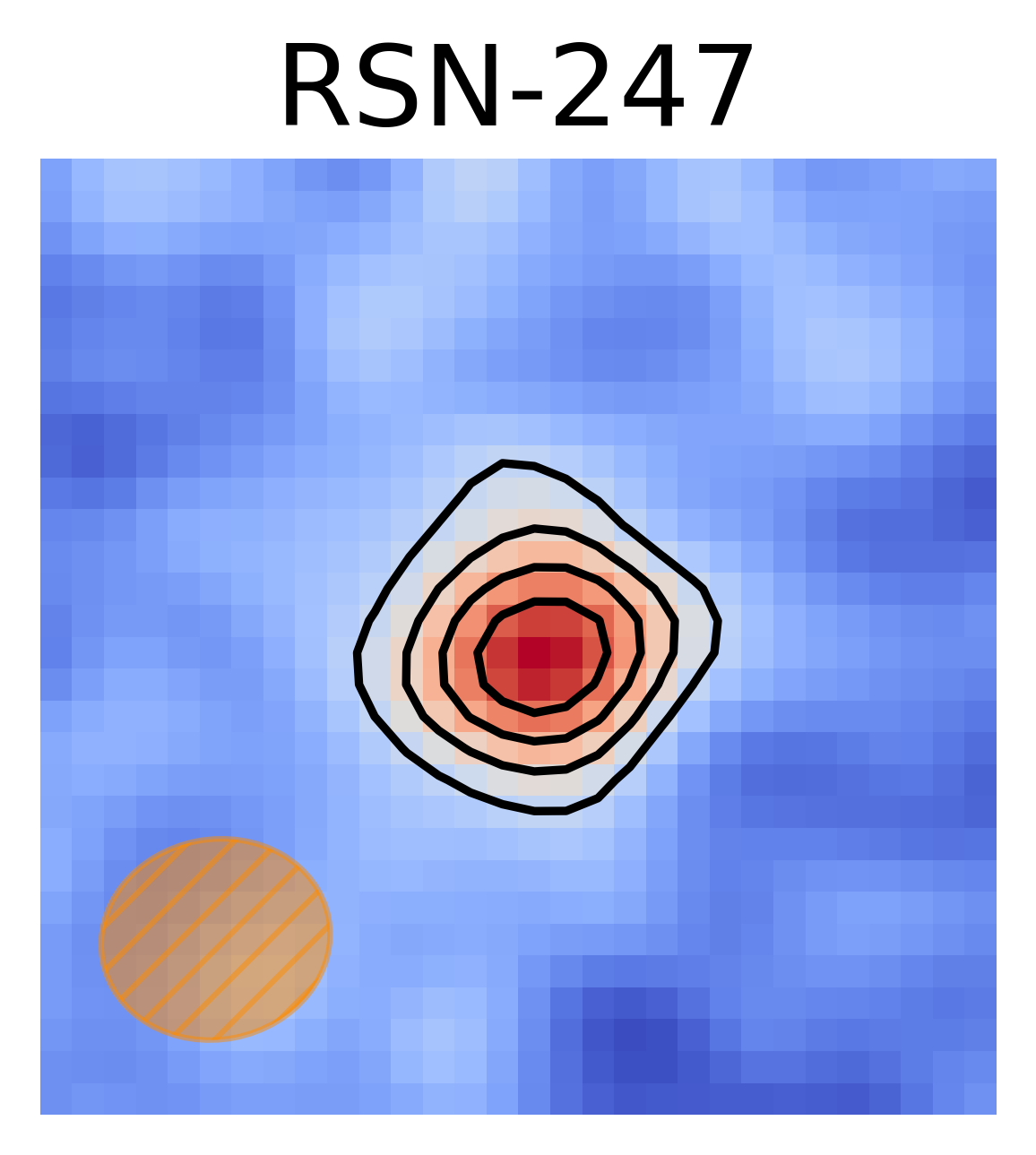} \includegraphics[width=0.16\textwidth]{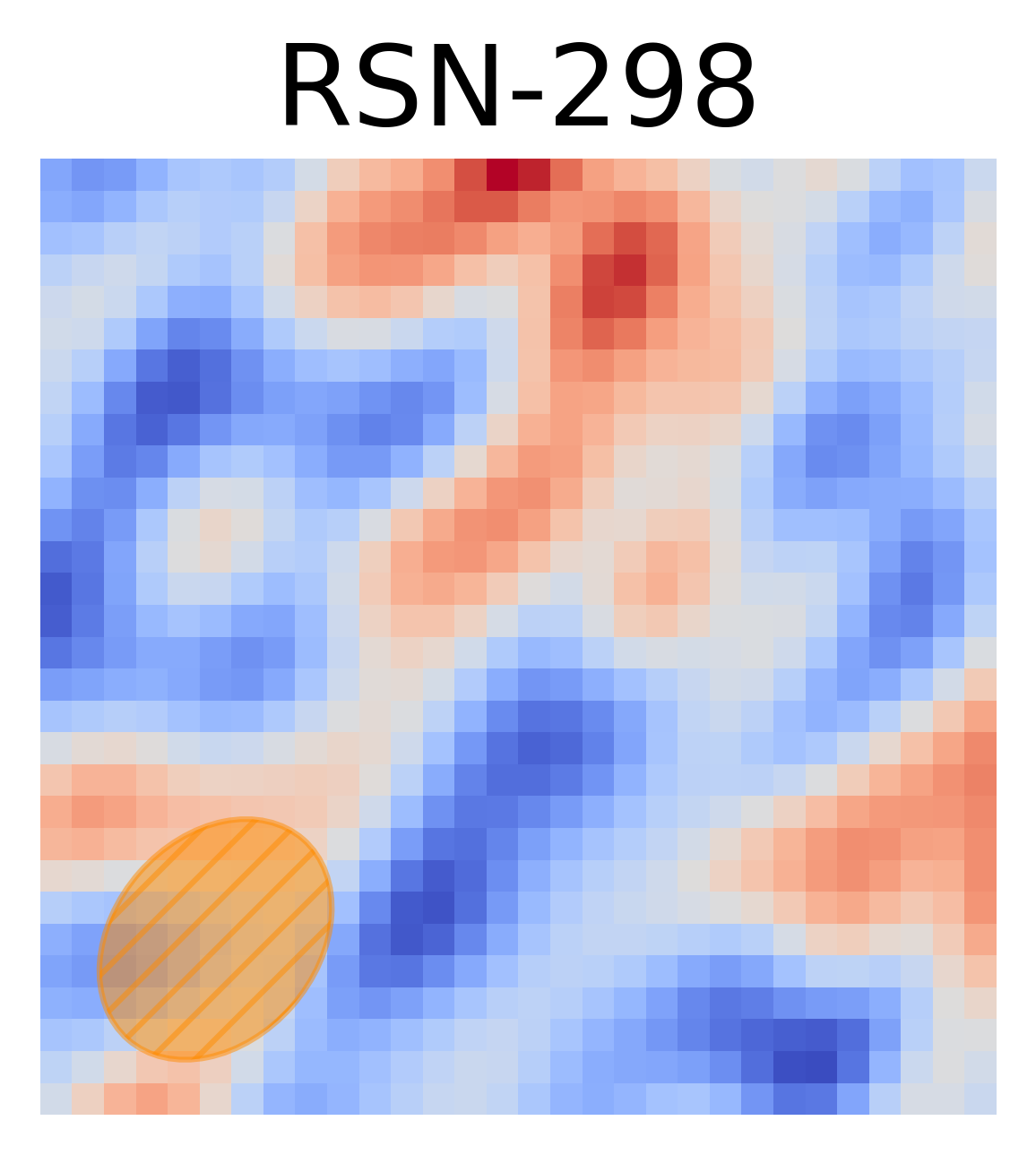} \includegraphics[width=0.16\textwidth]{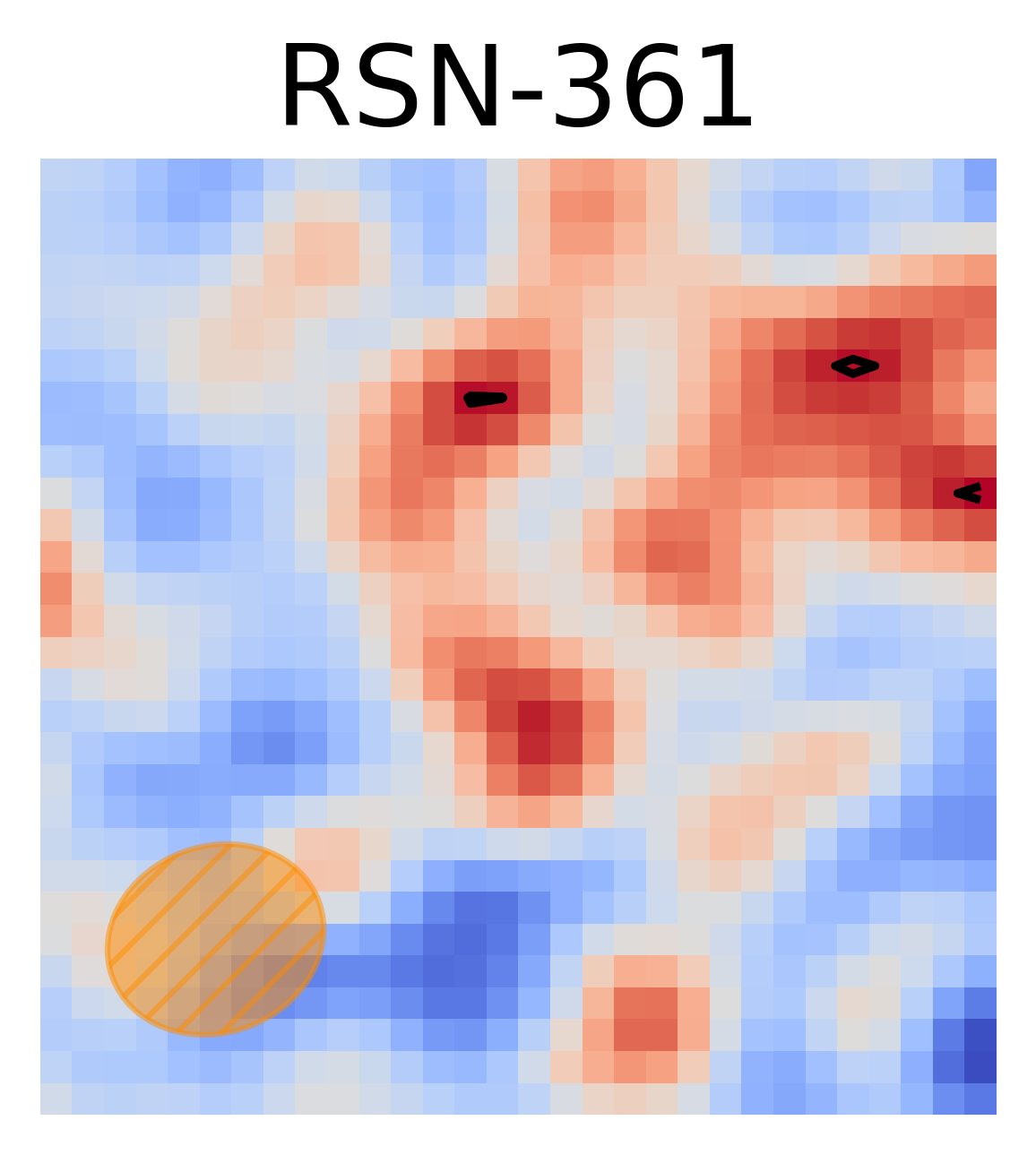} \includegraphics[width=0.16\textwidth]{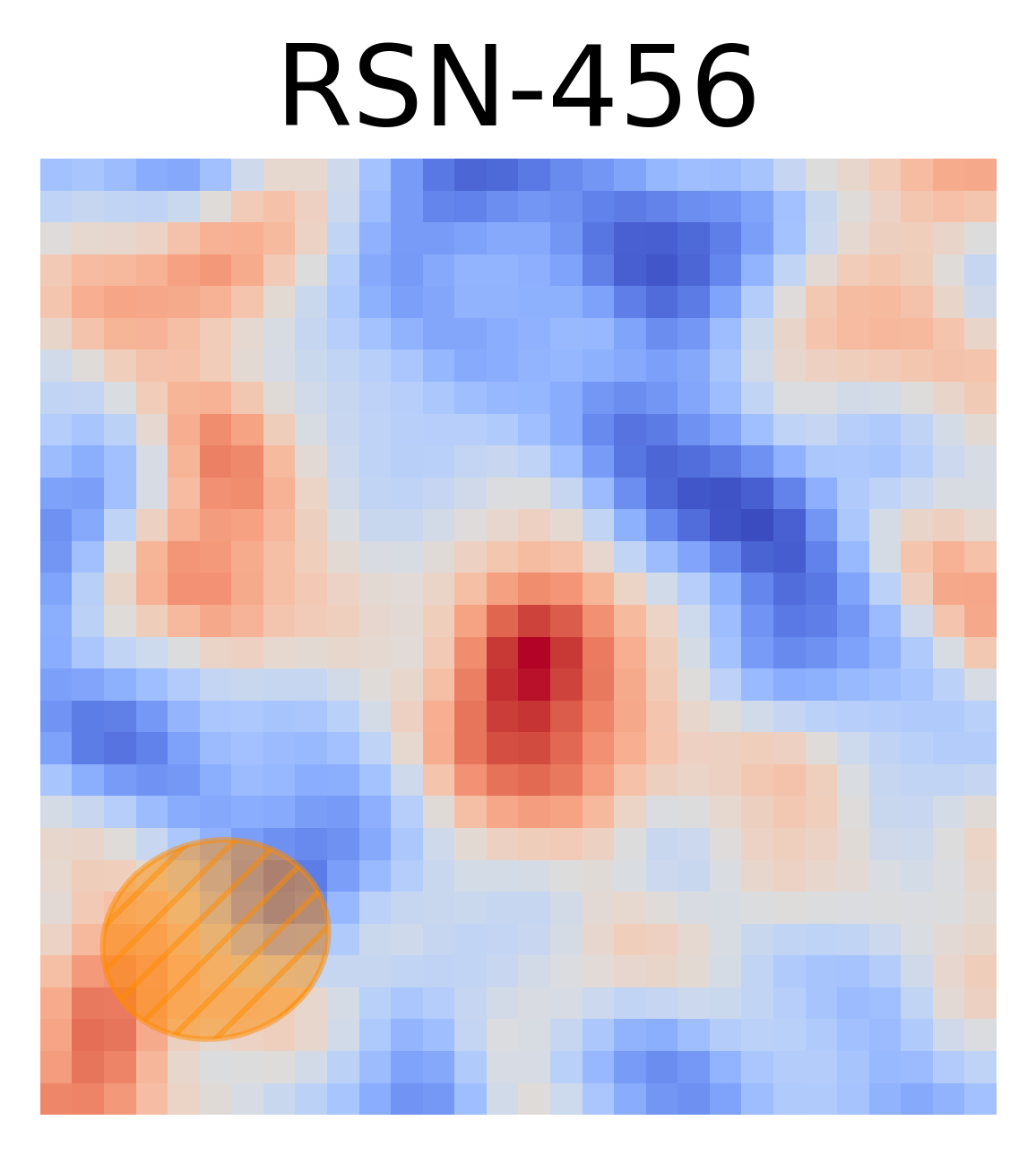} 
    \caption{Continuum maps of the nine targets analysed in this paper. The black contours are in step of $2\sigma$ starting from $3\sigma$. All the images have a 7.5" side, while the syntetized beam is reported in the lower left corner of each image.}
    \label{fig:continuum}
\end{figure*}

\subsection{Line identification and reliability}
\label{sec:lines}

\begin{table*}
\centering
\begin{threeparttable}
\renewcommand{\arraystretch}{1.1}
\caption{Lines detected in our targets following the procedure discussed in \Sec\ref{sec:lines} and related models (\Sec\ref{sec:redshifts}). In the first tier of galaxies, the two lines identified in the spectrum are employed to estimate the spectroscopic redshift. For the galaxies in the second tier (i.e. in those with a single line identified) we assume the spec-\textit{z} as the redshift allowed by the visible line with the best agreement with the photometry. For each galaxy, we also report the photometric redshift computed by \citet{Talia_21} and that computed through SED-fitting with \textsc{Cigale} following \citet{Gentile_24} once added the 3 mm continuum point.}
\label{tab:lines}
\begin{tabular}{ccccccccccc}
\toprule
ID & Line 1 & Freq & FWHM & SNR & Line 2 & Freq & SNR & $z_{\rm spec}$ & $z_{\rm phot}$ (G23) & $z_{\rm phot}$ (T21)\\
&  & (GHz) & (km/s)  & & & (GHz) &  &  &  & \\
\midrule
 RSN-84 & CO(4-3) & 102.05 & $584\pm96$ & 11.23 & [CI](1-0) & 108.91 & 5.39 & 3.518 & $4.5\pm0.3$ & $5.2\pm0.9$ \\
 RSN-121 & CO(4-3) & 106.64 & $683\pm77$ & 12.26 & [CI](1-0) & 113.74 & 4.89  & 3.323 & $3.3\pm0.2$& $5.2\pm0.1$\\
RSN-235 & CO(5-4) & 101.95 & $481\pm63$ & 13.24 & [CI](1-0) & 87.07 & 5.68 & 4.652 & $4.5\pm0.5$& $5.1\pm0.5$\\
 RSN-298 & CO(4-3) & 89.53 & $408\pm93$ & 8.44 & CO(5-4) & 111.94 & 4.93 & 4.150 & $3.5\pm0.3$& $5.4\pm0.8$\\
RSN-361 & CO(5-4) & 103.19 & $349\pm81$ & 6.30 & [CI](1-0)$^{(a)}$ & 88.13 & 3.47 & 4.585 & $4.3\pm0.5$ & $5.1\pm0.9$\\
\midrule
 RSN-41 & CO(3-2) & 91.50 & $627\pm60$ & 12.18 & - & - & - & \textit{2.779} & $3.0\pm0.2$ & $6.9\pm0.1$\\
 RSN-182 & CO(5-4) & 100.31 & $441\pm89$ & 9.59 & - & - & - & \textit{4.745} & $4.6\pm0.2$& $5.9\pm0.9$\\
RSN-247 & CO(3-2) & 88.02 & $905\pm285$ & 6.79&  - & - & - & \textit{2.929} & $3.4\pm0.4$& $5.0\pm0.1$\\
 RSN-456 & CO(3-2) & 86.74 & $497\pm71$ & 7.22 & - & - & - &  \textit{2.987} & $2.8\pm0.2$& $6.8\pm0.7$\\
 \bottomrule
\end{tabular}
\begin{tablenotes} 
\item{$^{(a)}$ Tentative second line, see the discussion in \Sec\ref{sec:redshifts}}.
\end{tablenotes}
\end{threeparttable}
\end{table*}

The presence of emission lines inside our datacubes is unveiled through a line-finding algorithm analogous to those employed in several previous studies \citep[e.g.][]{Daddi_15,Walter_16,Coogan_18,Puglisi_19,Jin_19,Jin_22} and summarized here:
\begin{enumerate}
    \item We obtain a continuum-subtracted MS for each source through the CASA task "\textsc{uvcontsub}". We model the continuum as a first-grade polynomial whose slope is fitted on the whole frequency range covered by our observations once masked any bright line that could contaminate the continuum estimation.
    \item We compute the 0th moment of the continuum-subtracted MS through the CASA task "\textsc{immoments}" to unveil the spatial region of the datacube with a significant line emission. In all the analyzed targets, the 0th moment significantly overlaps with the radio emission at 3GHz visible in the maps by \citet{Smolcic_17}. This result ensures that the mm emission can be safely associated to our targets.
    \item We perform the imaging of the continuum-subtracted visibilities through the CASA task "\textsc{tclean}". We employ a natural weighting to maximize the cleaned images' sensitivity.
    \item We convolve the cleaned datacube with a series of boxcar kernels with variable widths between 1 and 13 channels (i.e., between 60 and 780 km/s at a representative frequency of 100 GHz).
    \item For each convolved datacube, we produce an S/N cube by dividing each channel by the relative rms. This quantity is computed through a sigma clipping performed on the inner region of the primary beam to avoid possible biases due to the presence of significant emission and higher noise far from the phase center.
    \item Finally, we extract a S/N spectrum for each convolved datacube through the Python library \textsc{Interferopy} \citep{interferopy}. We employ as extracting region the 2$\sigma$ contour of the 0th moment map obtained in step number 2.
\end{enumerate}

\begin{figure*}
    \centering
    \includegraphics[width=0.32\textwidth]{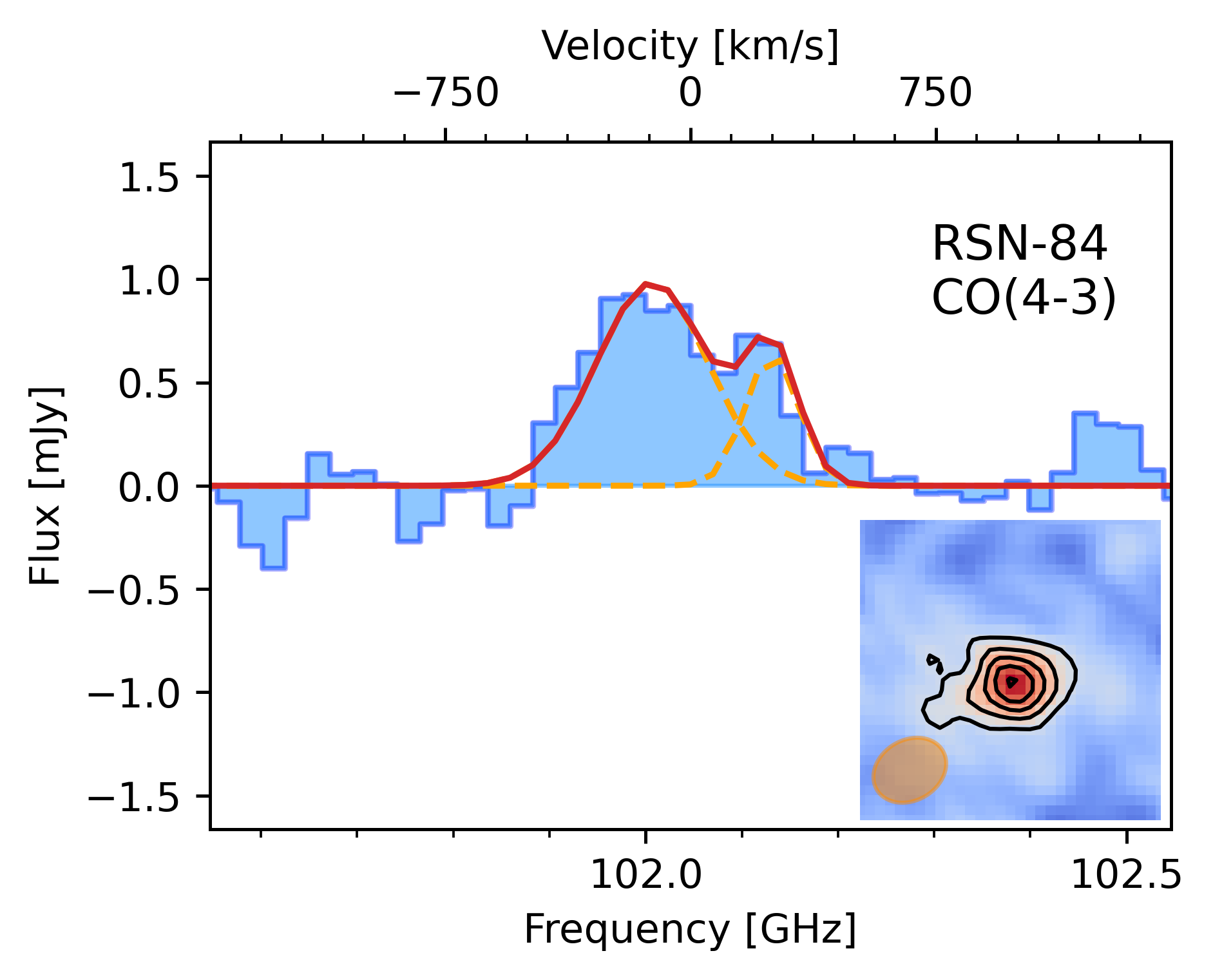} 
    \includegraphics[width=0.32\textwidth]{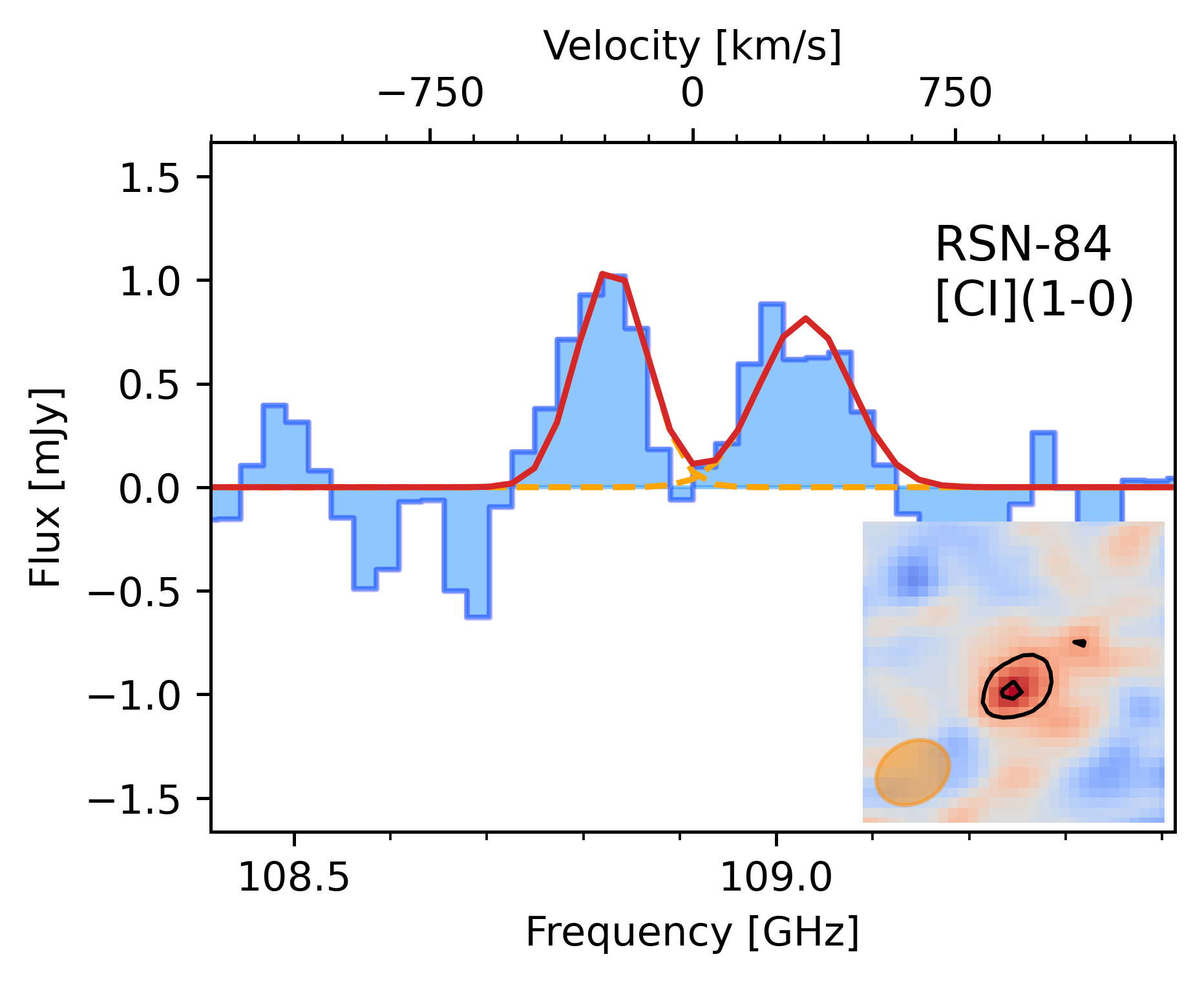}\\
    \includegraphics[width=0.32\textwidth]{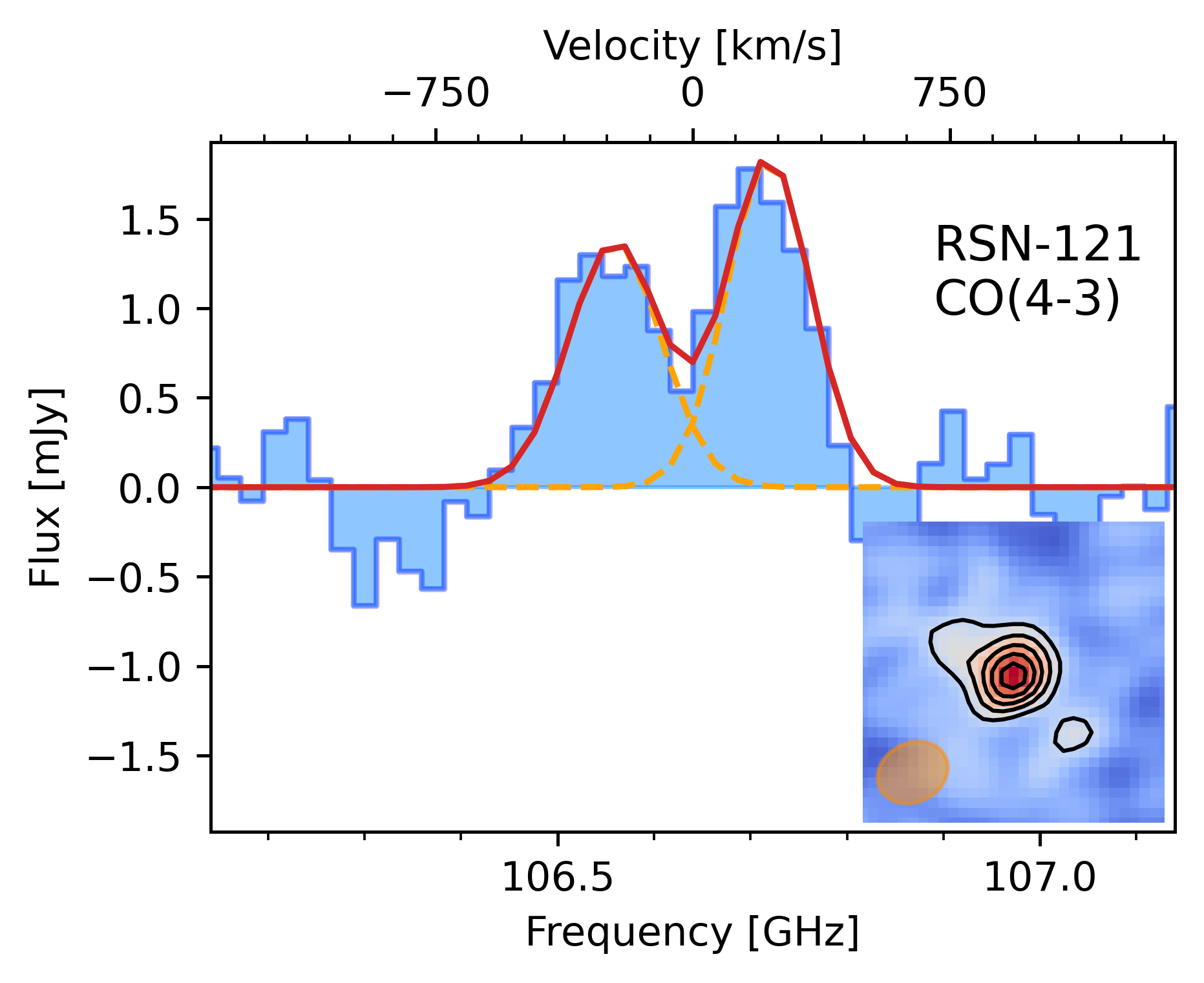} 
    \includegraphics[width=0.32\textwidth]{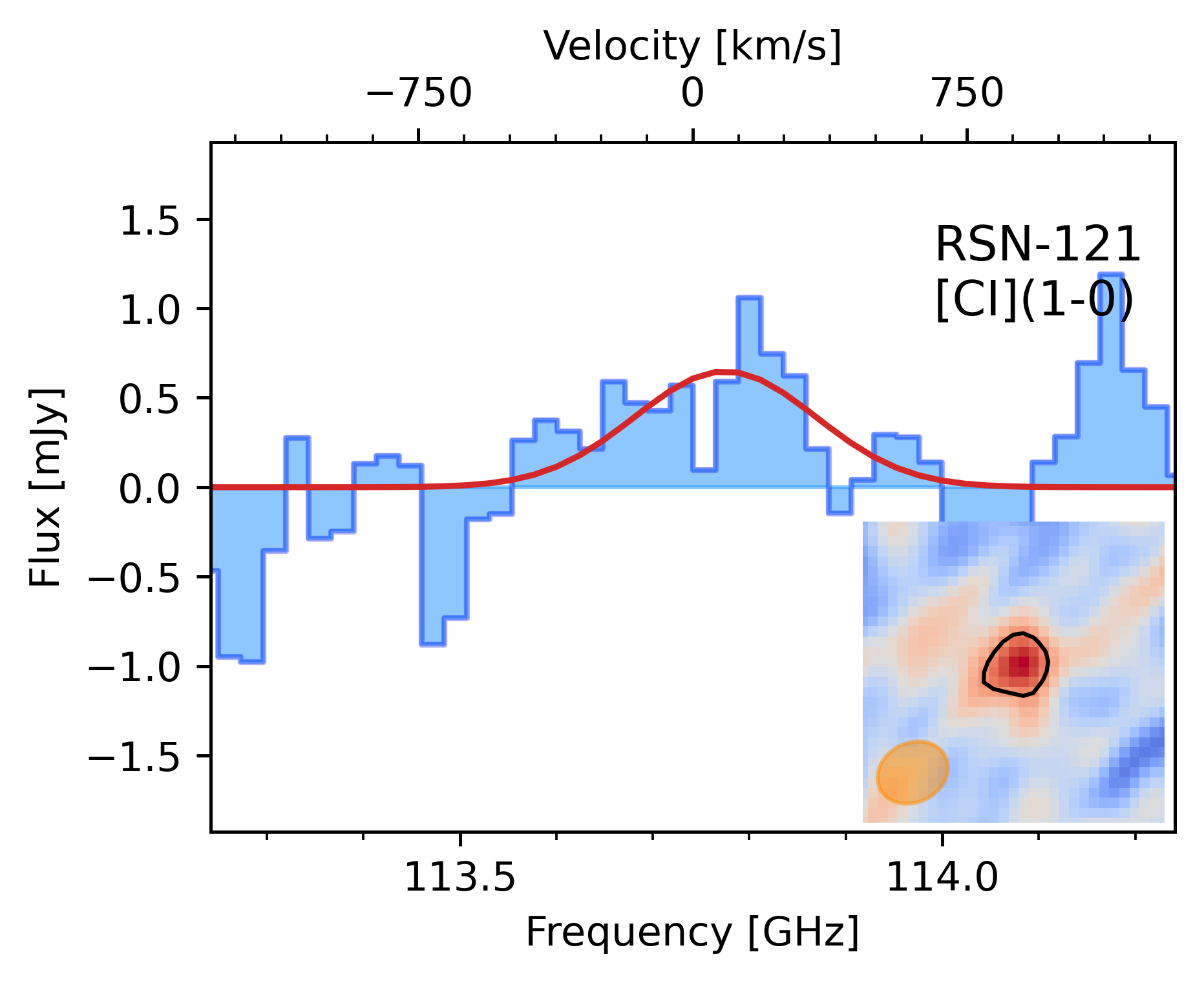} \\   \includegraphics[width=0.32\textwidth]{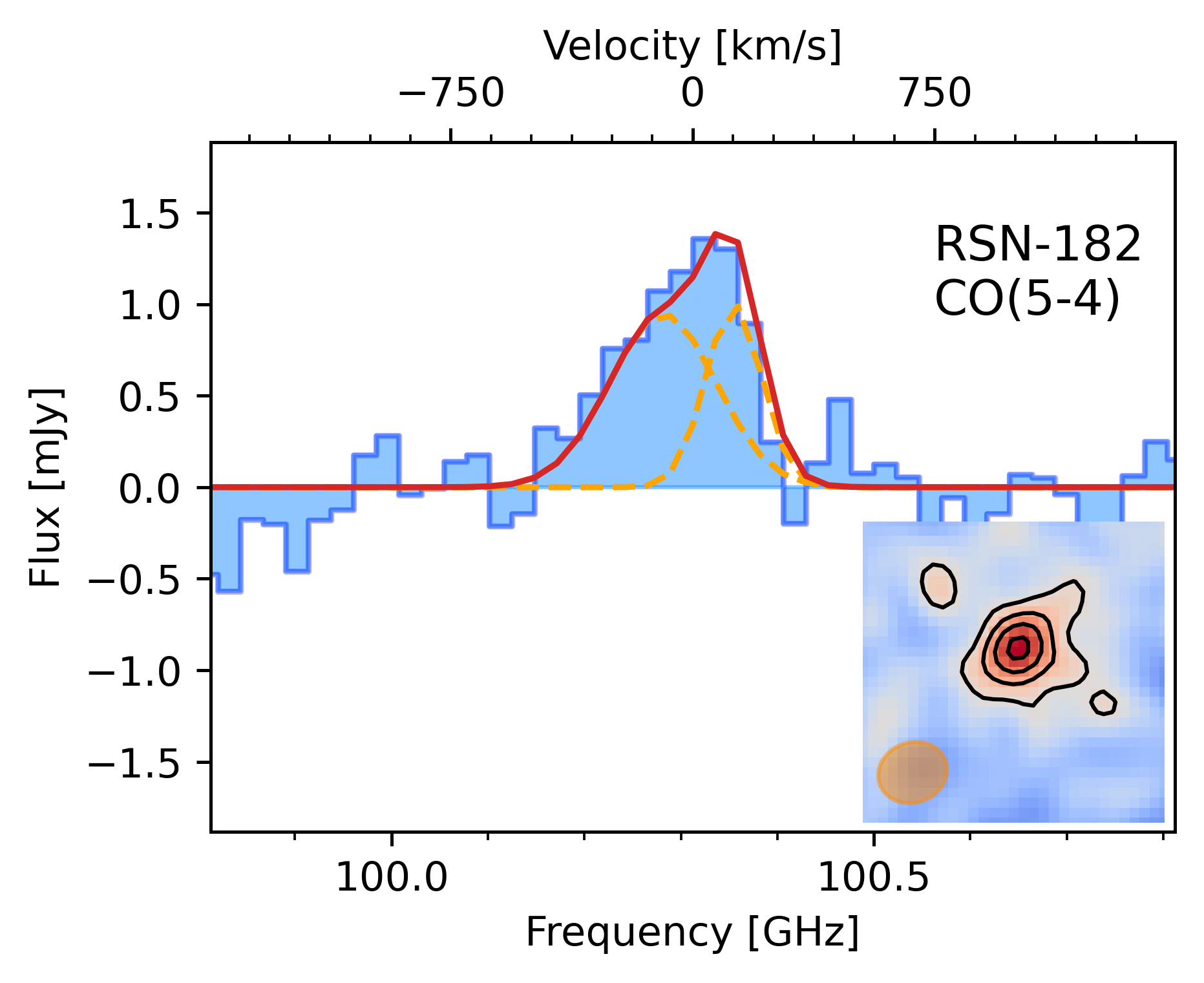}
    \includegraphics[width=0.32\textwidth]{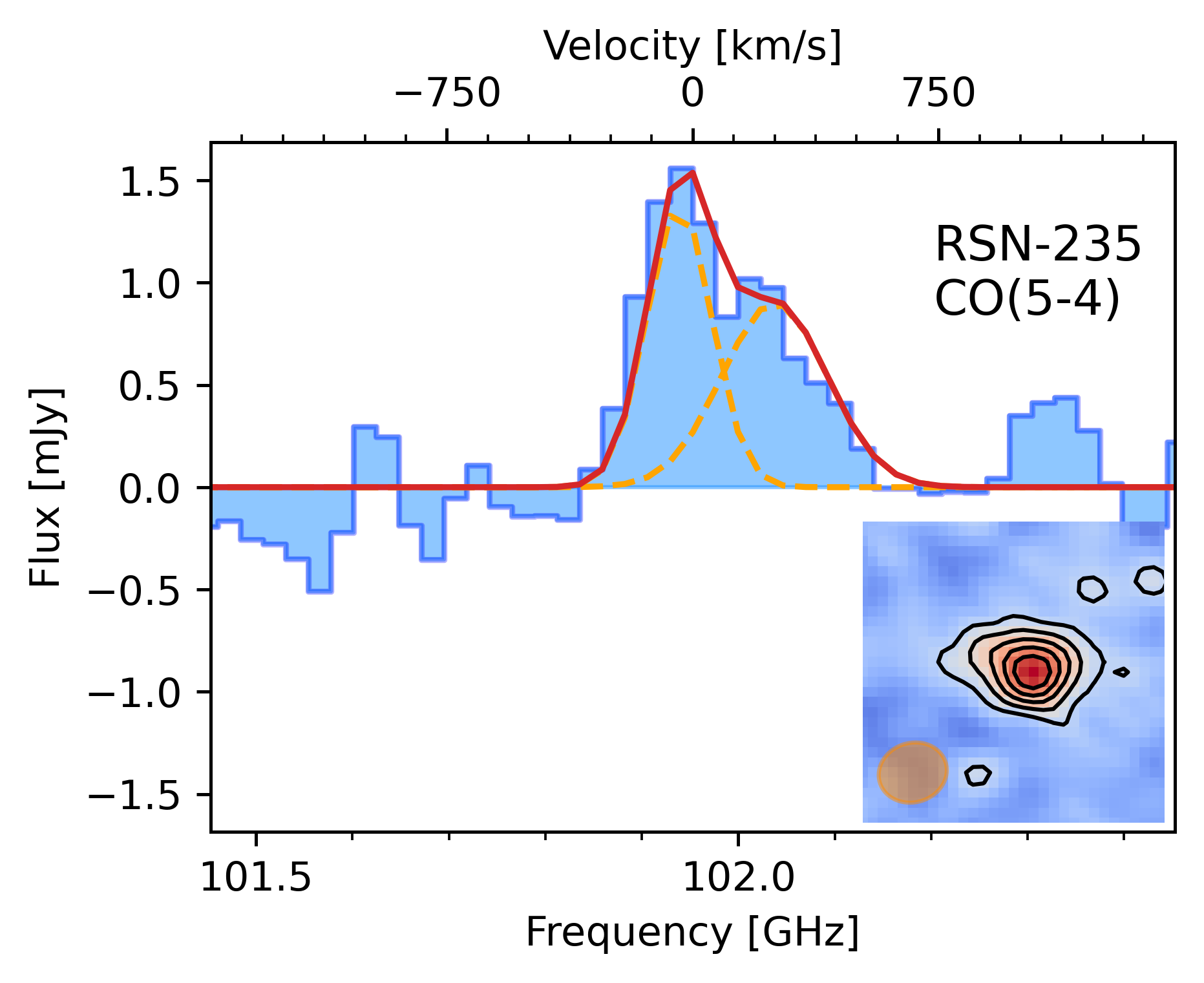}
    \includegraphics[width=0.32\textwidth]{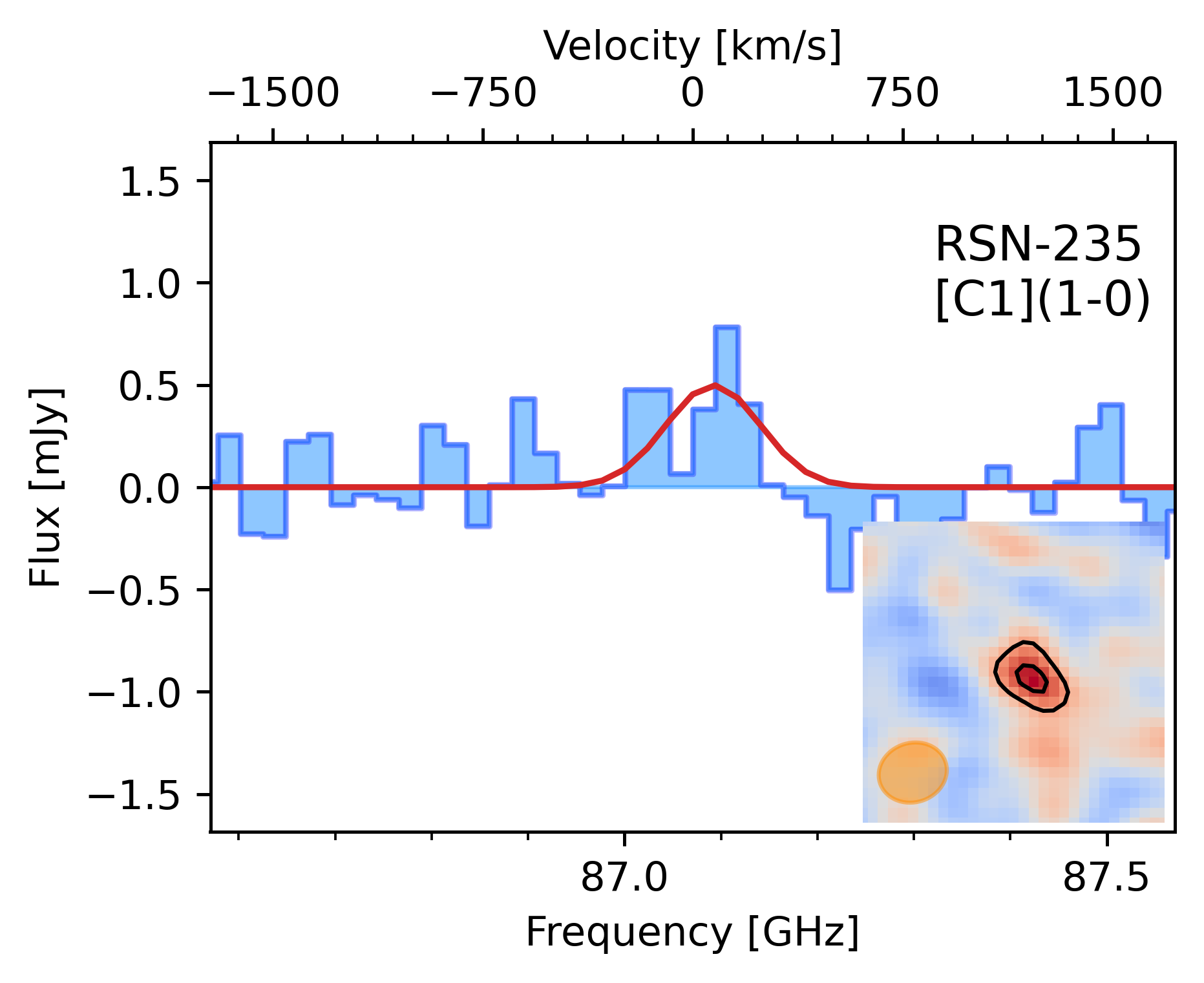}  
    \caption{Spectrum of the various lines identified through the procedure discussed in \Secs \ref{sec:lines} and \ref{sec:redshifts} in our targets. To increase the visibility of the lines, we resampled the original spectral resolution employed in the study up to $\sim180$ km/s. For each line, we report in the upper-right corner the ID of the galaxy and our modeling as CO/[CI] transitions. The insets show the moment 0th of each line {(7.5'' side) centered on the radio position measured from the 3 GHz maps}, with the contours being in steps of $2\sigma$ starting from $3\sigma$. On each line, we also report in red the Gaussian modeling with one or two components as described in \Sec\ref{sec:kinematics}. In the lines modeled with a double Gaussian, we also show the two sub-components with an orange dashed line.}
    \label{fig:spectra_lines}
\end{figure*}

\begin{figure*}
\centering
\renewcommand{\thefigure}{\arabic{figure}}
\addtocounter{figure}{-1}
\includegraphics[width=0.32\textwidth]{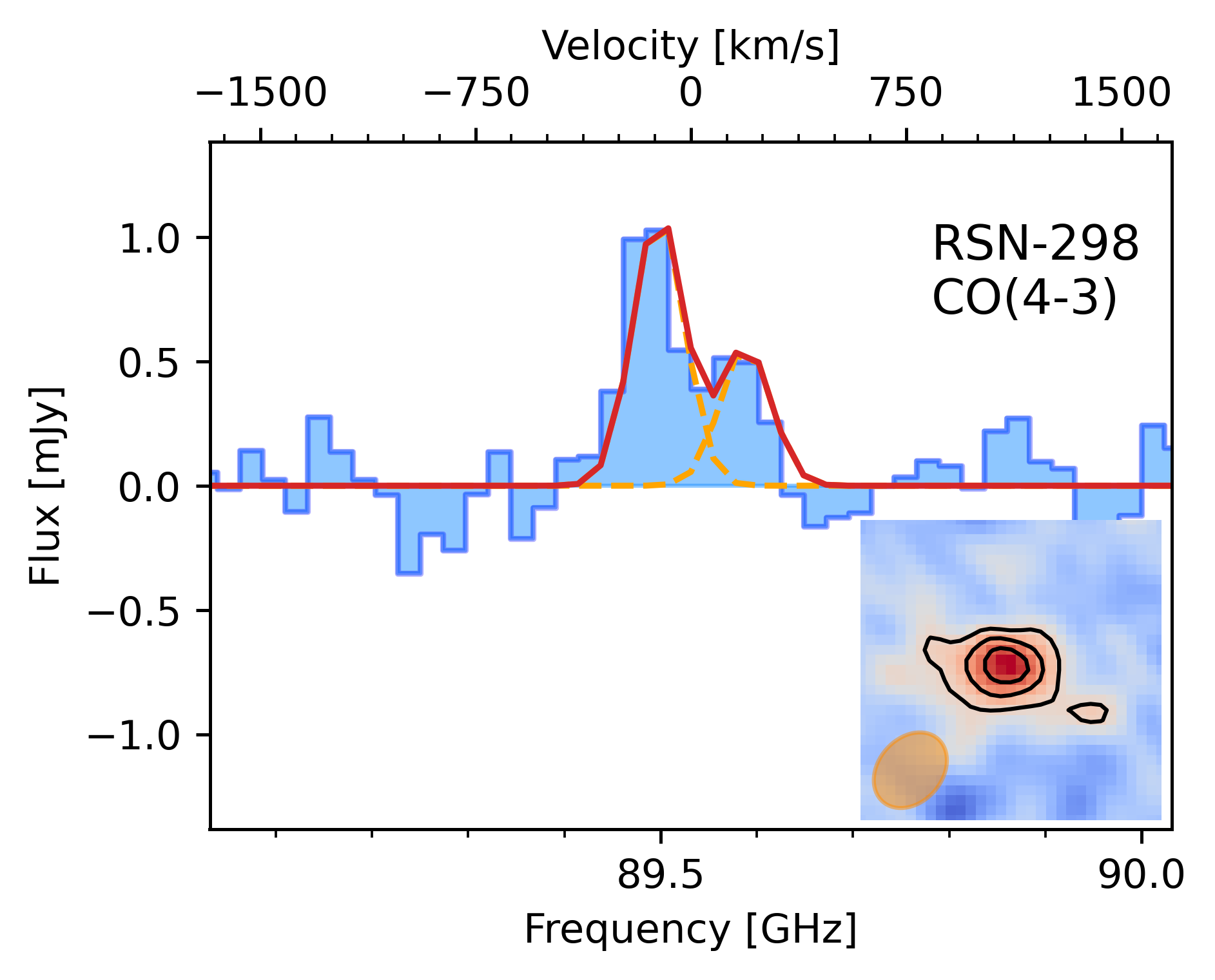} 
\includegraphics[width=0.32\textwidth]{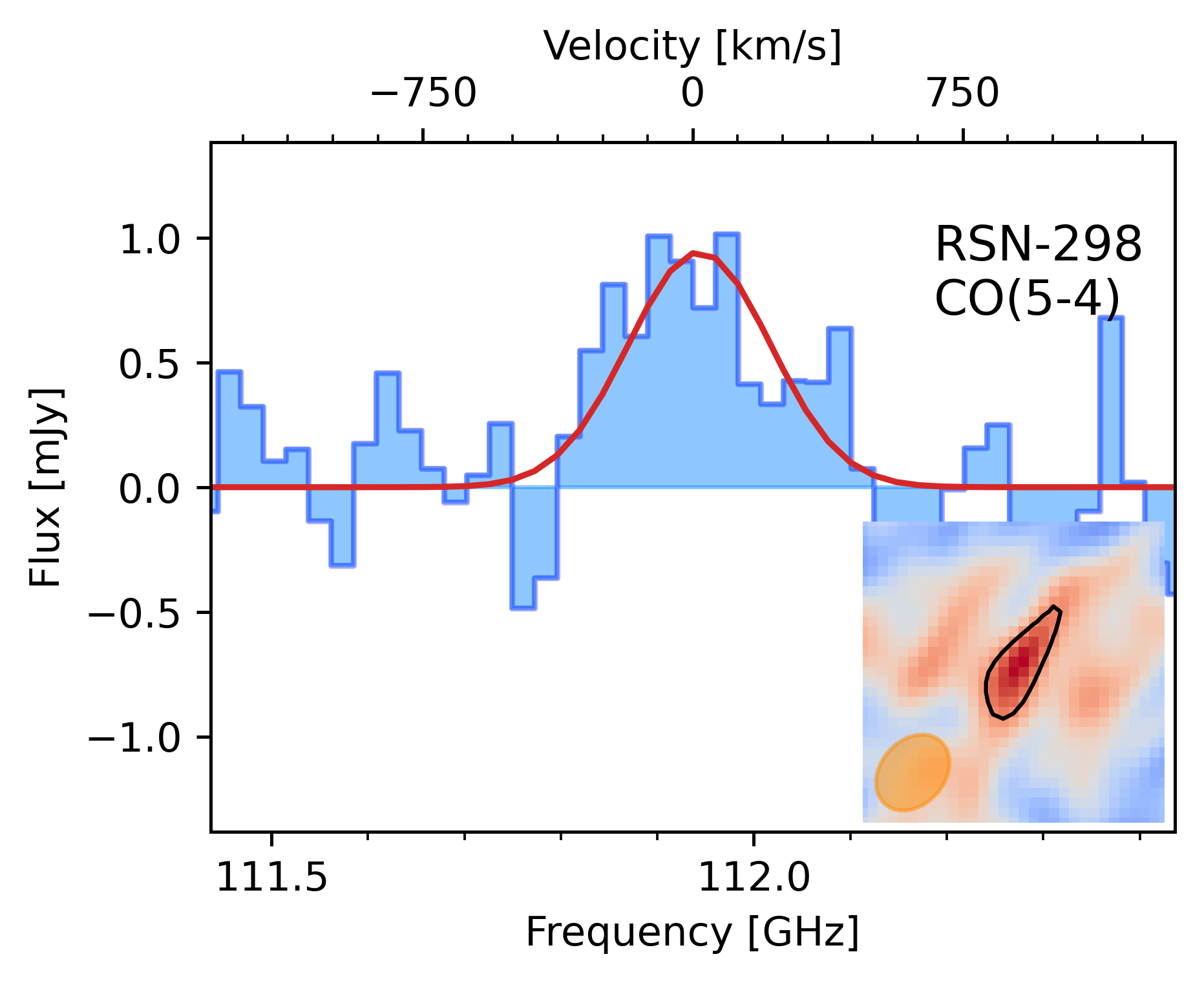} \\
\includegraphics[width=0.32\textwidth]{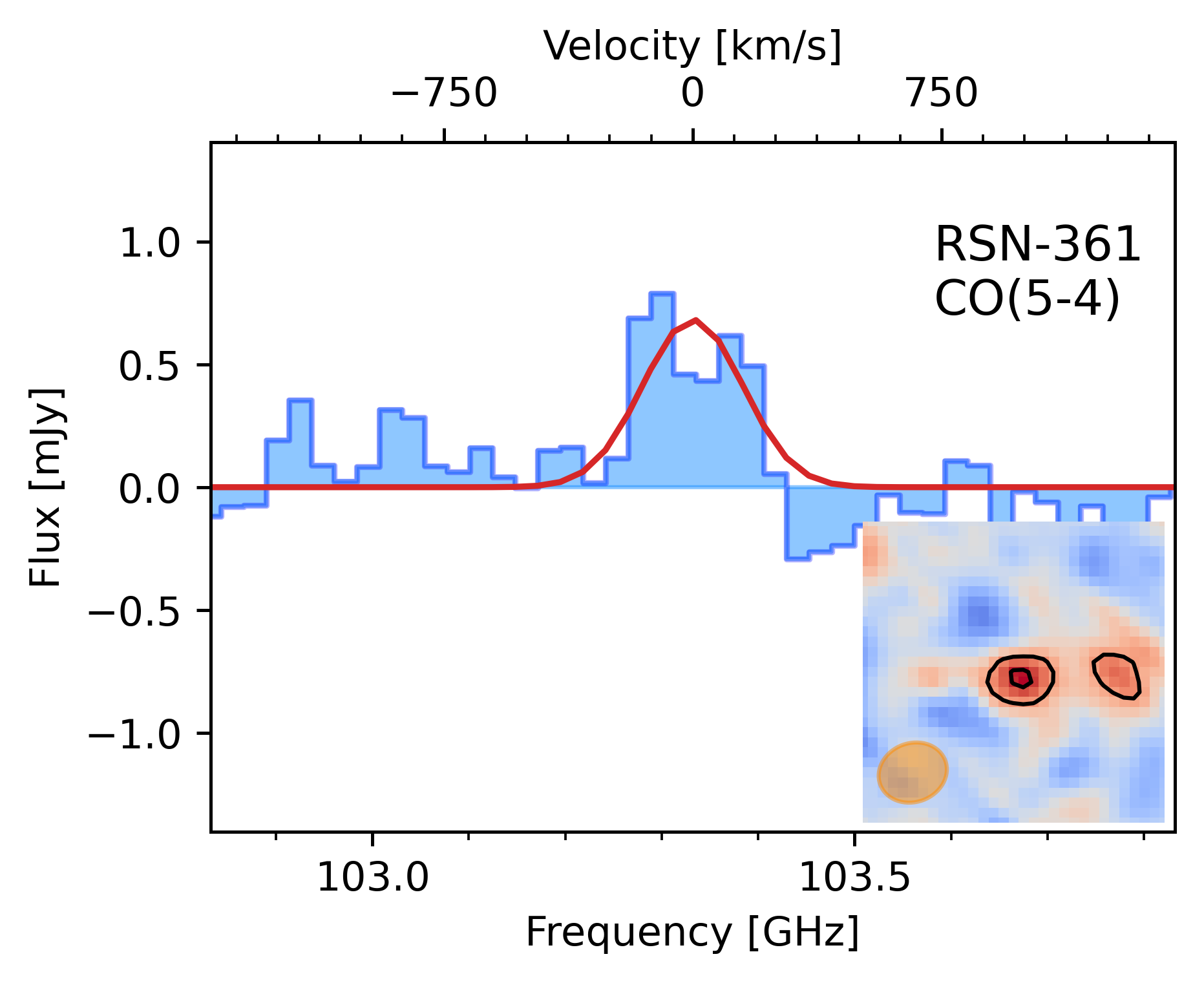}
\includegraphics[width=0.32\textwidth]{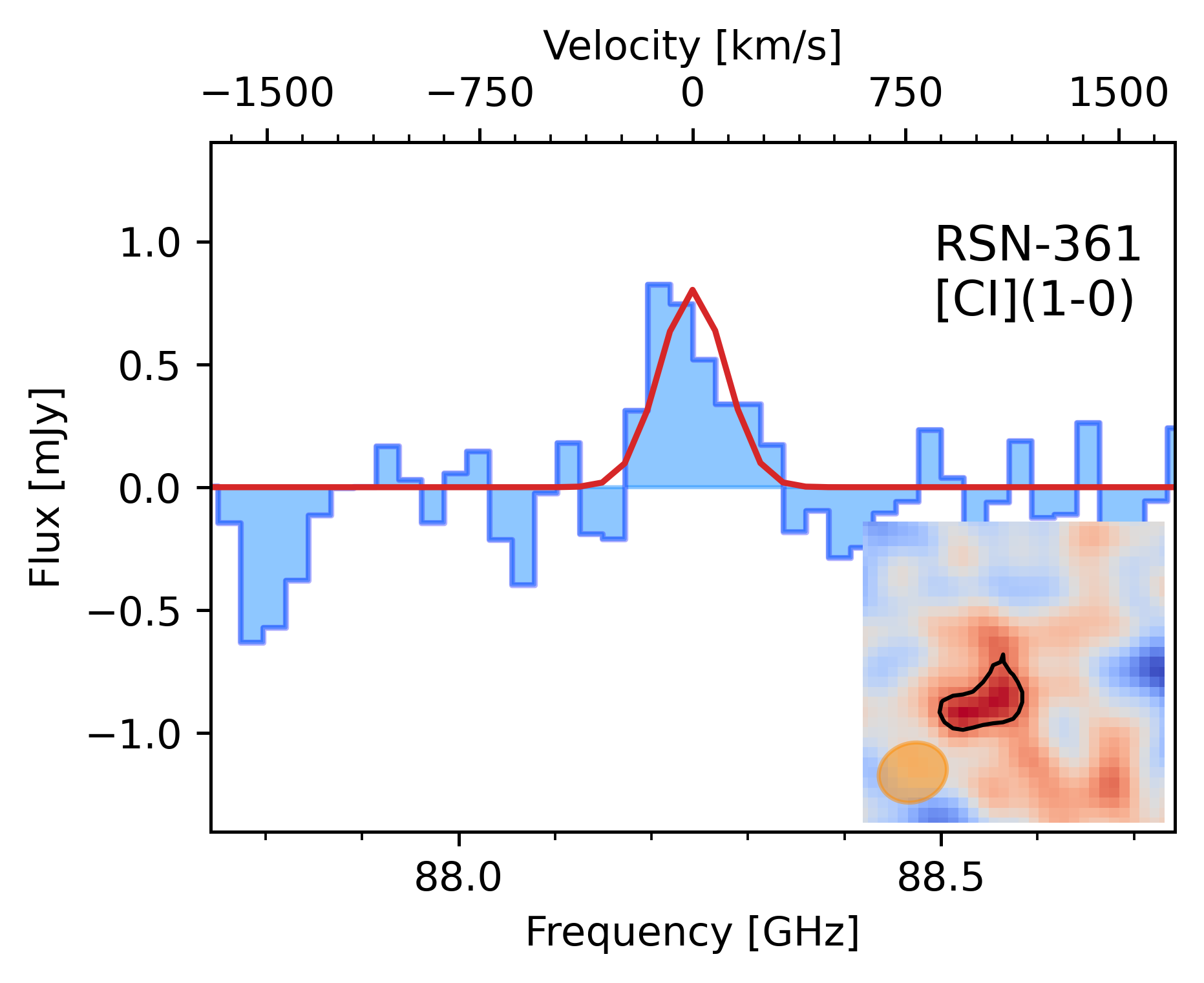} \\
\includegraphics[width=0.32\textwidth]{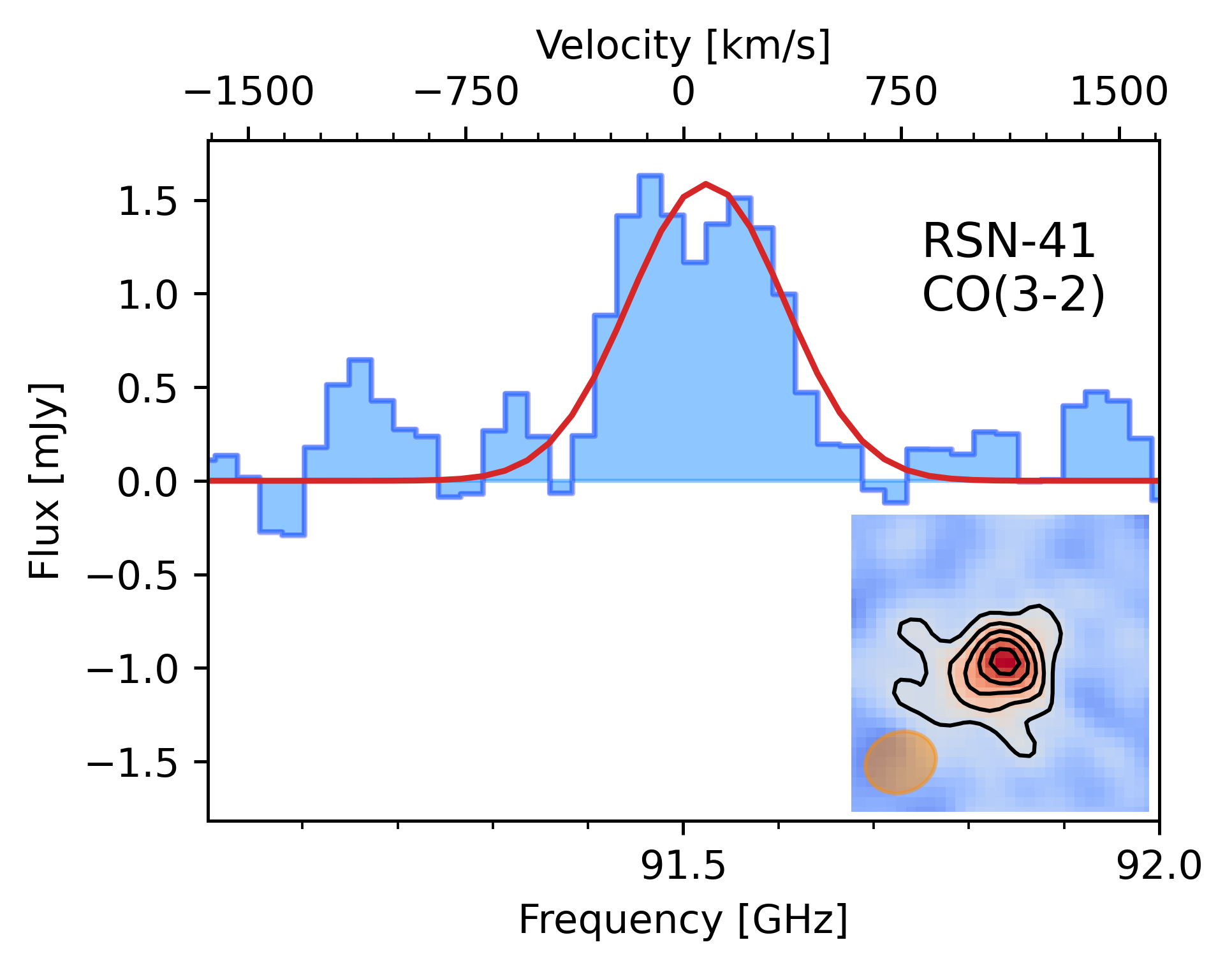}
\includegraphics[width=0.32\textwidth]{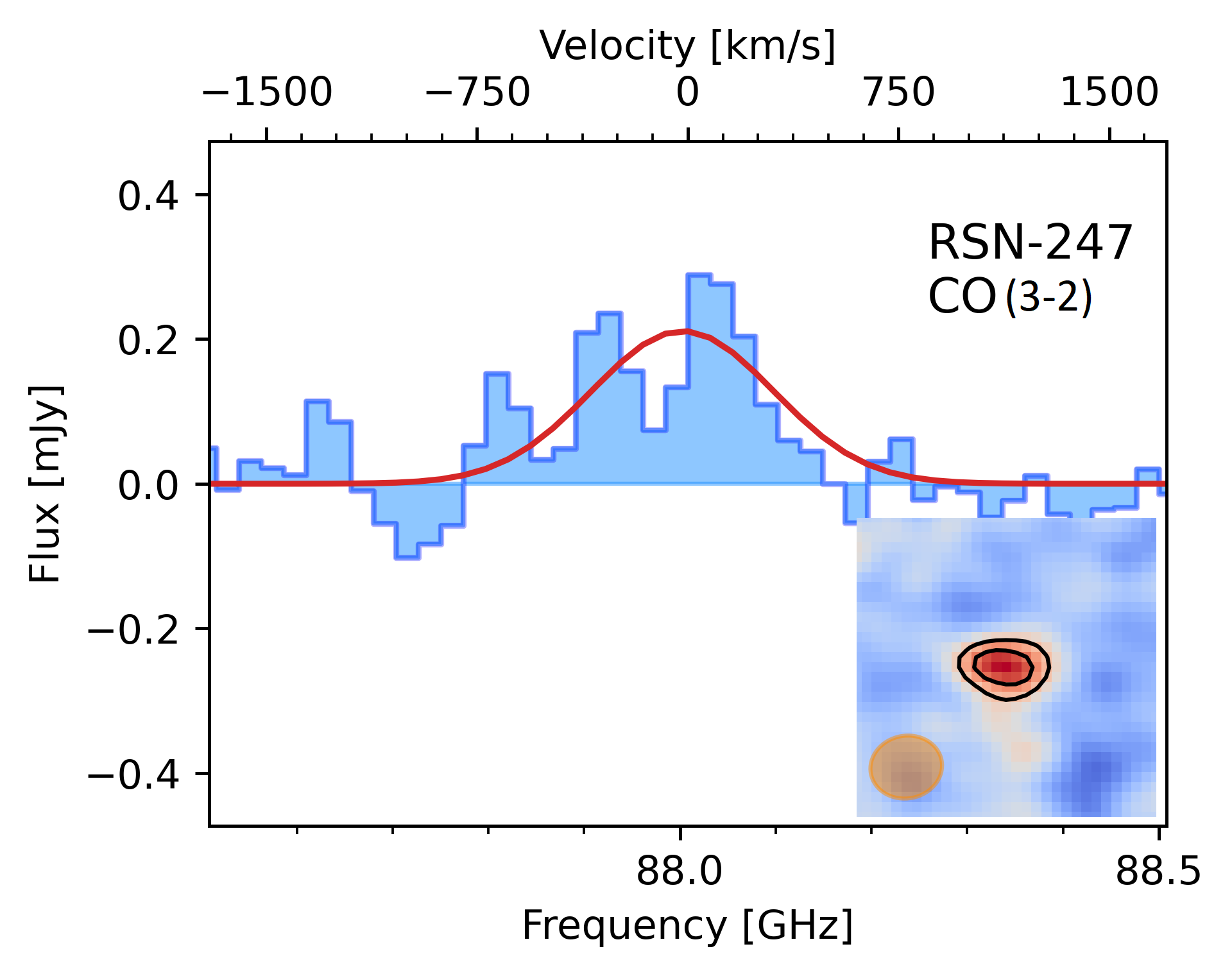} 
\includegraphics[width=0.32\textwidth]{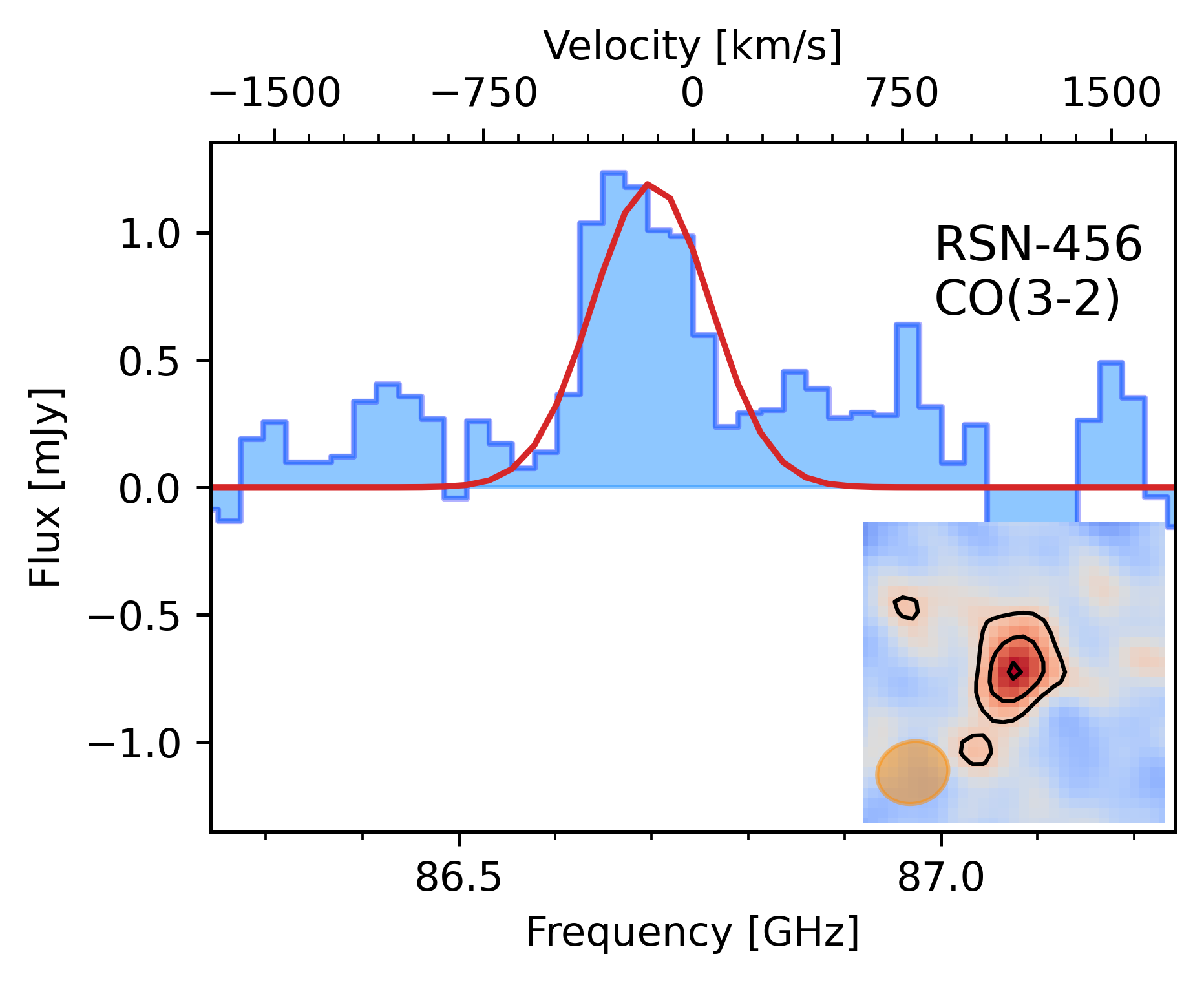}
\caption{\textit{(Continue)}}
\end{figure*}

This procedure results in a list of possible lines, with the related S/N and Full Width Zero Intensity (FWZI)\footnote{The FWZI is an immediate product of the procedure employed to identify the lines in the extracted spectra {since it corresponds to the binning maximizing the S/N of the line. However, in \Tab\ref{tab:lines}, we report the more common Full With at Half Maximum (FWHM) obtained through the Gaussian modeling described in \Sec\ref{sec:kinematics}}.}. However, given the nature of the noise in interferometric data, it is generally needed to establish the reliability of each line $R=1-p$ (where $p$ is the probability of a spurious detection). Since several methods exist to compute this quantity for interferometric data, in this study we follow two complementary approaches.
\begin{itemize}
    \item Following \citet{Jin_19}, we compute the spurious probability of each line as 
    \begin{equation}
    \centering
    \label{eq:jin}
        p({\rm S/N})=1-R_0({\rm S/N})^{N_{\rm Eff}}
    \end{equation}
    
    where $R_0$ is the reliability expected in the Gaussian case (and -- therefore -- approaching the unity towards higher S/N) and $N_{\rm Eff}$ is the "number of effective searches". Through an extensive series of simulations, \citet{Jin_19} estimated that this quantity can be approximated through the relation
    \begin{equation}
    \centering
        N_{\rm Eff}\sim10\frac{N_{\rm total, ch}}{N_{\rm line,ch}}N_{\rm Line,ch}^{0.58}\log\frac{N^{\rm max}_{\rm line,ch}}{N^{\rm min}_{\rm line,ch}}
    \end{equation}
    where $N_{\rm total, ch}$ is the total number of channels inside the datacube, $N_{\rm line, ch}$ is the number of channels in which the line is detected, and ${N^{\rm max}_{\rm line,ch}}$ and ${N^{\rm min}_{\rm line,ch}}$ are, respectively, the minimum and maximum width of the boxcar kernels employed during the line search (see point 4).
    While this approach is based on simulations and is not dependent on the properties of the actual cube, it relies on the hypothesis that the noise at the phase center of our observations is approximately Gaussian (reasonable assumption given the almost complete $uv$ coverage generally produced by ALMA).
    \item Following \citet{Walter_16}, we compute the reliability of each line through the \textsc{FindClumps} algorithm as implemented in the Python library \textsc{Interferopy}. This method estimates the reliability as 
    \begin{equation}
    \centering
        R({\rm S/N})=1-\frac{N_P ({\rm S/N})}{N_N ({\rm S/N})}
    \end{equation}
    where $N_P$ and $N_N$ are, respectively, the number of positive and negative peaks in the whole datacube in a given S/N bin. This approach does not rely on any assumption about nature of the noise in our datacubes, but it could be biased by the small statistics affecting the number of pixels in our observations.
\end{itemize}
The procedures described here allowed us to identify in all the targets at least one bright (S/N$>6$) line (see \Tab\ref{tab:lines}). Given the high S/N of all the detected lines, we can estimate for all of them a spurious probability lower than $10^{-6}$ following \citet{Jin_19}. Similarly, since they have a S/N higher than every negative peak in the analysed datacubes, we can estimate for all of them a 100\% reliability following \citet{Walter_16}. By producing the 0th moment of each (continuum-subtracted) line through the \textsc{CASA} task "\textsc{immoments}", we obtain the maps reported in the insets in \Fig\ref{fig:spectra_lines}. Moreover, by performing an aperture photometry with \textsc{CARTA} on the 2$\sigma$ contour of these maps, we measure the integrated line fluxes reported in \Tab\ref{tab:line_flux}.

\subsection{Redshift estimation}
\label{sec:redshifts}

Once detected the different lines present in our datacubes, we estimate the spectroscopic redshift of our sources following \citet{Jin_19}. For doing so, we consider the line with the highest S/N in each cube (i.e. that with the highest reliability) and model it as each of the CO transitions that should be visible in the redshift range $0<z<8$ (see \Fig\ref{fig:setup}). For all the redshifts higher than 3 –- for most of which a second line is expected -- we search for a detection at the expected frequency in the line list produced in \Sec\ref{sec:lines}. Through the S/N of each detection, we compute the reliability of the tentative second lines through \Eq\ref{eq:jin}. It is crucial to underline how -- for the second line -- the number of effective searches ($N_{\rm Eff}$) is much lower than those employed in \Sec\ref{sec:lines}. In this case, we are not performing an active search of the line throughout the whole spectrum, but we are only analyzing the frequencies allowed by the first line. Therefore, the $N_{\rm Eff}$ just corresponds to the number of possible CO transitions with which we can model the first line. For each redshift, we can finally estimate a joint spurious probability given by the product between the spurious probability of the first and the second line. The redshift with the highest reliability is assumed to be the spectroscopic redshift of our sources.

This approach is sufficient for all the galaxies in which two lines are robustly detected. However, for four of our targets, no second line is detected at a sufficiently high S/N in the analysed spectra. For these galaxies, we assess the redshift taking advantage of the additional information coming from the photometry. More in detail, once added the continuum datapoint at 3 mm (\Tab\ref{tab:cont}) to the photometry presented in \Sec\ref{sec:ancillary} (or an upper limit for the galaxies undetected in the continuum images), we perform a SED-fitting through the code \textsc{Cigale} (\citealt{Boquien_19}; see more details on the employed models in \Sec\ref{sec:SEDFitting}), fixing the redshift to all the spec-\textit{z} allowed by the single line identified in the spectrum. Hence, we assume as the final value for the spec-\textit{z} the redshift with the best agreement between the modeled SED and the photometry (i.e. that with the lowest value of the $\chi^2$). Remarkably, for one of the remaining galaxies (namely, RSN-182), the redshifts estimated through this procedure falls in a small range of frequency at $z\sim4.7$ where a single line (CO(5-4)) is expected. Similarly, three targets (namely, RSN-41, RSN-247, and RSN-456) have a redshift lower than three, where -- according to our spectral setup -- no second line is expected to be observed. We also report a "tentative" second line in RSN-361 at $\nu=88.13$ GHz. Even though this detection falls exactly where the [CI](1-0) line would be expected for a galaxy at $z=4.585$, the low $S/N\sim3.5$ and the spatial offset with the robustly detected line at $\nu=103.19$ GHz make the line identification unsure. Finally, we underline that continuum image (see \Fig\ref{fig:continuum}) shows a quite irregular morphology for this source, suggesting the possible presence of a major merger (that could explain the spatial offset of the tentative [CI](1-0) line). {It is important to notice how these redshifts based on a single detected line are clearly more uncertain than those relying on a double detection. We cannot exclude that a fainter line (e.g. a [CI](1-0)) would be observed in our frequency range with deeper observations. For instance, given the redshift estimated through CO(4-3) and CO(5-4) in RSN-298, we would expect a [CI](1-0) line at $\nu=95.57$ GHz in that source, even though nothing is detected at that frequency at S/N larger than 1$\sigma$. For the other galaxies where a single line is detected, however, these solutions would be disfavored by the photometry.}

\begin{figure*}
    \centering    \includegraphics[width=0.9\textwidth]{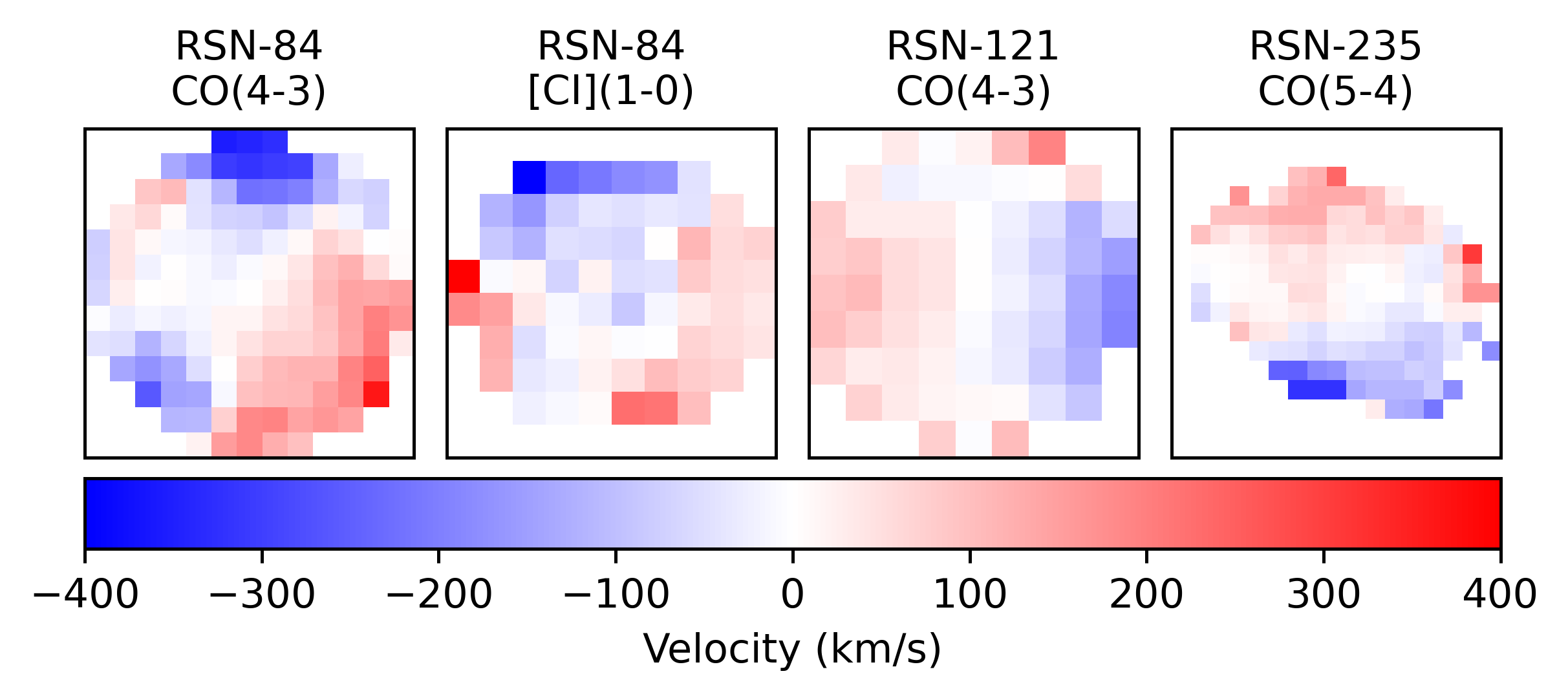}
    \caption{Moment 1 maps {(2.5'' $\times$ 2.5'')} of the lines detected in three of the targets. According to our modeling, the maps show clear evidence of a rotating structure or a late stage of a merger.}
    \label{fig:Mom1_84}
\end{figure*}

\begin{table*}
\centering
\begin{threeparttable}
\renewcommand{\arraystretch}{1.1}
\caption{Physical properties for our sources through an SED-fitting performed with the code \textsc{Cigale} \citep{Boquien_19} once assumed the spectroscopic redshifts retrieved in \Sec\ref{sec:redshifts}. The last two columns report the molecular gas mass and the depletion time as estimated in \Sec\ref{sec:gas_evolution}. The last two rows report the median value computed on the galaxies analysed in this sample and on the whole sample analysed in \citet{Gentile_24} {with \textsc{Cigale}.}}
\label{tab:properties}
\begin{tabular}{cccccccc}
\toprule
ID & log(M$_\star$) & log(SFR) & log(L$_{\rm IR}$) & A$_{\rm V}$ & log(M$^{CO}_{\rm H_2}$) & log(M$^{[CI]}_{\rm H_2}$) & $\tau_{\rm D}$$^{(a)}$ \\
& [M$_\odot$] & [M$_\odot$ yr$^{-1}$] & [L$_\odot$] & [mag] & [M$_\odot$] & [M$_\odot$] & [Myr]\\
\midrule
RSN-41 & $11.0\pm0.1$ & $2.72\pm0.06$ & $12.56\pm0.06$& $4.9\pm0.2$ & $11.10\pm0.06$ & - & $242\pm50$ \\
RSN-84 & $11.0\pm0.1$ & $2.76\pm0.03$ & $12.57\pm0.03$ & $4.8\pm0.2$& $10.86\pm0.08$ & $11.17\pm0.06$ & $258\pm53$ \\
RSN-121 & $11.0\pm0.1$ & $2.65\pm0.06$ & $12.47\pm0.06$ & $4.2\pm0.2$&  $11.11\pm0.07$ & $11.01\pm0.07$ & $229\pm60$\\
RSN-182 & $11.1\pm0.1$ & $2.92\pm0.03$ & $12.72\pm0.03$ & $3.8\pm0.2$ &  $11.02\pm0.08$ & - & $126\pm27$ \\
RSN-235 & $11.3\pm0.1$ & $3.04\pm0.03$ & $12.83\pm0.04$& $4.5\pm0.2$ & $11.07\pm0.08$ & $10.96\pm0.08$ & $83\pm18$ \\
RSN-247 & $10.7\pm0.2$ & $2.57\pm0.04$ & $12.46\pm0.04$& $4.6\pm0.4$ & $10.71\pm0.09$  & - & $140\pm31$ \\
RSN-298 & $11.0\pm0.1$ & $2.69\pm0.06$ & $12.58\pm0.06$ & $4.2\pm0.3$ & $10.81\pm0.08$ & - & $132\pm30$ \\
RSN-361 & $10.8\pm0.3$ & $2.79\pm0.03$ &$12.61\pm0.03$ & $3.9\pm0.6$ & $10.7\pm0.1$ & $11.06\pm0.06$ & $186\pm34$  \\
RSN-456 & $10.8\pm0.2$ & $2.41\pm0.07$ & $12.37\pm0.07$ & $4.6\pm0.3$ & $10.93\pm0.06$ & - & $288\pm63$ \\
\midrule
Median$^{(b)}$ & $11.0\pm0.1$ & $2.6\pm0.1$ & $12.6\pm0.1$ & $4.5\pm0.4$ & $10.9\pm0.2$ & $11.03\pm0.07$ & $263\pm86$\\
\midrule
G24$^{(b)}$ & $11.0\pm0.4$ & $2.8\pm0.4$ & $12.6\pm0.4$ & $3.9\pm0.3$ & $-$ & $-$ & $-$\\
\bottomrule
\end{tabular}
\begin{tablenotes}
    \item{$^{(a)}$ Estimated from the $M^{[CI]}_{H_2}$ when available.}
    \item{$^{(b)}$ The uncertainties on the median quantities are reported as half the interval between the 84th and the 16th percentile}
\end{tablenotes}
\end{threeparttable}
\end{table*}

\subsection{Initial insights on the ISM kinematics}
\label{sec:kinematics}

As shown in \Tab\ref{tab:lines}, most of our targets have quite broad lines (FWHMs of several hundreds of km/s). This result is familiar for what concerns high-\textit{z} DSFGs and ULIRGs in general (see e.g. \citealt{Jin_19,Jin_22,Cox_23}), and it is generally explained through an ISM much more turbulent than what is commonly observed in local galaxies. In this study – however – we can perform a more detailed analysis on some targets, since the FWHM of the lines is much larger than the spectral resolution requested in our observation. This property allows us to infer some initial insights into the ISM kinematics inside our galaxies. As visible in \Fig\ref{fig:spectra_lines}, most of the lines observed in our targets have a peculiar morphology, suggesting the possible presence of two peaks in the observed line spectrum. This result could be explained by the presence of a kinematically-decoupled component such as in a disk or in the later stage of a merger. To decide in a statistically motivated way if our lines should be modeled with a single or double component, we perform a test hypothesis. In our case, the null hypothesis consists in the modeling of the line with a single Gaussian, while the alternative hypothesis consists in the modeling with two Gaussians. Since we are employing two \textit{nested} models {(with four and seven free parameters, respectively, since we allow a residual continuum component)}, we perform a partial F-test \citep[e.g.][]{Bevington_03} employing a threshold of 0.05 for the level of significance to reject the null hypothesis. Considering only the highest S/N line in each spectrum, we obtain that the presence of a double component is statistically significant for five out of nine targets ($\sim 55\%$). {For most the galaxies where two lines are detected (with the notable exception of RSN-84), the lower S/N of the second line does not allow us to conclude that the additional component is statistically required for a correct modeling. For all the lines where the double model is statistically motivated, we report the best-fitting parameters in \Tab\ref{tab:decomposition}.} If we compare our fraction of double-peaked lines with other similar studies in the current literature presenting spectroscopic follow-up of SMGs at (sub)mm wavelengths, we find that our percentage is higher than the $\sim 30$\% reported by \citet{Bothwell_13} (detecting CO emission in a large sample of 32 SMGs in the redshift range $1.2<z<4.1$) and \citet{Aravena_16} (observing 17 lensed DSFGs at $2.5<z<5.7$). {and of the $\sim40\%$ of double-peaked profiled reported by \citet{Birkin_20} (studying 61 ALMA-detected SMGs)}. Unfortunately, the limited size of our sample does not allow us to unambiguously establish if this difference is due to the different selections or it is a consequence of the different S/N achieved by the different observations. The presence of two components with different velocities in our galaxies can be explained with both the presence of a rotating structure or as the signature of the late stage of a major merger. These hypotheses are also strengthened by the 1st moments of the CO/[CI] lines within three of our targets (those with the highest S/N in the CO-lines; namely, RSN-84, RSN-121, and RSN-235. In RSN-84 the same structure is visible in both the CO(4-3) and [CI](1-0) lines; see \Fig\ref{fig:Mom1_84}). Unfortunately, the spatial resolution of our observations is not sufficient to distinguish between the two possible models (i.e. presence of disk or merger). Similarly, the coarse spatial and spectral resolution does not allow us to perform a proper modeling of the possible disk (see e.g. \citealt{DiTeodoro_15,RomanOliveira_23}). Constraining the deprojected velocity of the gas and its velocity dispersion would allow us to determine if the structure is stable or not. This result would be of crucial importance to constrain some of the evolutionary models of massive galaxies (see \Sec\ref{sec:evolution}).

\begin{table}
\centering
\begin{threeparttable}
\renewcommand{\arraystretch}{1.1}
\caption{Integrated fluxes of the lines detected in our sources. All the values are obtained through aperture photometry on the 0th moment of each line with an aperture equal to the $2\sigma$ contour.}
\label{tab:line_flux}
\begin{tabular}{ccccc}
\toprule
ID & I$_{\rm CO(3-2)}$ & I$_{\rm CO(4-3)}$ & I$_{\rm CO(5-4)}$ & I$_{[CI](1-0)}$ \\
& [Jy km/s] & [Jy km/s] & [Jy km/s] & [Jy km/s]\\
\midrule
41  & $1.06\pm0.07$ & - & - & - \\
84 & - &$0.55\pm0.04$ & - & $0.53\pm0.08$\\
121 & - & $1.03\pm0.04$& - & $0.40\pm0.07$  \\
182 & - & - & $0.61\pm0.05$ & -\\
235 & - & - & $0.71\pm0.04$ &  $0.21\pm0.04$ \\
247 & $0.40\pm0.06$ & - & - &  - \\
298 & - & $0.37\pm0.03$ & $0.47\pm0.06$ & - \\
361 & - & - & $0.31\pm0.04$ & $0.27\pm0.04$ \\
456 & $0.63\pm0.03$ & - & -& -\\
\bottomrule
\end{tabular}
\end{threeparttable}
\end{table}

\begin{table}
\centering
\begin{threeparttable}
\renewcommand{\arraystretch}{1.1}
\caption{{Best-fitting parameters for the Gaussian modeling when two components are employed as described in \Sec\ref{sec:kinematics}. The three columns report the IDs of the galaxies, the velocity offset between the two Gaussian components, and the FWHM of each component.}}
\label{tab:decomposition}
\begin{tabular}{cccc}
\toprule
ID & $\Delta v$ & FWHM (red) & FWHM (blue) \\
& [km/s] & [km/s] & [km/s]\\
\midrule
84 - CO(4-3) & $(366\pm61)$ & $(191\pm99)$ & $(401\pm114)$ \\
84 - [CI](1-0) & $(552\pm39)$ & $(234\pm51)$ & $(305\pm77)$ \\
121 & $(449\pm33)$ & $(289\pm47)$ & $(319\pm68)$ \\
182 & $(219\pm321)$ & $(387\pm440)$ & $(203\pm214)$ \\
235 & $(290\pm132)$ & $(376\pm222)$ & $(233\pm78)$ \\
298 & $(302\pm58)$ & $(208\pm129)$  & $(208\pm66)$ \\
\bottomrule
\end{tabular}
\end{threeparttable}
\end{table}

\subsection{SED-Fitting}
\label{sec:SEDFitting}

\begin{figure*}
    \centering
    \includegraphics[width=0.48\textwidth]{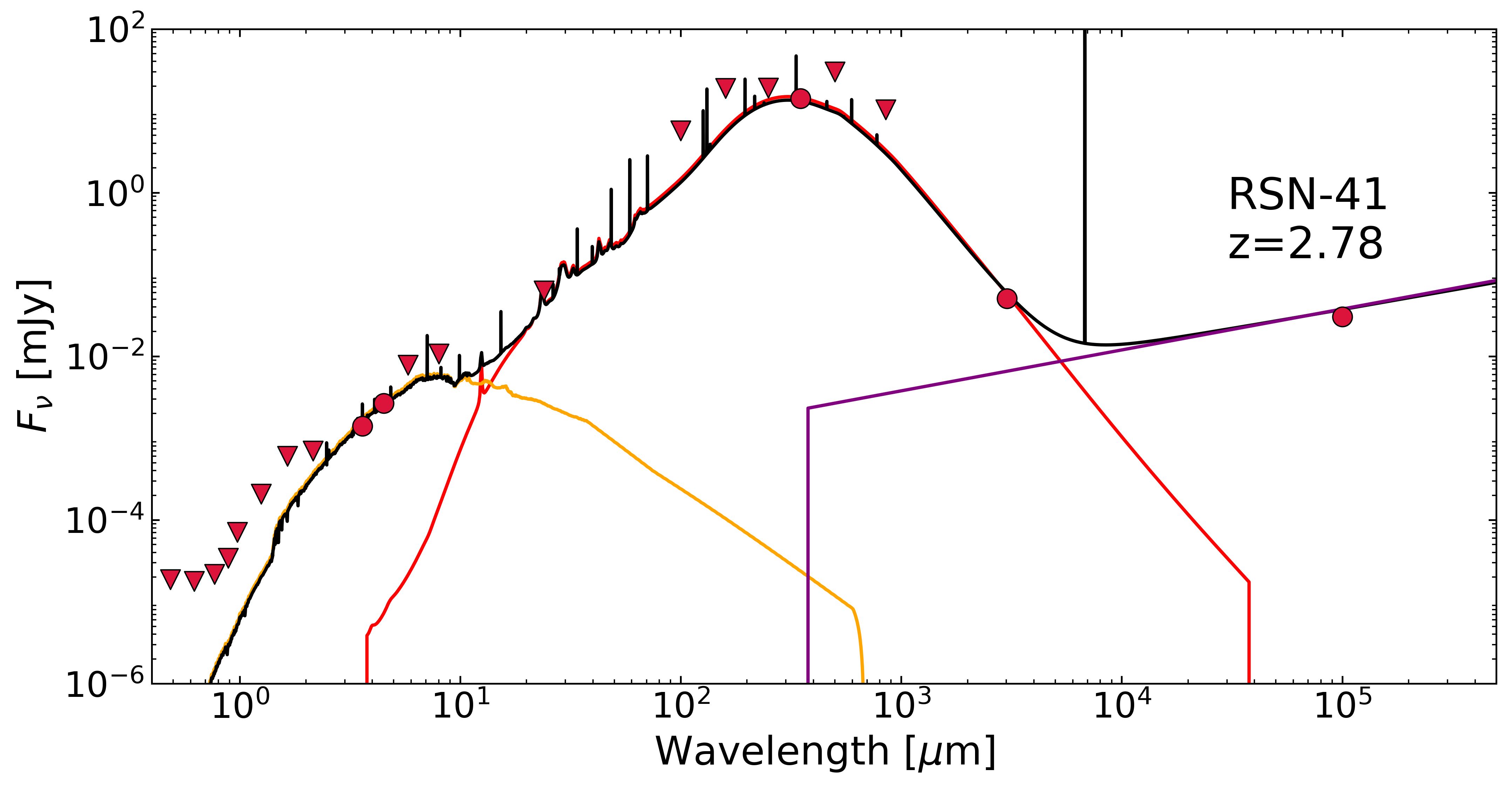} \includegraphics[width=0.48\textwidth]{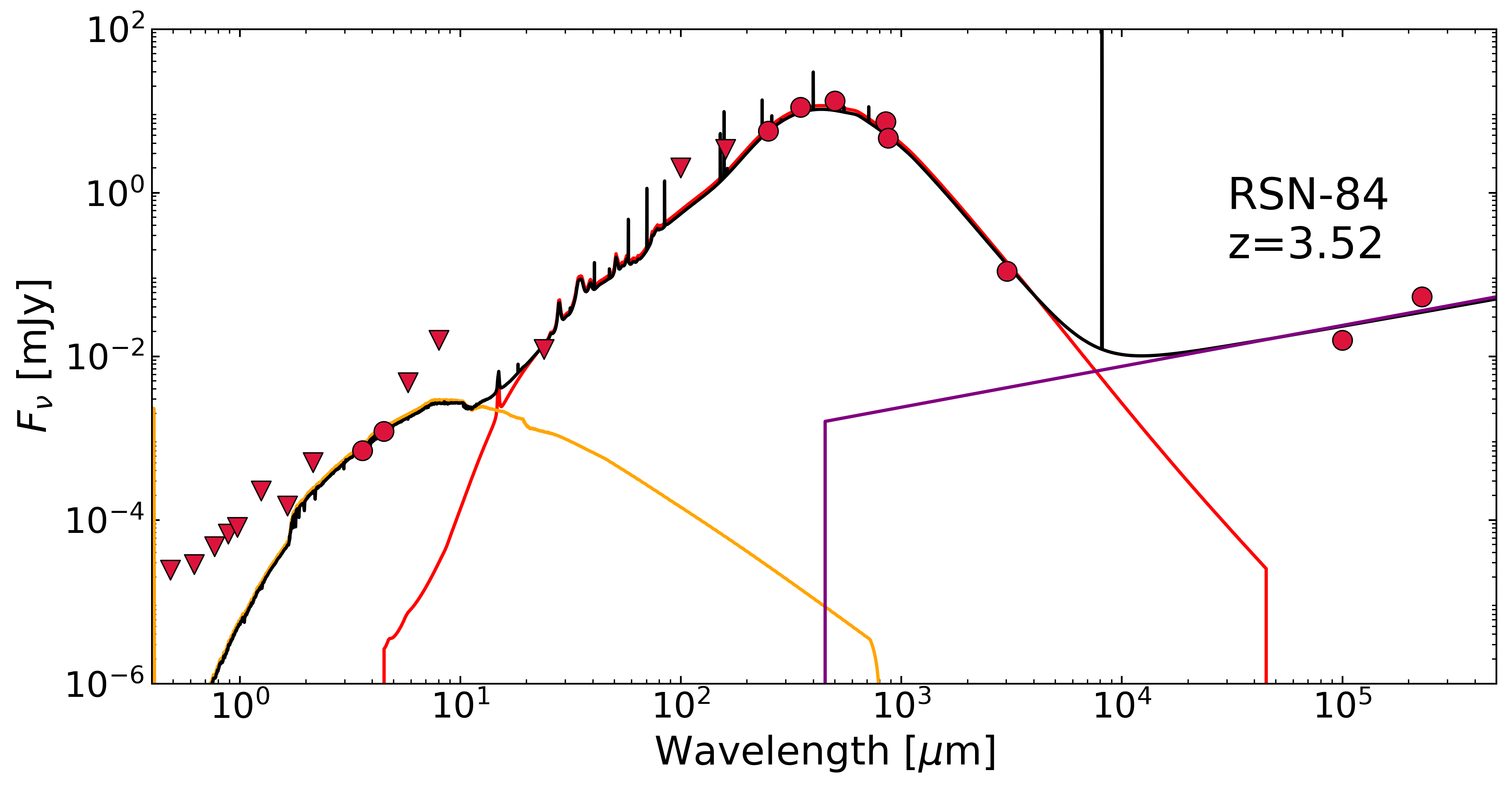} \\\includegraphics[width=0.48\textwidth]{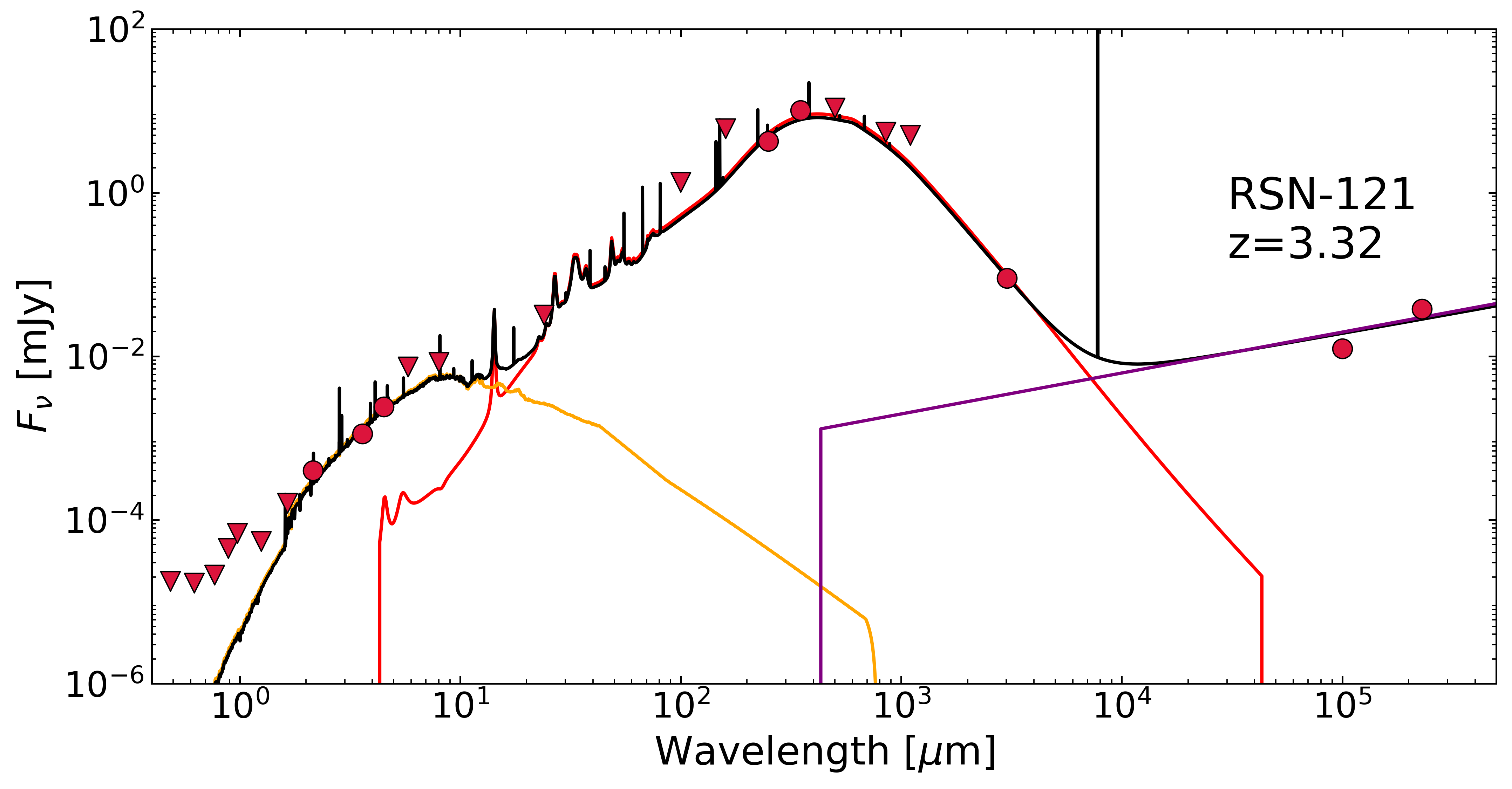} \includegraphics[width=0.48\textwidth]{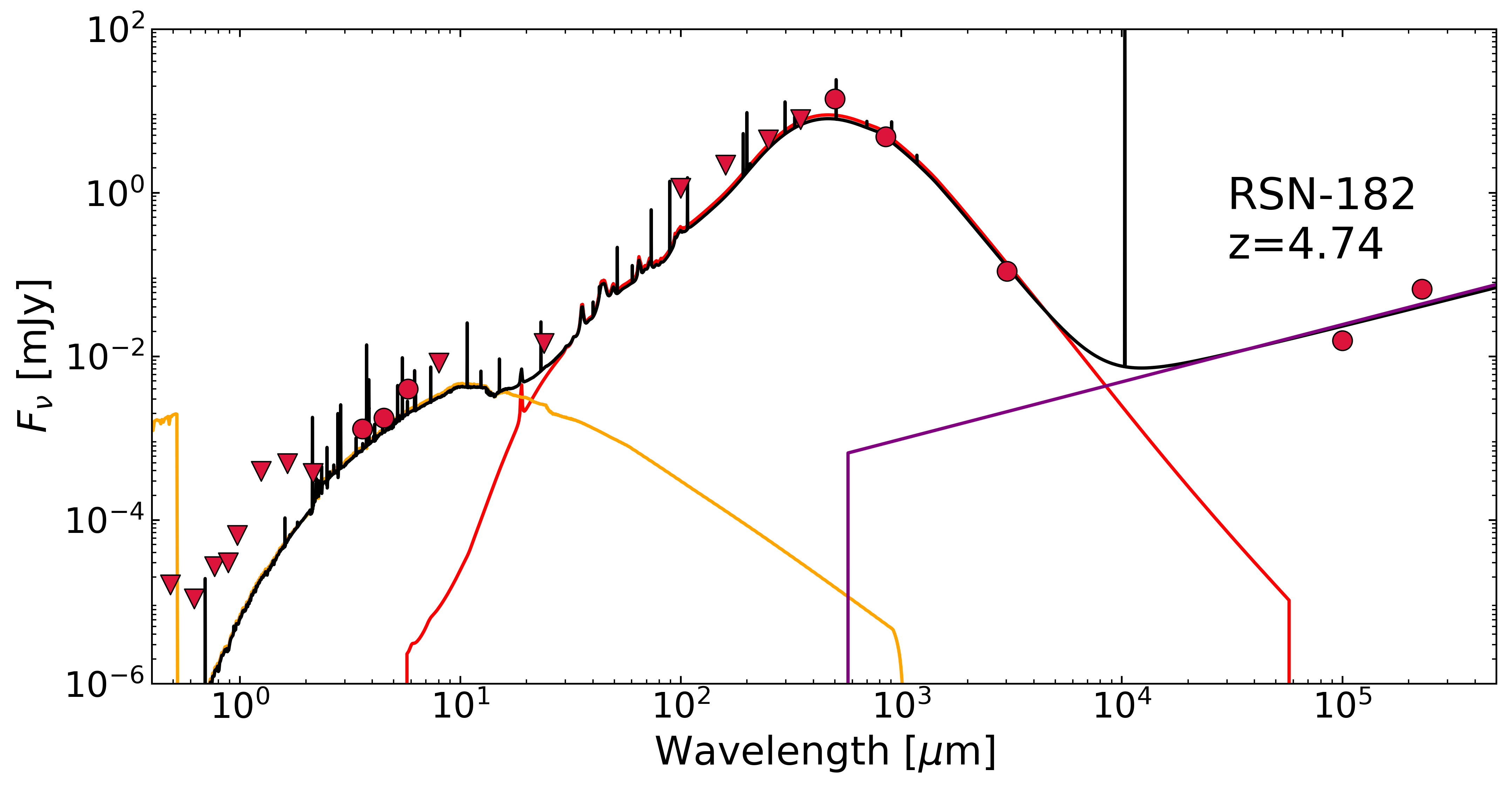} \\\includegraphics[width=0.48\textwidth]{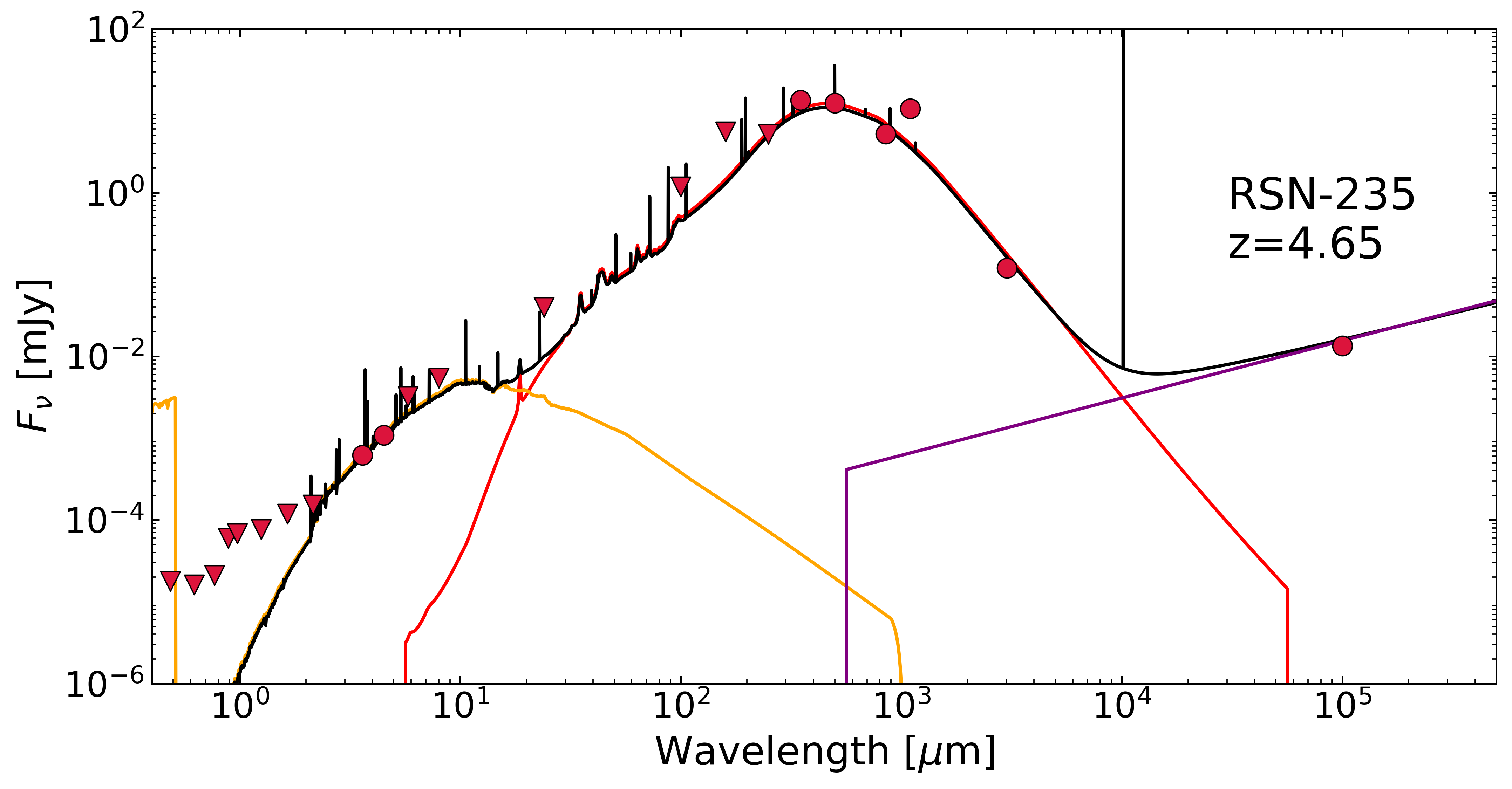}  \includegraphics[width=0.48\textwidth]{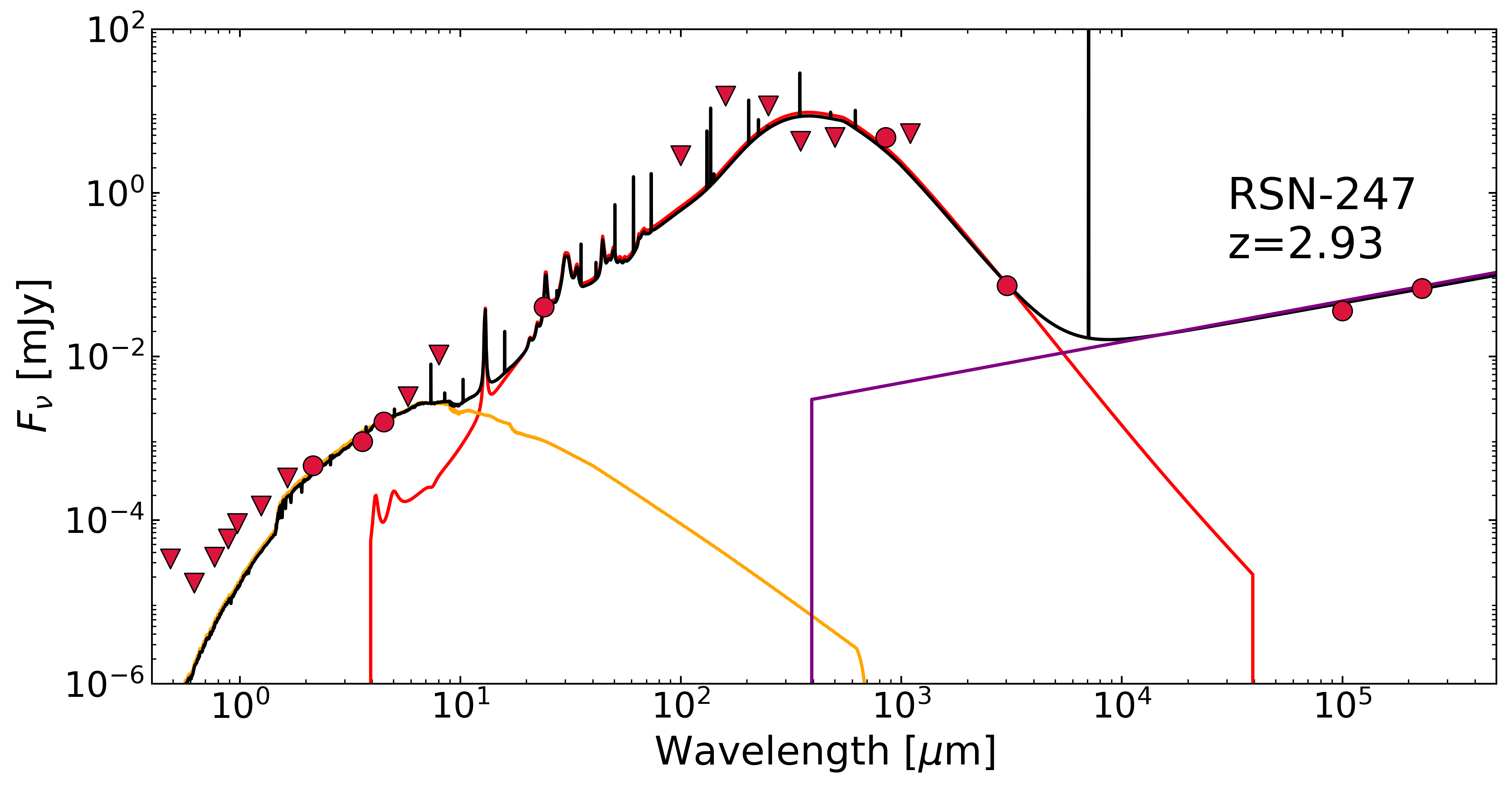} \\
    \includegraphics[width=0.48\textwidth]{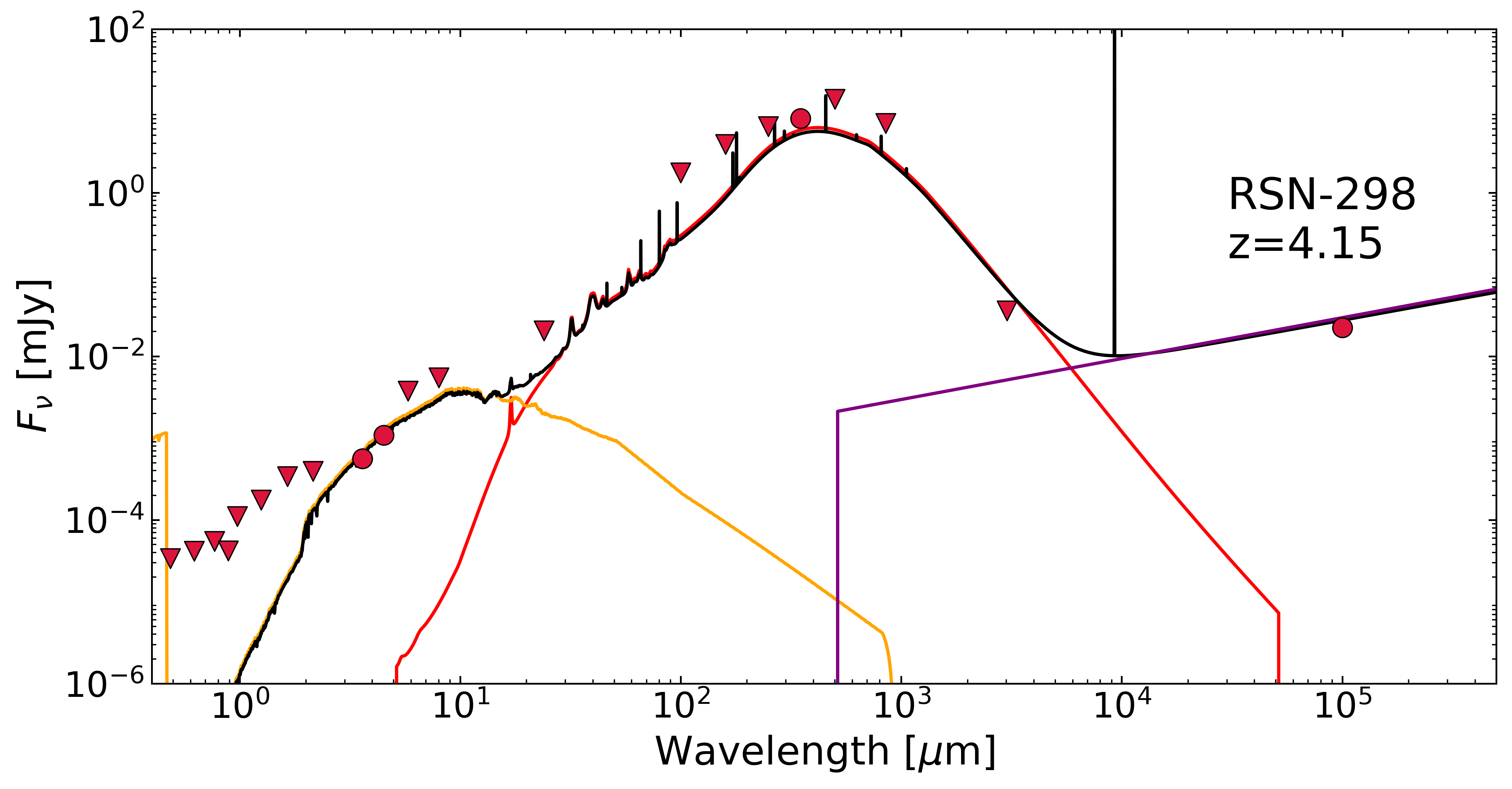} \includegraphics[width=0.48\textwidth]{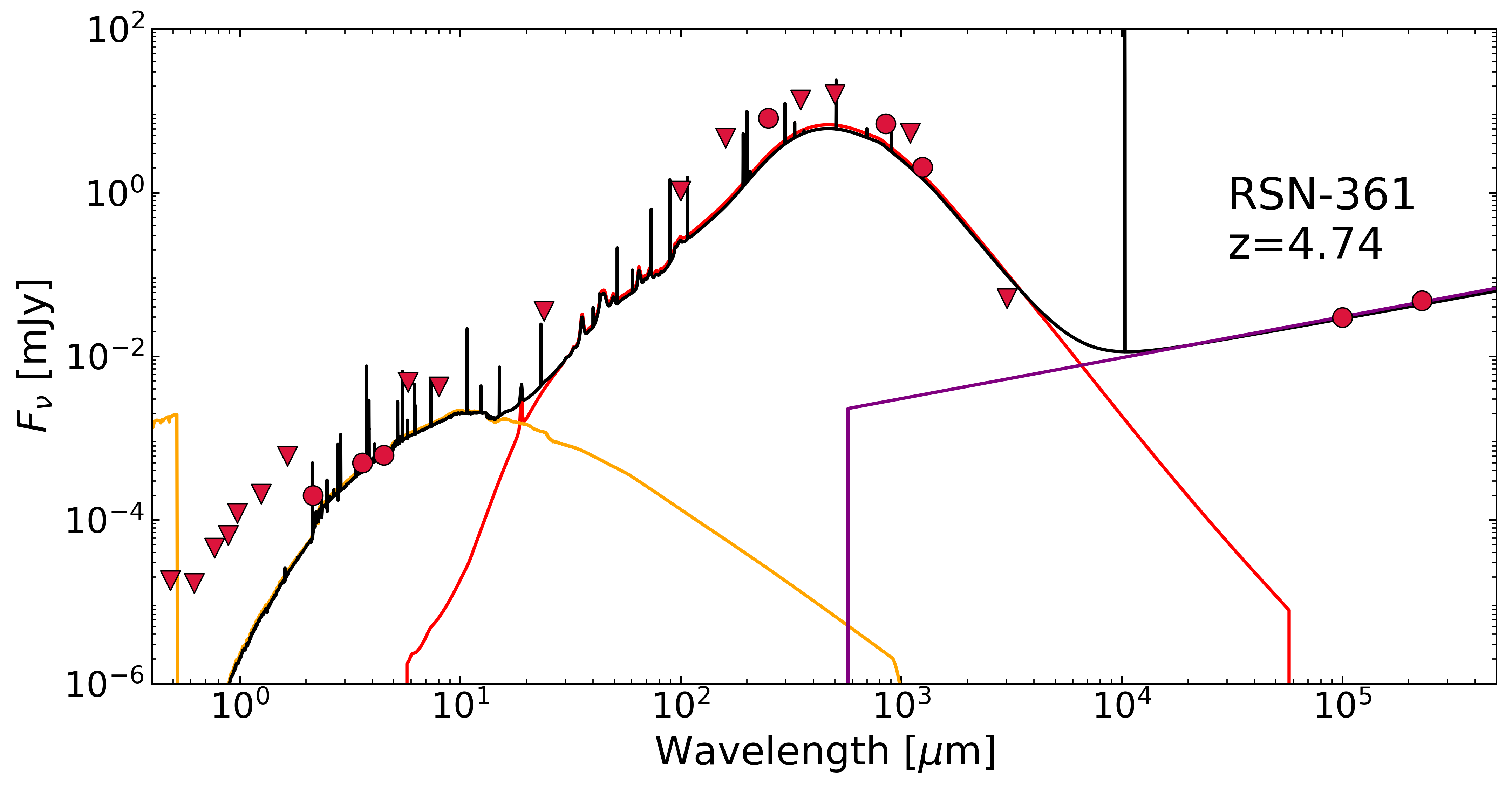} \\\includegraphics[width=0.48\textwidth]{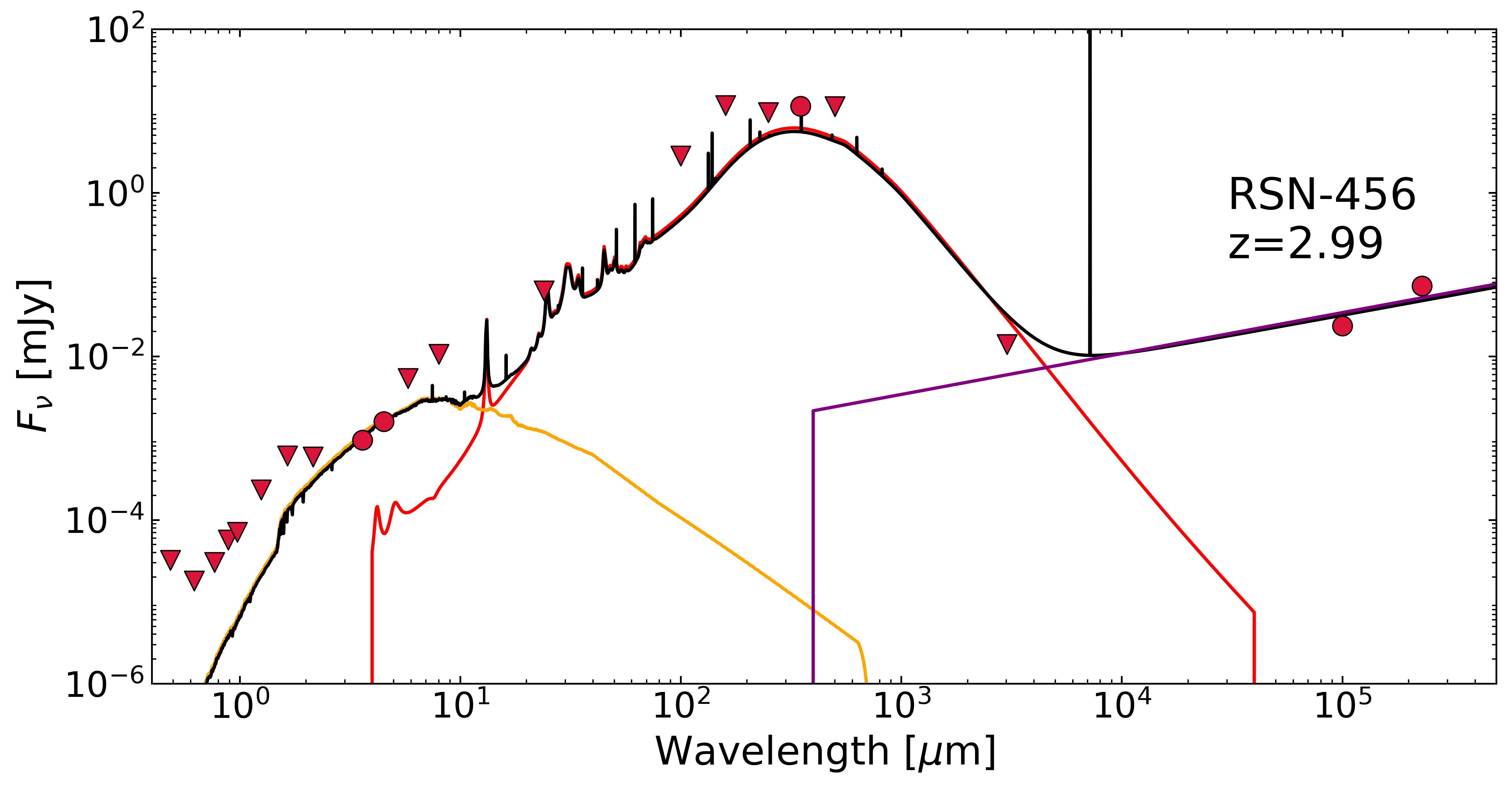}
    \caption{Best-fitting SEDs of our targets as computed by \textsc{Cigale} \citep{Boquien_19}. The different emissions in the galaxies are color-coded: the attenuated stellar emission is reported as the orange line. Similarly, the dust emission is reported in red, while the radio emission is reported in purple. Finally, the black solid line report the best-fitting SED.}
    \label{fig:SED}
\end{figure*}

Once assessed the spectroscopic redshift of our sources, we estimate their physical properties through a SED-fitting with the code \textsc{Cigale} \citep{Boquien_19}. To account for all the processes taking place inside our targets, we employ several libraries, following the same strategy employed in \citet{Gentile_24}. Firstly, we include the stellar emission through the stellar population libraries by \citet{Bruzual_03}, combined through an exponentially declining star formation history with random bursts of star formation. The stellar attenuation is modeled following \citet{Charlot_00}, while the dust thermal emission is included in the templates through the \citet{Draine_14} models. Finally, radio emission is treated as described in \citet{Boquien_19}. We also explore the possible presence of AGN within our sources, by adding a dusty torus component as modeled by \citet{Fritz_06}. For all the models, we employ the parameters used by \citet{Donevski_20} in the analysis of their sample of DSFGs with \textsc{Cigale}. The results of the SED-fitting and all the physical properties estimated with \textsc{Cigale} are summarized in \Fig\ref{fig:SED} and in \Tab\ref{tab:properties}. More in detail, through SED-fitting we estimate the stellar mass ($M_\star$), the infrared luminosity ($L_{\rm IR}$), and the dust attenuation ($A_{\rm v}$). Since the photometric coverage in the FIR/(sub)mm/radio for our targets is significantly better than that in the optical/NIR, we estimate the SFR from the $L_{\rm IR}$ through the relation by \citet{Kennicutt_12} rescaled to a \citet{Chabrier_03} IMF. A last interesting parameter estimated through SED-fitting is the AGN fraction ($f_{\rm AGN}$), defined as the ratio between the luminosity of the host galaxy and that of the AGN in the wavelength range $[5,40]\mu$m. Interestingly, \textsc{Cigale} reports a $f_{\rm AGN}=0$ for all our targets, therefore we can safely conclude that the photometry of our galaxies can be correctly reproduced without requiring a dusty torus component. {Another indication about the lack of powerful AGN contribution in our sample comes from the estimation of the $q_{\rm TIR}$ from the infrared luminosity (computed through SED-fitting) and the radio flux \citep[see e.g.][]{Helou_85} The latter is converted into a 1.4 GHz radio luminosity through the spectroscopic redshift and the radio slope measured through the radio fluxes at 3 GHz and 1.4/1.28 GHz. For the galaxies lacking a second radio detection (namely, RSN-41, RSN-235, and RSN-298), we assume the median slope computed on the rest of the sample. We obtain that all the $q_{\rm TIR}$ are in the range [2.45,2.55], in good agreement with what is commonly measured for star-forming galaxies \citep[e.g.][]{Yun_01}.}

\section{Results and discussion}
\label{sec:results}

\subsection{Analysis of the spec-\textit{z}}

\begin{figure}
    \centering
    \includegraphics[width=\columnwidth]{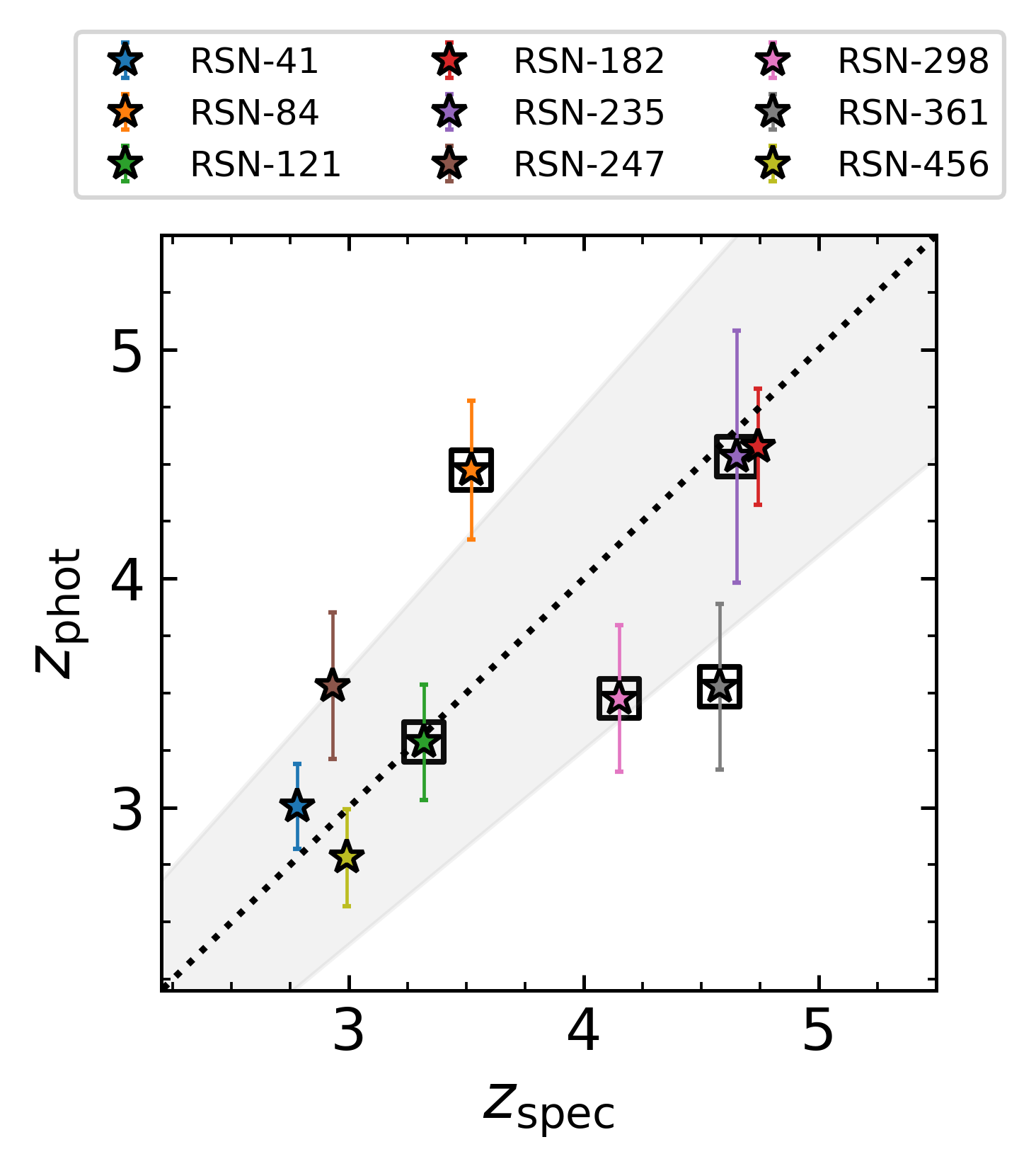}
    \caption{Comparison between the photometric redshift following \citet{Gentile_24} and the spectroscopic redshifts measured in this study. The one-to-one relation is reported as the dotted black line, while the gray shaded area reports the galaxies with $|\Delta z|/(1+z)<0.15$. The galaxies with a spec-\textit{z} obtained from the modeling of two lines are highlighted with an additional box.}
    \label{fig:comparison_zspec}
\end{figure}

As visible in \Tab\ref{tab:lines}, the nine galaxies analyzed in this paper have a spec-\textit{z} between 2.8 and 4.7, with a median value of $\sim 3.52$. Comparing these values with the photometric redshifts estimated following \citet{Gentile_24} (once added the continuum point at 3 mm obtained in \Sec\ref{sec:continuum}), we can notice quite a good agreement (see \Fig\ref{fig:comparison_zspec}). More quantitatively, we can measure the accuracy of our photo-\textit{z} as:
\begin{equation}
\centering
    {\rm median}\left(\frac{|z_{\rm phot}-\textit{z}_{\rm spec}|}{1+z_{\rm spec}}\right)=0.05
\end{equation}
when considering all the spec-\textit{z} assessed in \Sec\ref{sec:redshifts}. This result validates the procedure followed in \citet{Gentile_24} to estimate the photometric redshift of the RS-NIRdark galaxies and it is quite encouraging in planning future follow-ups for the high-\textit{z} candidates reported in that study. Finally, we underline that -- since for three of our galaxies with at least one line we have a counterpart in the COSMOS2020 catalog (see \Tab\ref{tab:coords}) -- we can retrieve three photo-\textit{z} from that catalog (namely $z=3.6\pm0.3$, $z=2.8\pm0.3$, and $4.6\pm0.3$ for RSN-84, RSN-247, and RSN-361, respectively). These quantities were computed by \citet{Weaver_22} with the two SED-fitting codes \textsc{Eazy} \citep{Brammer_08} and \textsc{LePhare} \citep{Arnouts_99,Ilbert_06} on the optical/NIR bands included in the COSMOS2020 catalog and on the first two channels of IRAC, and are in perfect agreement with the spectroscopic redshift estimated through (sub)mm spectroscopy (see \Tab\ref{tab:lines}). A last interesting comparison can be performed with the photometric redshifts computed by \citet{Talia_21} and employed to select the targets for these ALMA observations. Unfortunately, most the sources are located at a lower redshift than expected from that study (see \Tab\ref{tab:lines}). This difference can be explained by the several improvements in the photometry extraction and in the photo-\textit{z} estimation employed in \citet{Gentile_24} with respect to \citet{Talia_21}. {We expect most of the differences to arise because of the new deblending procedure based on the \textsc{PhoEBO} algorithm (allowing us to better extract the photometry from low-resolution maps; e.g. the four IRAC channels), the deeper IRAC images employed in \citet{Gentile_24}, and the more stringent upper limits employed in the photometric bands with no detections.} A full comparison between the two procedures can be found in \citet{Gentile_24}. {We expect this difference in the photo-\textit{z} computed by \citet{Gentile_24} and \citet{Talia_21} to produce different constraints on the contribution of the RS-NIRdark galaxies to the cosmic SFRD (especially at $z>4.5$, where most of the targets were located according to the previous photo-\textit{z}). This point will be addressed in detail through the analysis of the full sample of RS-NIRdark galaxies in COSMOS in a forthcoming paper (Gentile et al., in prep.)}

\subsection{Analysis of the physical properties}
\label{sec:properties_ms}

\begin{figure}
    \centering
    \includegraphics[width=\columnwidth]{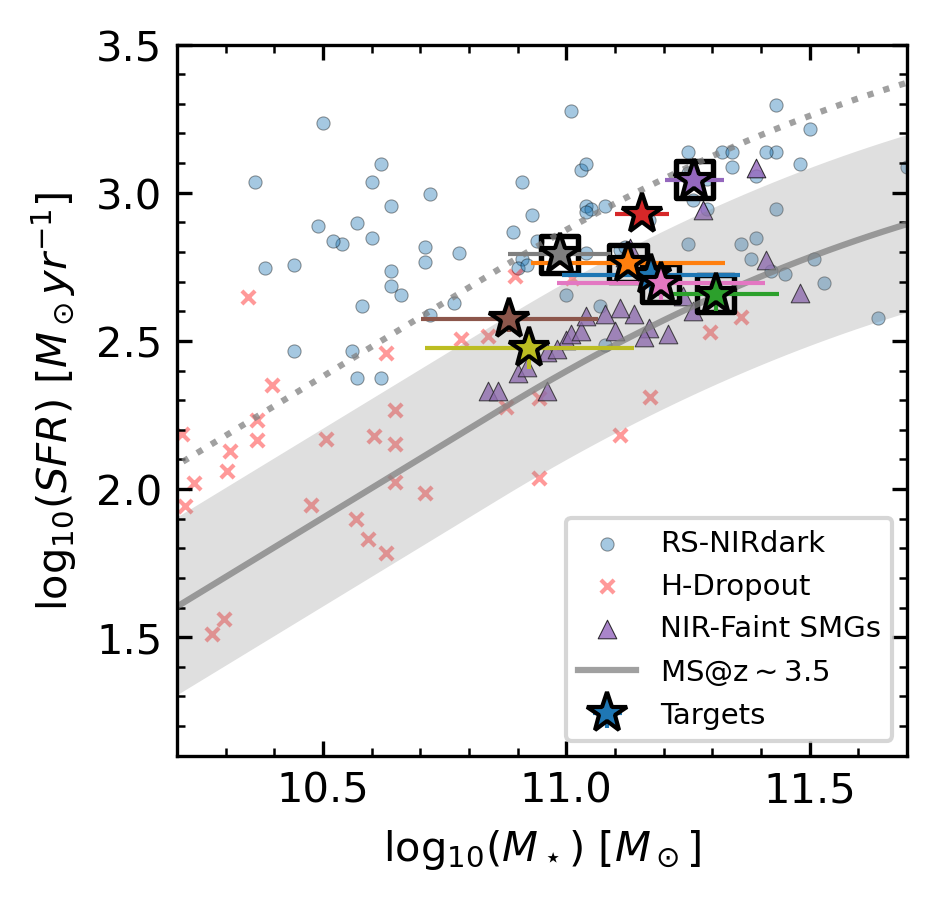}
    \caption{Comparison between the physical properties estimated through \textsc{Cigale} and the main sequence of star-forming galaxies by \citet{Schreiber_15} (solid gray line, with the shaded area being its $1\sigma=0.3$ dex its scatter. The targets are reported as colored stars, with the same color code employed in \Fig\ref{fig:comparison_zspec}. The gray dotted line reports our threshold for identifying the star-bursting galaxies (i.e. those with a SFR higher than three times that expected from a MS galaxy). For reference, we also report the location of the RS-NIRdark galaxies around $z\sim3.5$ studied in \citet{Gentile_24}, {the NIR-Faint SMGs by \citet{Smail_21}}, and the H-dropout galaxies by \citet{Wang_19}.}
    \label{fig:mainsequence}
\end{figure}

The results obtained through SED-fitting in \Sec\ref{sec:SEDFitting} allow us to confirm one of the main results established in \citet{Gentile_24}. In that study, we assessed that the RS-NIRdark galaxies selection produces a sample of star-bursting DSFGs. However, since this result was based on a SED-fitting where the redshift was a free parameter, the quantities estimated through this procedure were affected by significant uncertainties due to the several degeneracies between the shape of the SED and the redshift. By assuming the spec-\textit{z} measured through our ALMA observations, we can decrease the uncertainty on these quantities. First of all, we can see how the median properties estimated with the improved SED-fitting are broadly compatible with those estimated by \citet{Gentile_24} for the whole sample (see also the results discussed in \citealt{Talia_21} and \citealt{Behiri_23} regarding a smaller subset of the same sample). We underline – however – that since the galaxies in the proposal were selected among those with at least one secure detection in the FIR/(sub)mm regime, we do not expect the median properties of our sample to be necessarily similar to those of the whole sample of RS-NIRdark. A second interesting comparison between this study and that by \citet{Gentile_24} resides in the comparison between the SFR and stellar mass computed through the new SED-fitting and the main sequence of the star-forming galaxies. As visible in \Fig\ref{fig:mainsequence}, most of the targets are still located above the main sequence by \citet{Schreiber_15}, close to the star-bursting regime (i.e. galaxies with a SFR at least three times higher than what expected from a main sequence source with the same mass and in the same redshift bin).  We underline -- however -- that the location on the main sequence strongly relies on the estimated stellar mass, that is quite uncertain for our targets, being the rest-frame UV/optical continuum highly obscured by the dust. Nevertheless, the employment of the \textsc{Cigale} code, relying on the energy balance principle between the dust absorption and emission allows us to obtain some indirect constrain on the dust extinction from the infrared and radio coverage. More stringent constraints on the stellar mass will be provided -- for most of the galaxies in the whole sample of RS-NIRdark galaxies in COSMOS -- by the deep NIR imaging provided by JWST as part of the COSMOS-Web survey (\citealt{Casey_22}, Gentile et al., in prep.). {\Fig\ref{fig:mainsequence} also reports, for comparison, the location of other populations of \textit{"dark"} DSFGs in the stellar mass vs SFR plane: the H-dropouts by \citet{Wang_19} and the NIRfaint SMGs by \citet{Smail_21}. While the overlap between our RS-NIRdark galaxies and the H-dropouts has already been studied in \citet{Gentile_24}, the availability of the new (sub)mm data allows us to study more quantitatively how many of our sources would be selected with the criteria employed by \citet{Smail_21}. We remind that these sources are part of a sample of 707 SMGs collected in the Ultra Deep Survey by re-imaging with ALMA (in band 7, i.e. at a representative frequency of 870 $\mu$m; \citealt{Stach_19}) a sample of galaxies initially detected in the 850 $\mu$m maps produced with the SCUBA2 camera \citep{Geach_17}. The sample studied by \citet{Smail_21} contains all\footnote{The original sample of NIRfaint sources by \citet{Smail_21} would contain other 50 sources that are not included in that study due to a contaminated photometry at optical/NIR frequencies. Here, we suppose that the 30 galaxies analysed by \citet{Smail_21} are representative of the full sample} the sources with $Ks>25.3$ mag (at 5 $\sigma$). Since the $Ks$ limiting magnitude in the UDS and in COSMOS are similar, we can study the overlap between the two populations by comparing the (sub)mm flux at 850 $\mu$m. For our sources, only RSN-84 has an ALMA flux at the same frequency (see \Tab\ref{tab:cont}). Three of the other galaxies have analogous fluxes coming from the deblending of the SCUBA2 maps \citep{Jin_18}, while the other have only upper limits (S/N$<3\sigma$, since the uncertainties in the SuperDeblended catalog also account for the deblending procedure. For these sources, we employ the best-fitting fluxes at 850 $\mu$m computed with \textsc{Cigale}. On the other side, the sources in \citet{Smail_21} were initially selected for having a SNR$>4\sigma$ (equivalent to $S_{\rm 850\mu m}\ge3.8$ mJy) in the original SCUBA2 maps. { However, the higher resolution achieved by ALMA in the \citet{Stach_19} follow-up allowed the discovery of multiple fainter sources contributing to the original ones detected by SCUBA2, up to $S_{\rm 850\mu m}>1$ mJy and with a median value of $3.8\pm0.3$ mJy. Therefore, even though 4 out of 9 sources in our sample would not have been selected by the original SCUBA2 survey, being too faint for the limited sensitivity of that instrument (see \Tab\ref{tab:cont}), all of them would have been detected by the deeper ALMA follow-up.} }

\begin{figure}
    \centering
    \includegraphics[width=\columnwidth]{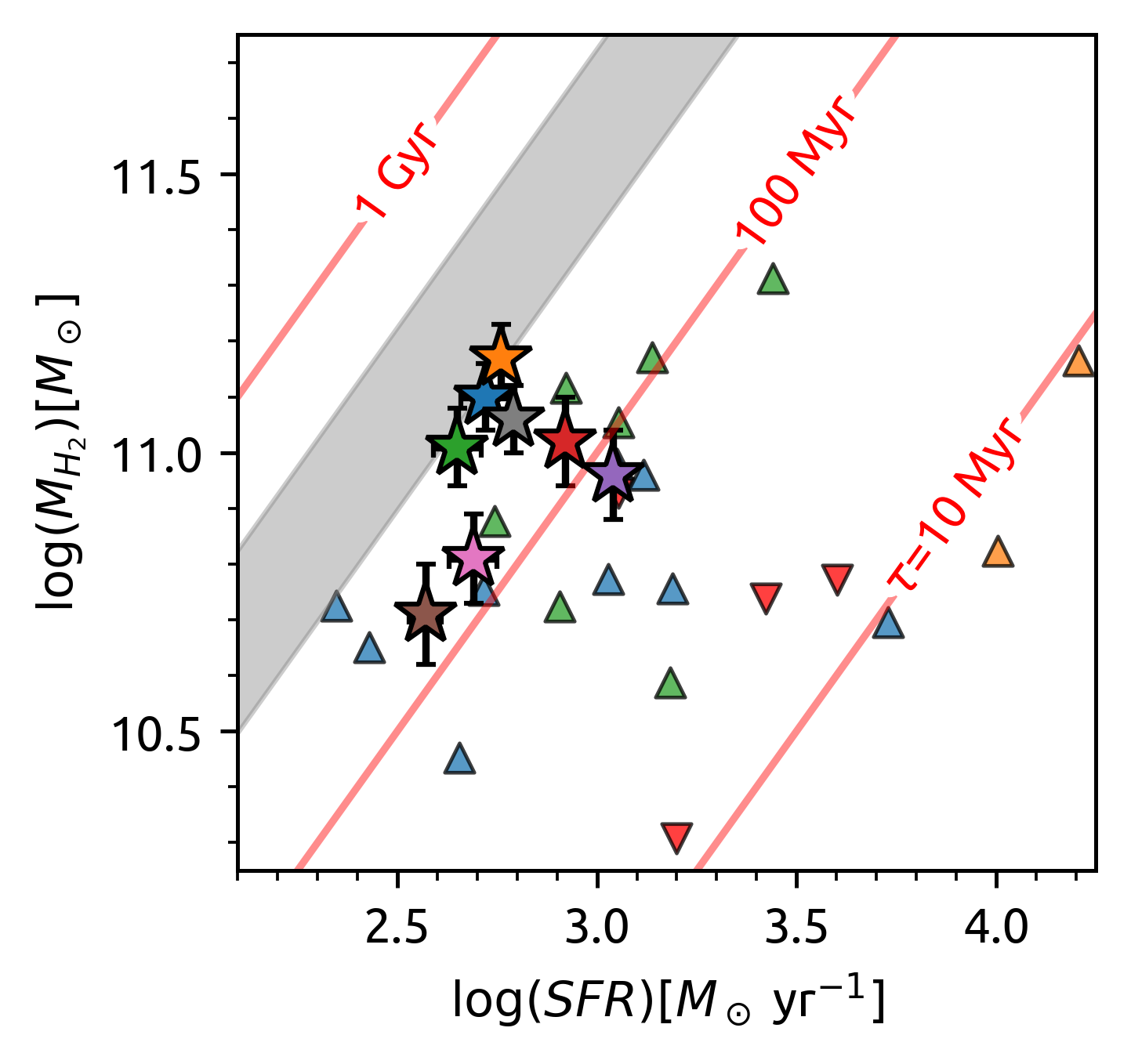}
    \caption{Depletion time and star formation rate for the targets analysed in this study. The RS-NIRdark galaxies are reported as colored stars, following the same color code as in \Fig\ref{fig:comparison_zspec}. The colored triangles are other populations of SMGs, namely those collected by \citet{Bothwell_17}, \citet{Canameras_18}, and \citet{Walter_11}, reported in blue, orange, and green, respectively. We also report some confirmed QSOs from the same studies as reversed red triangles. The gray shaded area reports the depletion time expected from main sequence galaxies at $z\sim3.5$ following the relation $\tau_{\rm D}=1.5(1+z)^{\alpha}$ found by \citet{Saintonge_13}, with $\alpha$ spanning from -1.0 \citep{Davè_12} to -1.5 \citep{Magnelli_13}, rescaled to a \citet{Chabrier_03} IMF.}
    \label{fig:depletion}
\end{figure}

\subsection{Gas mass and depletion time}
\label{sec:gas_evolution}

\begin{figure}
    \centering
    \includegraphics[width=\columnwidth]{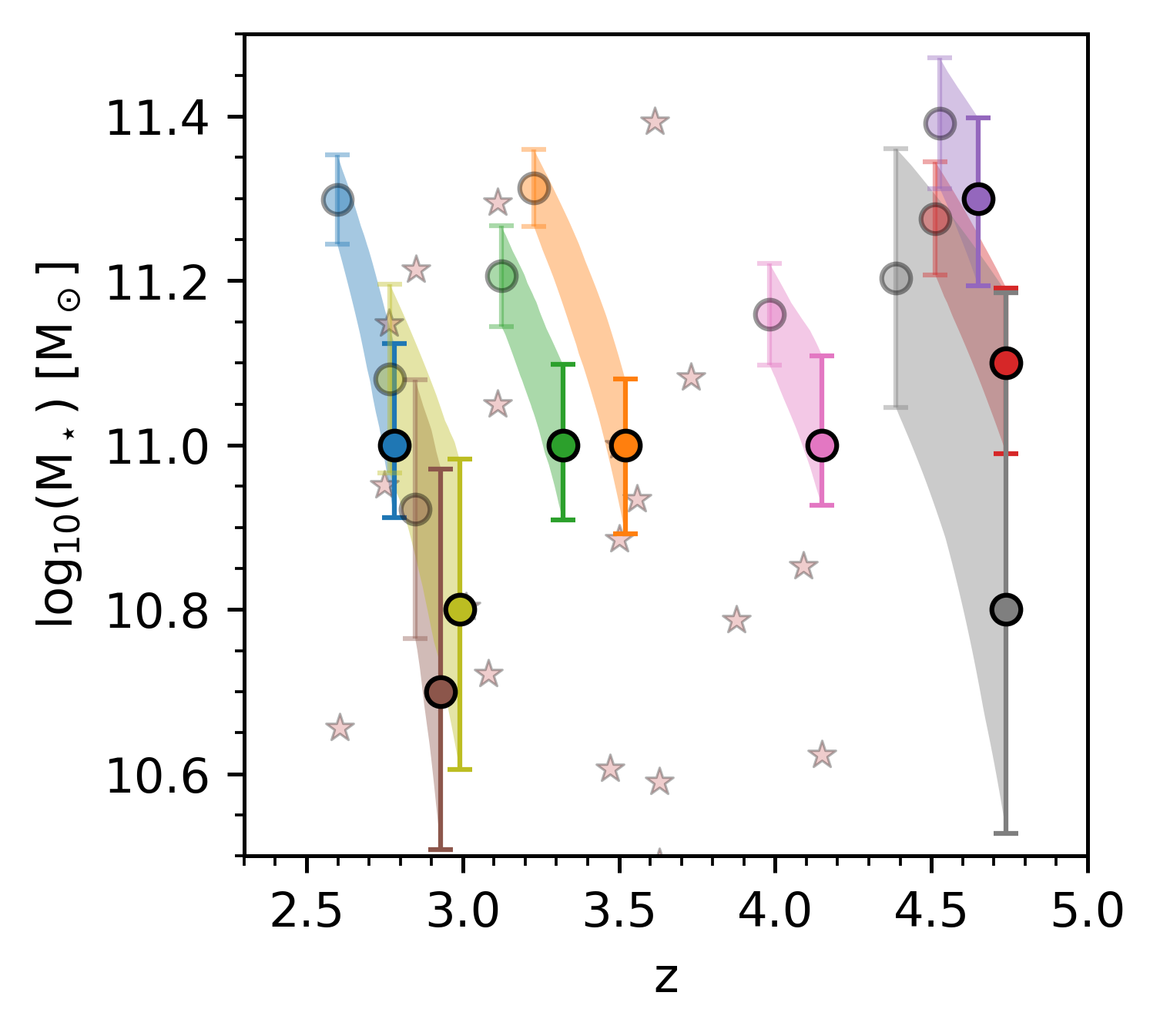}
    \caption{Possible evolutionary paths of our targets, assuming a simple evolutionary model with a constant star formation and a final stage where all the molecular gas has been transformed into stars. The uncertainties are considered through a MonteCarlo integration. The color map is the same employed for \Fig\ref{fig:comparison_zspec}, For reference, the shaded red stars report the stellar mass and the redshifts for the massive and passive galaxies at $z\sim3$ discovered by \citet{Schreiber_18}.}
    \label{fig:evolution}
\end{figure}

Several studies highlighted how the [CI](1-0) and the CO(1-0) lines can be employed as good tracers of the molecular gas inside galaxies \citep[e.g.][]{Papadopoulos_04,Valentino_20CI,Gururajan_23}. For four galaxies in our sample (see \Tab\ref{tab:lines}), we observed the [CI](1-0) line: for them, we can directly estimate the molecular gas mass in our objects by employing the relation by \citet{Papadopoulos_04}:
\begin{equation}
\centering
\label{eq:ci}
M(H_2)^{[CI]}=1375.8\times10^{-12}\frac{D_L^2 I_{[CI](1-0)}}{(1+z)A_{10}Q_{10}X_{CI}}[M_\odot]
\end{equation}
where $D_L$ is the luminosity distance of our target expressed in Mpc, $I_{\rm CO}$ is the integrated line flux, and $A_{10}=0.793\times10^{-7}$ s$^{-1}$ is the Einstein coefficient. $Q_{10}$ and $X_{10}$ are, respectively, the [CI] excitation factor and the [CI]/H$_2$ abundance ratio. For these quantities, we employ literature-standard values of $X_{CI}=3\times10^{-5}$ and $Q_{10}=0.6$ \citep{Papadopoulos_04,Bothwell_17}. Through \Eq\ref{eq:ci}, we estimate the gas masses for RSN-84, RSN-121, RSN-235, and RSN-361 reported in \Tab\ref{tab:properties}. For all the other targets, in which we did not detect the [CI](1-0) line, we derive the gas mass from the CO(1-0) line through the relation
\begin{equation}
\centering
\label{eq:alpha_co}
    M(H_2)^{\rm CO}=3.25\times10^7\alpha_{\rm CO} I_{\rm CO} \nu_{\rm obs}^{-2}D_L^2(1+z)^{-3}
\end{equation}
(see e.g. \citealt{Bolatto_13} and references therein). It is well known that the value of $\alpha_{\rm CO}$ is highly uncertain and strongly dependent on the specific property of each galaxy. In this study, we choose a literature value of 0.8 $M_\odot$ [K km s$^{-1}$ pc$^{-2}$]$^{-1}$ usually employed for star-bursting galaxies (see e.g. \citealt{Bolatto_13,Gururajan_23}). In order to use \Eq\ref{eq:alpha_co}, we need to rescale the measured fluxes of our CO lines to the CO(1-0) transition by assuming a CO-Spectral Line Energy Distribution (SLED). Several studies highlighted how the CO-SLED of galaxies is strongly affected by the presence of AGN \citep[see e.g.][]{Vallini_19} and evolves with redshift \citep[e.g.][]{Boogard_20}. Since the previous test performed on our targets by \citet{Talia_21} and \citet{Gentile_24} (together with the null AGN fraction obtained through SED-fitting with \textsc{Cigale}; \Sec\ref{sec:SEDFitting}) excluded the presence of strong nuclear activity, we choose the CO-SLED obtained by \citet{Bothwell_13} for a sample of DSFGs in the redshift range $1.2-4.1$ (i.e. compatible with the spec-\textit{z} of our targets). More in detail, for our targets, we employ $R_{31}=2.3\pm0.3$, $R_{41}=3.0\pm0.4$, and $R_{51}=3.8\pm0.7$, where $R_{n1}$ is the ratio between the {integrated line flux in the} nth CO transition and the CO(1-0). The gas mass obtained through these relations are reported in \Tab\ref{tab:properties}. For the galaxies with both a CO and a [CI] line, we report both the estimation of the gas mass. However, since the values estimated from the [CI](1-0) line rely on less assumption than those based on the CO lines ({i.e. they do not rely on the assumed CO-SLED, albeit they still depend on the conversion factor between the [CI](1-0), as uncertain as the $\alpha_{\rm CO}$ value}), we employ these gas masses in the following analyses. The information about the gas content of our galaxies can be combined with the SFR estimated in \Sec\ref{sec:SEDFitting} and reported in \Tab\ref{tab:properties} to assess the depletion time of our galaxies. This quantity is defined as the amount of time in which each object would transform its whole gas mass in stars assuming a constant SFR:
\begin{equation}
\centering
    \tau_{\rm D}=\frac{M(H_2)}{\rm SFR}
\end{equation}
For our galaxies, we obtain the depletion times reported in the last column of \Tab\ref{tab:properties}. {The} galaxies in our sample have values in the range $80-300$ Myr. These quantities can be compared with other population of galaxies in the current literature, as shown in \Fig\ref{fig:depletion}. More in detail, we can compare the depletion time of our targets with that expected from main sequence galaxies. \citet{Saintonge_13} reported a $\tau_{\rm D}$ evolving with redshift as $\tau_{\rm D}=1.5(1+z)^{\alpha}$, with several collaborations finding different values for the exponent, spanning from $\alpha=-1.0$ \citep{Davè_12} to $\alpha=-1.5$ \citep{Magnelli_13}. Our targets result to have a shorter depletion time than main sequence galaxies. This result represents a further confirmation (independent from the more uncertain stellar mass) of the star-bursting nature of our sources. Similarly, we can compare the gas mass and the SFR of our RS-NIRdark galaxies with several SMGs in the current literature (namely, those analysed by \citealt{Walter_11}, \citealt{Bothwell_17} and \citealt{Canameras_18}). For all these sources, we retrieve the infrared luminosity and the [CI](1-0) line fluxes from the study by \citet{Valentino_20CI}. Following the relations by \citet{Kennicutt_12} and \citet{Papadopoulos_04}, we estimated the SFR and gas mass in a consistent way with those derived from our targets. We obtain that our RS-NIRdark galaxies are -- on average -- {more gas rich than the SMGs analyzed in those studies, in the low-SFR tail of their distribution, and -- therefore -- with a longer depletion time}.

\subsection{Possible evolutionary path}
\label{sec:evolution}
The gas mass and depletion times estimated in \Sec\ref{sec:gas_evolution} allows us to forecast a possible evolutionary path for our sources. We employ the same simplistic model used to define the depletion time: we assume that the SFR will stay constant inside our sources until all the gas mass is converted into stars. Clearly, this model does not account for any quenching mechanism (e.g. those due to possible AGN feedback; see e.g. \citealt{Fabian_12}) or gas accretion from the inter-galactic medium \citep[e.g.][]{Sancisi_08}. With this model, we can assume that our galaxies will evolve from an initial state characterized by $z_0=z_{\rm spec}$ and $M_{\star,0}=M_\star$ to a final state with $z_f=z_{\rm spec} - \Delta z$ and $M_{\star, f}=M_\star + M_{\rm H_2}$, where $\Delta z$ is the difference in redshift elapsed during the depletion time. The results of this simplistic model applied to our targets is shown in \Fig\ref{fig:evolution}, where we consider all the uncertainties on the involved quantities through a MonteCarlo integration. We can compare the final stage of our galaxies with the redshift and stellar mass of the massive and passive galaxies discovered by \citet{Schreiber_18} at $z\sim3$, we can notice a significant overlap between the two populations. This result, combined with the number densities estimated by \citet{Talia_21}, \citet{Behiri_23} and \citet{Gentile_24} suggest that the RS-NIRdark galaxies could represent a significant fraction of the progenitors of the massive and passive galaxies at $z\sim3$. In the broader context of galaxy evolution studies, this result confirms the idea that the dust-obscured galaxies play a significant role in the evolution of the most massive galaxies in the Universe \citep[see e.g.][]{casey_14,Toft_14,Valentino_20}. In addition, the components with different velocities detected in our targets in \Sec\ref{sec:kinematics} could support two possible evolutionary scenarios for our galaxies. In one case, they could be the signature of a significant fraction of major mergers within our sample. This result would confirm the so-called \textit{merger-driven} scenario for the formation of massive galaxies, where these objects are formed through a series of major mergers (see e.g. \citealt{Hopkins_18a,Hopkins_18b}). On the other case, the double component in our galaxies could be due to the presence of a rotating disk. In this case, this result would support the so-called \textit{in-situ} scenario, where the build-up of massive galaxies occurs via the rapid compaction of a gaseous outer disk triggering a huge burst of star formation, and via the subsequent stellar and AGN feedback processes quenching it within a relatively short timescale ($<1$ Gyr; see e.g \citealt{Lapi_14}, \citealt{Lapi_14}, \citealt{Pantoni_19}). The possibility of shedding new light on these alternative scenario increases the scientific interest in the RS-NIRdark galaxies and the need for high-resolution follow-up for these sources.

\section{Summary}
\label{sec:conclusions}
{In this paper, we presented the first spectroscopic follow-up at mm wavelengths for a pilot sample of 9 RS-NIRdark galaxies in the COSMOS field. {These sources were initially selected by \citet{Talia_21} as radio-detected sources at 3 GHz lacking an optical/NIR counterpart in the COSMOS2015 catalog, even though three of them (see \Tab\ref{tab:coords}) were subsequently detected in the deeper COSMOS2020.} Through a new series of ALMA observations, we managed to identify at least one bright emission line in all the targets and two lines in five out of nine objects. {From the analysis of the new ALMA data, we obtain the following results:}
\begin{itemize}
    \item Modeling the detected lines as CO/[CI] transitions, we estimated a spectroscopic redshift for all the galaxies in our sample. These values are in good agreement with those estimated through SED-fitting in \citet{Gentile_24}.
    \item The new availability of a spectroscopic redshift allowed us to decrease the degeneracies in the SED-fitting procedure. This improved SED-fitting confirmed one of the main results concerning the RS-NIRdark galaxies: they represent a population of highly-obscured ($A_{\rm v}\sim4$), massive ($M_\star\sim10^{11}M_\odot$), and star-forming (SFR$\sim 500$ M$_\odot$ yr$^{-1}$) galaxies.
    \item The same improved SED-fitting, together with pre-existing data, allowed us to estimate the flux of our sources at 870 $\mu$m. This value is analogous to that reported by \citet{Smail_21} for their sample of NIR-faint SMGs, suggesting that our radio selection is able to provide a similar population of DSFGs as those obtained from (sub)mm selections. A similar conclusion is reached by observing the overlap between the physical properties (stellar mass and SFR) computed through SED-fitting for the two samples.
    \item Thanks to the good spectral resolution of the new ALMA observations, we assessed the presence of a high fraction ($\sim 55$\%) of double-peaked profiles in the lines detected in our targets. We explained this result with the possible presence of a rotating structure within our galaxies or with the presence of major mergers in our sample. High-resolution follow-up with ALMA or JWST are needed to discriminate between these two possibilities.
    \item Thanks to the CO/[CI] lines detected in our targets, we estimated the gas mass and the depletion time of our galaxies. These results allowed us to forecast a possible evolutionary path for our objects, strongly suggesting that the RS-NIRdark galaxies could represent a significant fraction of the progenitors of the massive and passive galaxies at $z\sim 3$ and excellent probes to test galaxy evolution models. 
\end{itemize}

}
\section*{Acknowledgments}
{We thank the anonymous referee for their comments that improved the initial version of this paper}. FG thanks Gayathri Gururajan for the insightful conversations about the gas tracers, Anton Koekemoer {and Ian Smail} for the careful reading of the manuscript and for the useful comments, and Roberto Decarli for the precious feedback on an early version of this paper. FG, MT, AL, MM and AC acknowledge the support from grant PRIN MIUR 2017-20173ML3WW\_001. ‘Opening the ALMA window on the cosmic evolution of gas, stars, and supermassive black holes’. ID acknowledges support from INAF Minigrant "Harnessing the power of VLBA towards a census of AGN and star formation at high redshift". MV acknowledges financial support from the Inter-University Institute for Data Intensive Astronomy (IDIA), a partnership of the University of Cape Town, the University of Pretoria and the University of the Western Cape, and from the South African Department of Science and Innovation's National Research Foundation under the ISARP RADIOSKY2020 and RADIOMAP Joint Research Schemes (DSI-NRF Grant Numbers 113121 and 150551) and the SRUG HIPPO Projects (DSI-NRF Grant Numbers 121291 and SRUG22031677). This paper makes use of the following ALMA data: ADS/JAO.ALMA\#2021.1.01467.S. ALMA is a partnership of ESO (representing its member states), NSF (USA) and NINS (Japan), together with NRC (Canada), MOST and ASIAA (Taiwan), and KASI (Republic of Korea), in cooperation with the Republic of Chile. The Joint ALMA Observatory is operated by ESO, AUI/NRAO and NAOJ

\bibliographystyle{aa}
\bibliography{sample631}

\begin{appendix}

\begin{figure*}
\centering
    \includegraphics[width=\textwidth]{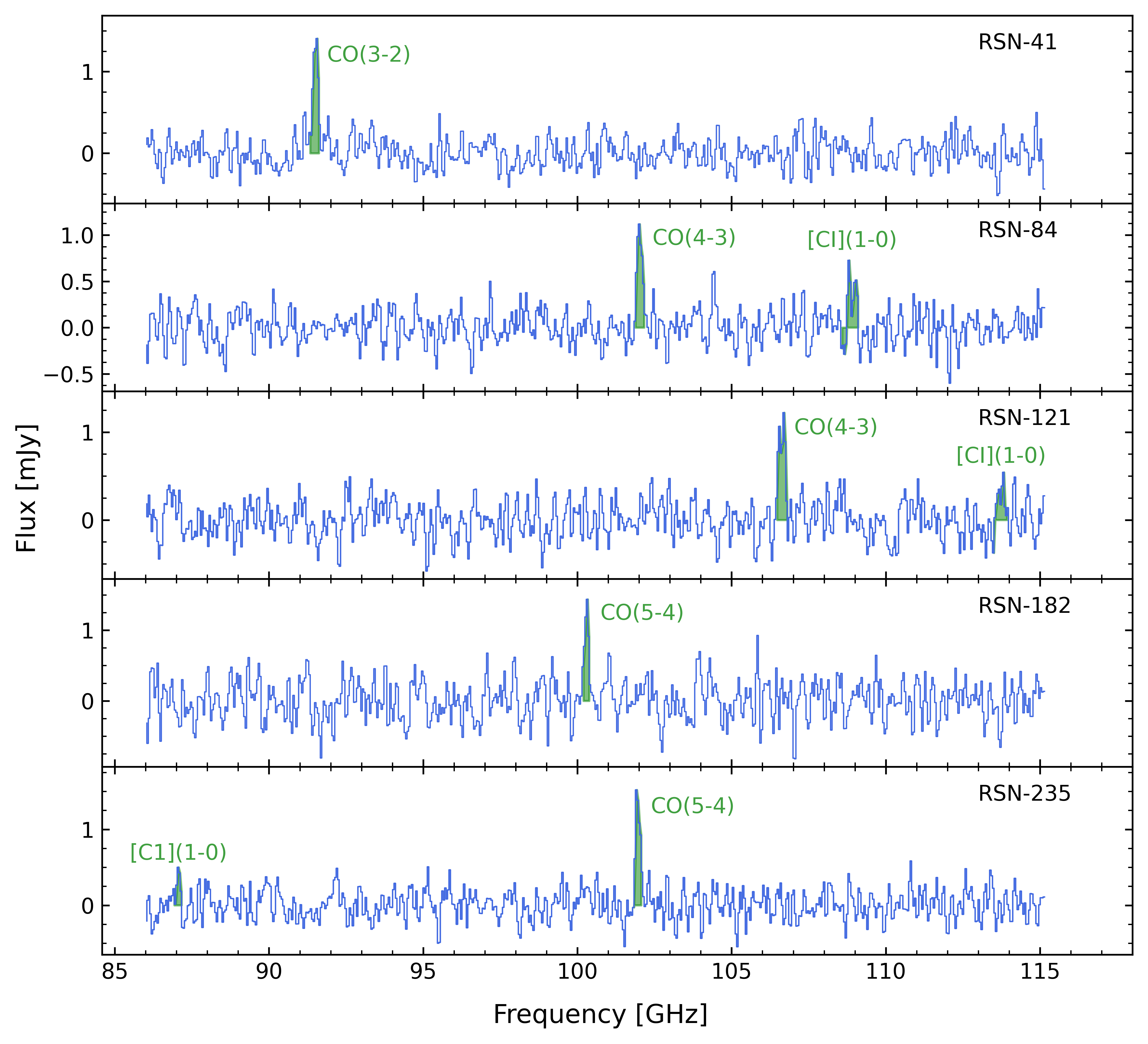}
    \caption{Full ALMA spectra for the nine targets analysed in this study. For each source, we highlight the detected lines and report for each of them our modeling as CO/[CI] transitions.}
    \label{fig:spectra}
\end{figure*}

\begin{figure*}
\centering
\renewcommand{\thefigure}{\arabic{figure}}
\addtocounter{figure}{-1}
\includegraphics[width=\textwidth]{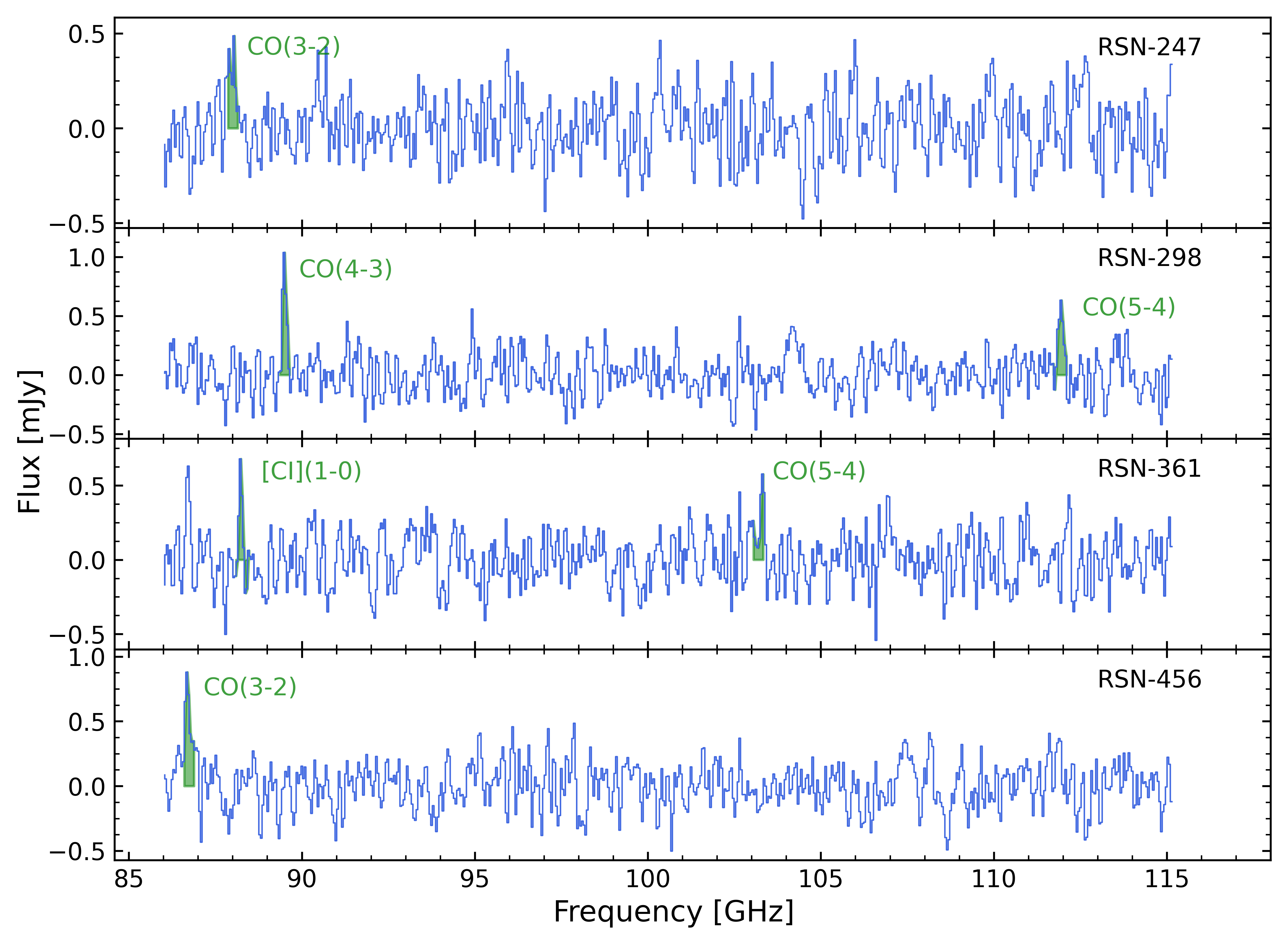}
\caption{\textit{(Continue)}}
\end{figure*}

\begin{figure*}
    \centering
     \includegraphics[scale=1]{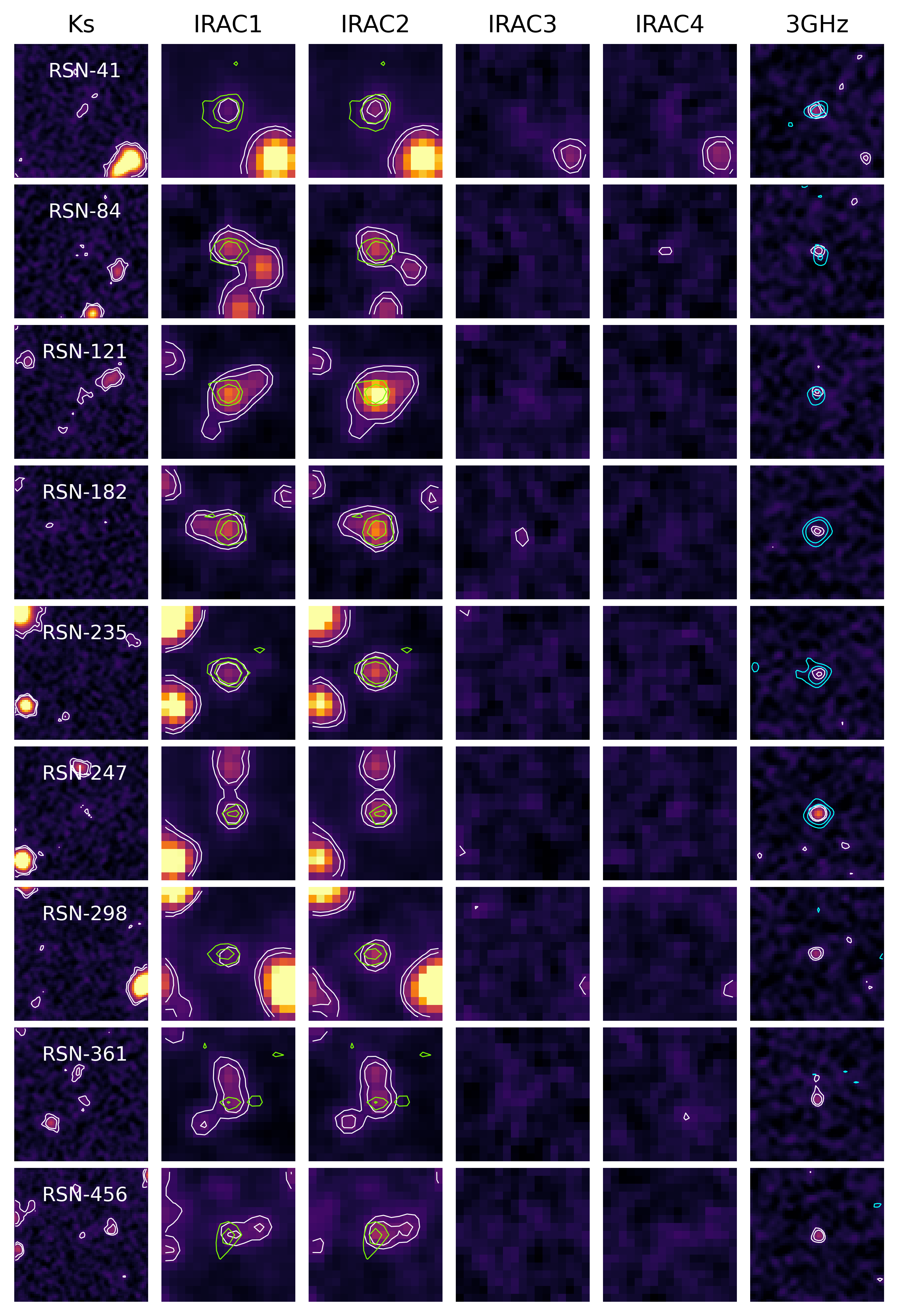}
    \caption{{Cutouts (10'' $\times$ 10'') in the main NIR-to-MIR bands for the 9 targets analyzed in this work. On the images, we report the 3 and 5 $\sigma$ contours. {Moreover, for each galaxy, we overplot in green the 3 and 5 $\sigma$ contour of the brightest line on the IRAC ch1 and ch2 images. Similarly, the same contours of the continuum emission are overplotted in cyan on the 3 GHz images. }}}
    \label{fig:cutouts}
\end{figure*}

\end{appendix}

\end{document}